\documentclass[12pt]{article}
\usepackage[graphicx]{realboxes}
\usepackage[textwidth=8em,textsize=small]{todonotes}
\usepackage{amsmath,mathtools}
\usepackage{natbib,float}
\usepackage{bm}
\usepackage{amsfonts}
\usepackage{booktabs}
\usepackage[colorlinks,citecolor=blue,urlcolor=blue,linkcolor = blue]{hyperref}
\usepackage{threeparttable}
\usepackage{multirow}
\usepackage{pdflscape}
\usepackage{etoolbox}
\newtheorem{remark}{Remark}

\DeclareMathOperator*{\argmax}{arg\,max}
\title{Do financial regulators act in the public's interest? A Bayesian latent class estimation framework for assessing regulatory\\ responses to banking crises 
}
\author{Padma Sharma\footnote{The views expressed here are the opinions of the authors and should not be attributed to the Federal Reserve Bank of Kansas City or the Federal Reserve System.}~~and Trambak Banerjee$^1$\\
	*Federal Reserve Bank of Kansas City and $^1$University of Kansas}
\date{}
\begin{document}	
\maketitle
\begin{abstract}
\noindent When banks fail amidst financial crises, the public criticizes regulators for bailing out or liquidating specific banks, especially the ones that gain attention due to their size or dominance. A comprehensive assessment of regulators, however, requires examining all their decisions, and not just specific ones, against the regulator's dual objective of preserving financial stability while discouraging moral hazard. In this article, we develop a Bayesian latent class estimation framework to assess regulators on these competing objectives and evaluate their decisions against resolution rules recommended by theoretical studies of bank behavior designed to contain moral hazard incentives. The proposed estimation framework addresses the unobserved heterogeneity underlying regulator’s decisions in resolving failed banks and provides a disciplined statistical approach for inferring if they acted in the public interest. Our results reveal that during the crises of 1980’s, the U.S. banking regulator’s resolution decisions were consistent with recommended decision rules, while the U.S. savings and loans (S\&L) regulator, which ultimately faced insolvency in 1989 at a cost of \$132 billion to the taxpayer, had deviated from such recommendations. Timely interventions based on this evaluation could have redressed the S\&L regulator's decision structure and prevented losses to taxpayers.
\end{abstract}
\noindent%
\section{Introduction}
\label{sec:intro}
During financial crises when a large number of banks fail, actions of financial regulators receive substantial public scrutiny. The global financial crisis of 2008 is one such recent example that led to widespread bank failures, reviving debate over how financial regulators might preserve immediate financial stability while also safeguarding against future moral hazard. {This debate was reignited more recently in 2023 as U.S. policymakers  protected funds
of uninsured depositors of Silicon Valley Bank and Signature Bank when the two banks were on the brink of failure.} During a financial crisis, regulators determine and administer the bailout, sale or liquidation of failed banks. Regulators bail out banks when they place greater emphasis on preserving financial stability and liquidate institutions when they are more attentive to the curtailment of moral hazard incentives.
Critical assessments of these actions are essential to ensuring regulators balance the competing concerns in a manner that serves the public interest. However, the public typically criticizes specific regulatory decisions instead of evaluating how closely their overall decision framework serves the public interest. Individuals disfavor bailouts because they represent transfers from taxpayers to shareholders. {For instance, \cite{bernanke2019firefighting} note: \textit{``The firms we rescued were usually not gracious about the terms of their rescues, while the overwhelming sentiment among the public was that they shouldn't have been rescued at all."}} The public also criticizes bank liquidations because they are costly for depositors and loan customers whose banking relationships are disrupted \citep{isaac2010senseless}. How can the public and their elected representatives comprehensively assess the actions of regulators against their competing objectives of preserving financial stability and restraining moral hazard? 

Theoretical studies provide a benchmark for evaluating regulators by deriving decision rules that resolve the trade-off between the two objectives in a manner that preserves public interest. Broadly, these studies recommend using distinct decision rules in response to different economic and industry-wide conditions that accompanied bank failures (see for example \cite{cordella2003bank,acharya2007cash,deyoung2013theory} and the references therein). For instance, \cite{cordella2003bank} recommend that regulators adopt a forbearing rule that allows for the bailout of banks that failed in the midst of economic distress, and a more stringent rule that typically involves the liquidation of banks that failed in normal economic conditions. {Other theoretical studies similarly consider two disparate states in which bank failures occur, and derive decision rules for each of the two states.} However, assessing regulators based on the extent to which they conformed to these theoretical recommendations is not straightforward. Although we can measure economic distress through observable indicators like the unemployment rate, and growth in output, the threshold regulators may have used to distinguish between regions of high and low distress is unobserved. {Consider two banks in our sample that failed in 1988, one bank in Kansas and another in Kentucky, where unemployment rates were 5\% and 8.5\% respectively. It is not apparent from the data and historical records, whether regulators considered unemployment in Kansas and Kentucky to be high enough to classify both states as regions of high distress, whether they deemed only Kentucky, or neither state to be undergoing economic distress. Thereby, regulators' evaluation of economic conditions and assignment of banks into potentially distinct or common rules, which is a criterion for evaluating whether they acted in the public interest, is unobserved.}

In this paper, we develop a novel Bayesian latent class estimation framework to compare regulators' decisions on failed banks against theoretical recommendations. Our framework directly addresses the empirical challenge in performing this comparison--theoretical studies recommend adopting distinct rules that vary with underlying economic and banking conditions, but regulators' true mechanism for assigning distinct rules to different groups of banks is unobserved. The latent class structure disentangles the unobserved nature of regulators' assignment process by incorporating the researcher's uncertainty on this mechanism into the model structure. This model assigns banks into classes with probability, rather than certainty and every latent class corresponds to a segment of banks that receives a distinct decision rule.  In estimating this model, we have developed enhancements to the MCMC sampling process. We construct a  collapsed Gibbs sampler algorithm, which outperforms a standard Gibbs sampler in terms of autocorrelations across posterior draws and thereby enhances inference using credible intervals and model selection using Bayes factors. 
\begin{figure}[!t]
	\centering
	\includegraphics[width=0.75\textwidth]{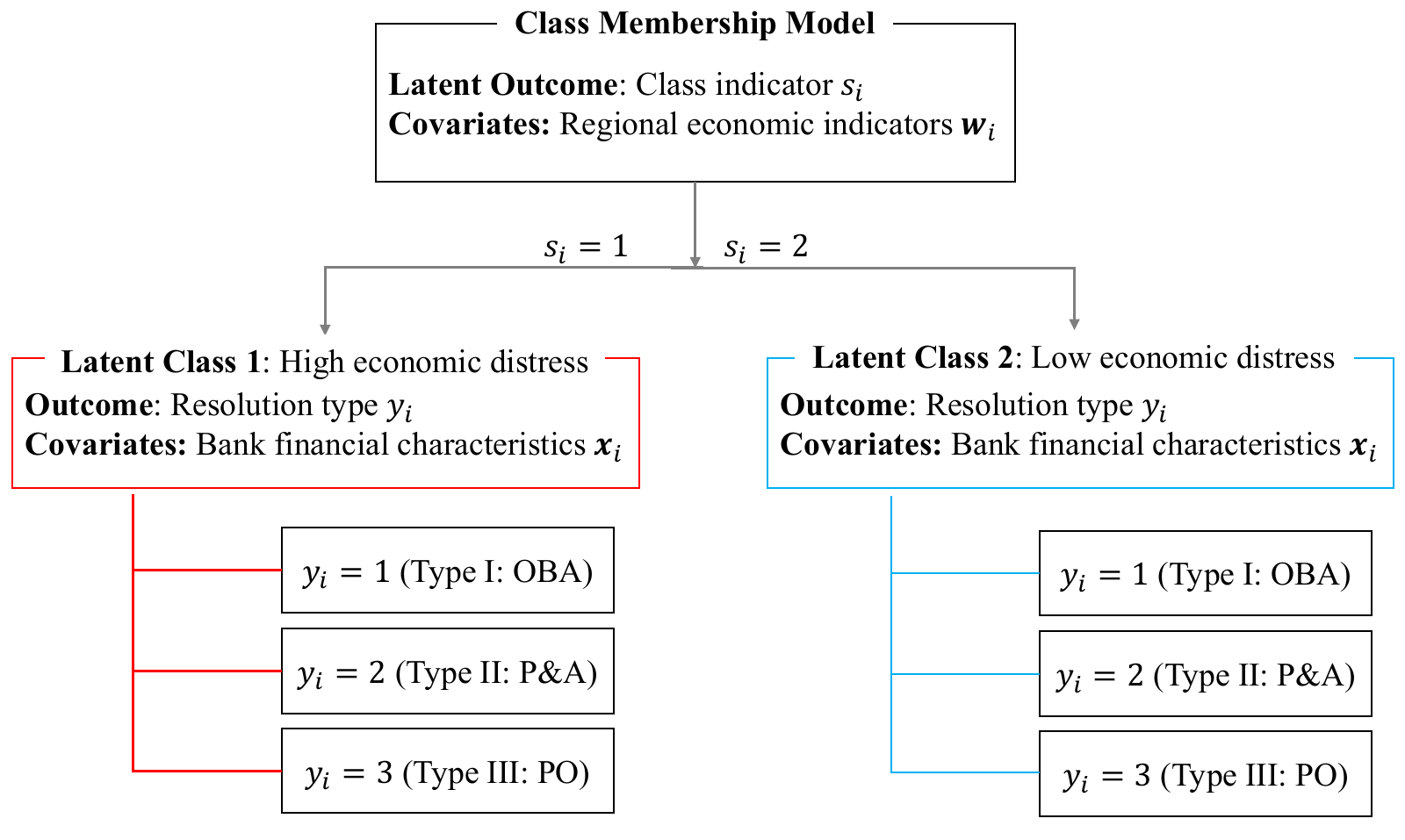}	
	\caption{Structure and components of the latent class model for bank resolutions. Here  index $i$ refers to a representative failed bank (or a S\&L as defined in Section \ref{sec:resoultion}) $i$, {$\bm w_i$ is a $p-$dimensional vector of economic indicators pertaining to the state and the county in which bank $i$ operates and $\bm x_i$ is a $q-$dimensional vector of bank $i$'s characteristics that are representative of its financial health and importance}. 
 }
 \label{fig:flowFinal}
\end{figure}

Figure \ref{fig:flowFinal} provides a pictorial representation of our latent class model {for a representative failed bank $i$}. A precise formulation and formal mathematical statements are deferred until Section \ref{sec:model}. The proposed model consists of a hierarchical structure in which the first layer is a probabilistic class-membership model that assigns the failed bank $i$ to {one of two} classes that
correspond to the two states of nature considered in theoretical studies, such as high or low underlying economic distress. {To measure economic distress, the covariates $\bm w_i$ used in this model include economic indicators 
pertaining to the state and the county in which bank $i$ operates. Note that the outcome variable $s_{i}$ in this class membership model is latent since the true assignment of the failed banks into distinct classes by regulators is not observable as the data record only the regulator's final resolution decisions but not the rationale that motivated each decision.} Conditional on class membership, the second layer specifies the relationships between the resolution type $y_{i}$, namely assistance (Type I), sale (Type II) or liquidation (Type III) of bank $i$, and bank-specific financial characteristics $\bm x_i$, such as size and asset quality. These relationships are homogeneous within and heterogeneous across the latent classes when the classes are statistically different from each other.

{We use our estimation framework to assess two regulators from the U.S. banking industry, the Federal Deposit Insurance
Corporation (FDIC), and the Federal Savings and Loans Insurance Corporation (FSLIC), during the banking crises of the 1980's when more banks failed relative to any other period since the Great Depression. The FDIC is the regulatory authority for commercial banks and the FSLIC was the counterpart for Savings and Loans (S\&L) institutions until its failure in 1989. Our analyses uncover specific weaknesses in the FSLIC’s decision structure and
the FDIC’s relative strengths that are likely to have contributed to the former’s failure in 1989
and the latter’s continued survival. For instance, we find that the FDIC, in line with theoretical recommendations, provided bailouts
to banks that failed amid macroeconomic and banking industry distress and withheld such assistance for failures in normal economic conditions. The FSLIC, in contrast, deviated from this theoretical norm as it did not distinguish across S\&L's by the level of economic distress accompanying their failure. Moreover, while political support for the banking industry played a limited role in the FDIC's decisions, our analyses reveal that the FSLIC assigned assistance to S\&L's that likely received a higher degree of political support and failed amid lower economic distress. 
}

{Our proposed framework is widely applicable in settings where interest centers on modeling covariate-outcome relationships  that vary across classes, when true class assignment is unobserved but can be modeled using an additional set of covariates. This approach expressly models the researcher's uncertainty by assigning units into classes with probability. In addition, the model incorporates information from multiple covariates in estimating the probability of belonging to a latent class. Alternative modeling strategies entail determining classes by setting arbitrary thresholds based on some percentile of class-membership covariates and deterministically assigning units into classes. The latent class model forgoes the implicit assumptions about class assignment in such approaches and estimates the classes in a data-driven manner.  Beyond our application in assessing financial regulators, this modeling approach can be utilized to determine latent classes of consumers, patients, financial instruments such as stocks and bonds and evaluate heterogeneity in their choices, outcomes or valuations.  }

The rest of the article is organized as follows. {In Section \ref{sec:resoultion} we collect background information on the two U.S. banking regulators, review the recommended decision rules from theoretical studies and summarize the main contributions of our work}. In Section \ref{sec:data} we discuss the data used in this article. Section \ref{sec:model} presents the proposed Bayesian latent class estimation framework for evaluating the regulatory decision rules for resolving bank failures against recommended decision rules from theoretical studies.  Sections \ref{sec:results_fdic} and \ref{sec:results_fslic}, respectively, discuss the results of our analysis pertaining to the resolution decisions of the FDIC and the FSLIC. The article concludes with a discussion in Section \ref{sec:discuss}. Additional technical details, numerical results and background information are relegated to the supplementary materials. {All MATLAB and R codes for reproducing the empirical analyses in this paper are available at \texttt{https://github.com/trambakbanerjee/latent\_class\_bank\_resolution}.}
\section{Regulators of the U.S. banking sector, their resolution methods and the crises of 1980's}\label{sec:resoultion}
We assess regulators from two sub-sectors of the U.S. banking industry, commercial banks and Savings and Loans (S\&L) institutions, during their simultaneous crises of the 1980's. {\cite{fdiccrisis} define a S\&L institution as a financial institution 
that is chiefly organized and primarily operates to promote savings and home mortgage lending rather than commercial lending.
} The Federal Deposit Insurance Corporation (FDIC) serves as the regulatory authority for commercial banks while the Federal Savings and Loans Insurance Corporation (FSLIC) was the counterpart for S\&L's until its failure in 1989. The two regulators are comparable on account of the fundamental similarities across banks and S\&L's in that both institutions offer loans and deposits, undertake maturity transformation, monitor information and offer liquidity and payments services \citep{freixas2008microeconomics}.
\begin{figure}[!t]
	\centering
	\includegraphics[width=0.65\linewidth]{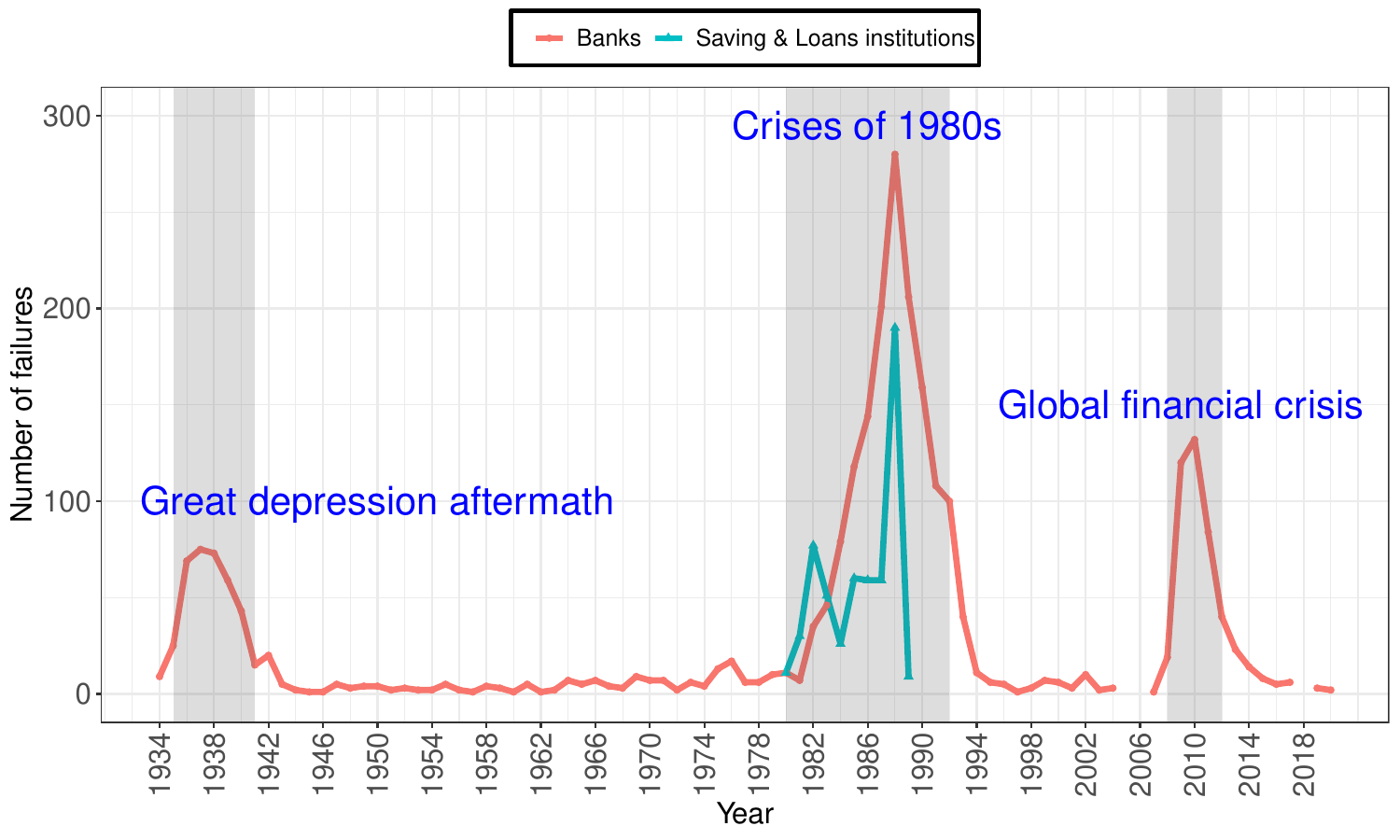}
	\caption{Number of failed banks resolved by the FDIC and number of failed S\&L's resolved by the FSLIC from 1934 to 2020.\label{fig:failedInst_1}}
\end{figure}
During the crises of 1980's, these two related sectors of the U.S. banking industry, namely commercial banks and S\&L's, witnessed the highest number of failures since the Great Depression as depicted in Figure \ref{fig:failedInst_1}. Notably, the FDIC and FSLIC underwent contrasting trajectories following the crises. While the FDIC survived the crisis, albeit with depleted insurance funds, the FSLIC faced insolvency by the end of the crisis and was closed in 1989 at a cost of \$132 billion to the taxpayer \citep{fdiccrisis}. 

When banks and S\&Ls failed, the FDIC and FSLIC applied one of the following three resolution methods \citep{walter2004closing}: 
\begin{enumerate}
	\item Type I: Open Bank Assistance (OBA) - Under this resolution method, the regulator provides financial assistance to acquirers toward the purchase of a failing bank or grants direct assistance to the failing bank. 
	\item Type II: Purchase and Assumption (P\&A) - Resolutions under this category consist of acquiring a part of the assets and liabilities of a failed bank by a participating institution. 
	\item Type III: Deposit Payout (PO) - Under this resolution category, the regulator liquidates the failed institution and pays out its insured depositors from the insurance fund. 
\end{enumerate}
Each of the three resolution methods described above involve a progressively more severe breakdown of relationships between the bank and its customers \citep{ashcraft2005banks}. For instance while a Type I resolution method ensures continuity of banking relationships, a Type III resolution terminates all such relationships. 

The crises of the 1980's are particularly suitable for comparing the resolution decisions of the FDIC and FSLIC against theoretically recommended state-dependent decision rules for several reasons. First, the simultaneous crises in banking and S\&L industries in the 1980's provided a basis to compare the two regulators, and to identify the stronger of the two approaches to resolving failed institutions. Second, bank and S\&L failures in this period occurred against the backdrop of shocks in specific sectors, namely, agriculture, real-estate and energy that resulted in regional crises \citep{fdiccrisis}. For instance, the major sectoral crises that occurred during this period were the recessions following the collapse of energy prices in Texas, Louisiana and Oklahoma, the agricultural recession in Kansas, Iowa and Nebraska and the real-estate-led downturns in California, the Southwest and the Northeast \citep{hanc1997history}. Third, banks were subject to varying levels of branching restrictions and operated either within state borders or across states that had entered into reciprocal arrangements \citep{kroszner1999drives}. Specifically, this period predates the elimination of the interstate branching restrictions mandated by the 1994 Riegle-Neal Interstate Banking and Branching Efficiency Act \citep{medley2013riegle}. Thus, the combination of sectoral crises that were regionally contained and branching restrictions that limited the geographic scope of banking markets entailed that certain bank failures occurred amid economic and financial distress, and others, in relatively normal economic conditions. 
This provided an ideal setting for the two regulators towards implementing the theoretically recommended state-dependent decision rules for resolving failed banks during this crisis.
\subsection{Recommended decision rules from theoretical studies and testable hypotheses}
\label{sec:hypo}
We first discuss the decision rules that are recommended by three different branches of theoretical literature for the resolution of failed banks. {Each of these branches of theoretical studies derive two distinct decision rules for regulators based on underlying states accompanying bank failure. The first branch of this literature derives appropriate regulatory actions when banks fail amid high versus low economic distress, the second, amid high versus low banking industry distress and the third branch considers how the presence or absence of political influence affects regulators' decisions on failed banks.} Thereafter, we identify the specific testable hypotheses based on each of these scenarios. 
\\[1.5ex]
\noindent\textbf{Recommended decision rule under economic distress - }\cite{cordella2003bank} determined a resolution strategy, or a decision rule, in which the regulator provides bailouts to banks if their failure occurred under macroeconomic distress, when bank failures are less likely to have arisen due to their unsound portfolio decisions and more likely to have arisen due to exogenous factors. Correspondingly, in the event of bank failures under normal economic conditions, their theoretical model recommended liquidating such banks.
\\[1ex]
\textit{Hypothesis $H_{1}$}: The testable hypothesis is that the FDIC and FSLIC applied different decision rules for banks that failed in normal economic conditions and those that failed amid macroeconomic distress. Conditional on the presence of two distinct rules, the subsequent statistical inference centers on testing the hypothesis that the probability of receiving a Type I resolution was higher for banks that failed amid high economic distress relative to those that failed amid low distress.
\\[1.5ex]
\noindent\textbf{Recommended decision rule under banking industry distress - }
\cite{acharya2007too,acharya2007cash} propose resolution strategies for the ``too-many-to-fail" problem or, equivalently, the simultaneous failure of many banks. Their recommended decision rule consisted of facilitating acquisitions of failed banks when such failures were small in number but providing bailouts and financial assistance when there were a large number of failures.
\\[1ex]
\textit{Hypothesis $H_{2}$}:  The first testable hypothesis is that regulators applied distinct rules in the presence and absence of banking industry distress. Second, the decision rule employed in the presence of banking industry distress designated a higher proportion of resolutions as Type I compared to the rule applied in its absence. 
\\[1.5ex]
\noindent\textbf{Political influence on regulators' decision rule - }{Several empirical studies \citep{igan2012fistful,duchin2012politics} have revealed the evidence of political and lobbying influence on regulators' decisions for bank resolutions.} 
For instance, \cite{deyoung2013theory} find that when a regulator experiences political pressure to place greater emphasis on maintaining current liquidity, they will provide more bailouts than when the regulator prioritizes the prevention of future moral hazard. 
\\[1ex]
\textit{Hypothesis $H_{3}$}: The primary hypothesis is that the presence of political influence induces a separate decision rule that is distinct from the rule applied in its absence. Subsequently, inferences center on whether the decision rule utilized under political influence resulted in a higher probability of receiving a Type I resolution relative to the rule applied in the absence of political influence.
\subsection{Our contributions and connections to existing works}
We consider the three hypotheses from Section \ref{sec:hypo} and test for the presence of state-dependent resolution strategies, either recommended by theory or arising from political interference, in the decision rules of the FDIC and FSLIC during the crises of 1980's. Our findings reveal that regular assessments of financial regulators by lawmakers and the public can uncover gaps between observed and recommended resolution rules and provide guidelines for corrective actions. For instance, timely assessments of the FSLIC could have revealed that the agency had provided excessive assistance to institutions that failed amid relatively normal economic conditions and to institutions that received political support (Section \ref{sec:results_fslic}). Interventions based on these assessments could have potentially prevented both, the failure of FSLIC in 1989 and the ensuing costs to the taxpayer. Conversely, the decision structure of the FDIC identified in this paper (Section \ref{sec:results_fdic}) provides a road-map for newer resolution agencies that face widespread failures from systemic shocks. The ensuing discussion summarizes our main contributions.
\begin{enumerate}
	\item 
 We statistically evaluate how regulators' decisions align with recommended decision rules from alternative theoretical models, as well as the extent to which political economy factors interfere with those recommended decisions. {By assessing the two regulators against theoretical rules, we can then compare the FDIC and FSLIC with each other.} 
 To the best of our knowledge, this is the first article to compare between the decision rules of the FDIC and FSLIC during the simultaneous crises in the banking and S\&L industries during the 1980's.
	\item To evaluate the three hypotheses in Section \ref{sec:hypo}, we develop a new methodology for assessing regulators in the form of a Bayesian latent class estimation framework for ordered outcomes. The proposed framework detects unobserved heterogeneity in regulators' resolution decisions based on underlying economic and political conditions. For estimating this model and conducting inference, we design a novel collapsed Gibbs sampler algorithm that provides a technique for efficient sampling from the posterior distribution relative to standard approaches by reducing the autocorrelations across successive draws. Our method provides a statistical framework to compare parameters across the latent classes, 
	and additionally allows for inferences on all estimated quantities of interest, including marginal effects and the probability of class membership. 
	\item We consider hypothesis $H_1$ and evaluate whether the FDIC and FSLIC provided bailouts to banks or S\&L's that failed amid macroeconomic distress and withheld such assistance for failures in normal economic conditions. 
	Our results reveal that the FSLIC deviated from and the FDIC adhered to this theoretically recommended rule. Specifically, banks that failed amid high economic distress received financial assistance, or bailouts from the FDIC with an average probability of 25\% compared to 3\% for banks that failed amid low distress. The FSLIC, on the contrary, assigned Type I assistance with probabilities 68\% and 70\% to the two groups of S\&L institutions that did not statistically differ across measures of macroeconomic distress. 
	\item We examine hypothesis $H_2$ on whether the two regulatory agencies experienced a too-many-to-fail problem and responded to it in the form of a greater reliance on bailouts and financial assistance to acquiring institutions. Our results show that the FDIC's decision rules aligned with the theoretical rules as the agency provided Type I assistance with a probability of 27\% for failures amid economic and banking industry distress and a statistically lower probability of 4\% for failures amid low levels of such distress. 
	The FSLIC assigned Type I assistance with probabilities of 76\% and 70\% among groups of institutions that did not statistically differ by industry and economic distress. 
	\item We evaluate hypothesis $H_3$ and assess the extent to which political pressures influenced the resolution decisions of the two agencies. Whereas the previous two hypotheses examined the extent to which regulators followed recommended theoretical rules, this assessment examines potential institutional weaknesses. 
	We find that political support for the banking industry played a limited role in the FDIC's decisions, but a more salient role in the FSLIC's decisions.  Notably, the FSLIC assigned assistance to S\&L's that likely received a higher degree of political support and failed amid lower economic distress at a statistically higher probability of 92\% relative to 59\% for S\&L's that failed amid low political support and in a climate of higher economic distress.
\end{enumerate}
A rich literature examined the financial costs incurred by the FDIC \citep{bennett2014effects, balla2015did} and the weaknesses of the FSLIC \citep{kane1989high, akerlof1993looting, romer1991political, greene2003s} in the course of resolving failed institutions. However, prior literature has not formally assessed the two agencies against resolution rules recommended by theoretical models. Relatedly, the decision rules of the two agencies have not been compared with each other. Our paper addresses both these gaps in the literature. 

{To assess regulators based on theoretical guidelines, we provide a new interpretation of latent class models as an evaluation tool and develop a novel Bayesian framework to estimate these models. Although our application considers regulatory decisions during a major crisis from the 1980s, this empirical framework is general and can be applied in evaluating regulators in the event of future crises. The model developed in this paper is an extension of a finite mixture model \citep{aitkin1985estimation} and traces its origins to  \cite{heckman1984method} who proposed the use of latent classes in duration models as a nonparametric alternative to random coefficients models in addressing unobserved heterogeneity in the population. Latent class models have since been developed for a broad range of outcomes including multinomial \citep{greene2003latent, burda2008bayesian}, count \citep{wang1998analysis, wedel1993latent, deb1997demand, nagin1993age} and ordered \citep{greene2014latent,greene2010modeling} responses. These studies apply latent class models to study heterogeneity in fields ranging across healthcare, marketing and transportation. The main function of this category of models is to uncover heterogeneity in the relationship between an outcome $Y$ and a vector of  covariates $x$. We recognize that a separate branch of literature on latent class models has venerable foundations in item response theory and consists of uncovering latent traits $s$ through multiple dichotomous indicators $Y$ \citep{goodman1974analysis, mccutcheon1987latent}. Notably, the models in this branch of literature do not incorporate covariates for predicting  outcomes $Y$. We note that while both categories of latent class models address unobserved heterogeneity, our model belongs to the branch that is founded in uncovering  heterogeneous relationships between outcomes and covariates, which is distinct from the models that uncover subgroups of the population exhibiting heterogeneous traits.}
\section{Data}
\label{sec:data}
We examine bank resolutions by the FDIC between 1984 and 1992, and S\&L resolutions by the FSLIC from 1984 until the agency's closure in 1989. 
The period between 1984 and 1992 is particularly suited to evaluate the decisions of the FDIC since the agency was subject to restrictions in applying Type I resolutions before 1982 and after 1993 {\citep{walter2004closing}}. 
The FSLIC, on the contrary, retained this authority from the start of the sample period until its closure in 1989. 
The data used in this paper includes variables that can be broadly divided into the following seven categories: 
\\[1ex]
\noindent\textbf{Data on resolution types -} the data on resolution types applied to failed banks and S\&L's are obtained from the Historical Statistics on Banking (HSoB) maintained by the FDIC. The sample consists of 1385 banks, of which there are 118, 1175 and 92 institutions resolved under resolution types I, II and III respectively. There are 389 S\&L institutions in the sample of which 270, 104 and 15 institutions underwent resolution methods I, II and III respectively.
\\[1ex]
\noindent\textbf{Bank and S\&L-level characteristics -} failed banks from the HSoB are matched with call report data from the Federal Reserve Bank of Chicago to obtain information from the financial statements of each institution. We aggregate the call reports by certificate number, which the FDIC uniquely assigns to each head office of depository institutions, and use this identifier to merge the two datasets. To allow for the duration of 90 to 100 days \citep{fdiccrisis} between the FDIC receiving notification of an institution that is in danger of failing and determining the resolution method, call reports from two quarters prior to the date of failure are used in the study. 
The failed S\&L institutions in the sample are similarly matched with Thrift Financial Reports from the Research Information System (RIS) of the FDIC as of two quarters prior to failure. The data on S\&L institutions is less extensive than the corresponding bank-level data due to differences in the reporting requirements for banks and S\&L's. Specifically, data on Agricultural loans, Nonperforming loans and Core Deposits are not available for S\&L institutions for the period under study. 
\\[1ex]
\noindent\textbf{Insurer characteristic -} we obtain the year-end data on outstanding balances on the deposit insurance fund of the FDIC from annual reports from the agency's website {(see \url{https://www.fdic.gov/about/financial-reports/reports/index.html}).} 
These balances remain constant across banks and S\&L's, and vary by year. We use this information to calculate the amount of deposit insurance fund as a percentage of total insured deposits for both FDIC and FSLIC. This characteristic measures the extent of insurance funds available to the two agencies relative to the maximum value of their potential insurance payouts.
\\[1ex]
\noindent\textbf{State characteristics -} our data hold several dimensions of information pertaining to the states in which the banks and S\&L's operated. Specifically, the data on quarterly housing starts at the state level have been obtained from IHS Global Insight, the data on annual unemployment at the state level were obtained from the Iowa Community Indicators Program of Iowa State University and the information on branching deregulation laws was collated using the table in \cite{strahan2003real}. 
\\[1ex]
\noindent\textbf{County characteristics -} the data pertaining to county economic characteristics in which the banks and S\&L's operated have been collated from the Bureau of Economic Analysis. These characteristics consist of per capita growth in gross domestic product (GDP) and the share of employment in each sector, which measure the economic output of each county and the importance of each sector to the county's economy respectively. 
\\[1ex]
\noindent\textbf{County-level characteristics of bank distress -} county-level statistics on the banking industry are obtained by aggregating bank-level data from the Research Information System (RIS) of the FDIC, which is available starting from 1984. 
\\[1ex]
\noindent\textbf{State-level political economy characteristics -} {
we measure political support for financial institutions by way of the percentage of Congressional representatives from each state who voted for a bill that is favorable to the banking or S\&L industry. This approach recognizes votes in favor of legislation that benefits the banking industry as indicative of lobbying efforts by the industry, which is consistent with the theoretical model of \cite{becker1983theory} in which pressure groups compete for political favors. This approach also follows from \cite{kroszner1999drives} and \cite{economides1996political}, who provide evidence of private interest groups influencing the voting behavior of elected representatives on legislation pertaining to the banking industry.} Congressional voting data were obtained from the website of GovTrack (\url{https://www.govtrack.us}) and converted into state-level percentages of representatives who voted in favor of each bill evaluated in this study. The description of these bills is available in Section \ref{sec:data_app} of the supplement. 
\begin{table}
\caption{\label{tab:data_role} A summary of the role of the seven categories of data used in this article with reference to Figure \ref{fig:flowFinal} and the three hypotheses discussed in Section \ref{sec:hypo}.}
\centering
\scalebox{0.85}{
\begin{tabular}{clccccc}
\hline
\multicolumn{1}{l}{}  &   Data  categories   & \multicolumn{1}{c}{$H_1$} & \multicolumn{1}{c}{$H_2$} & \multicolumn{1}{c}{$H_3$} & \multicolumn{1}{c}{Covariate} & \multicolumn{1}{c}{Outcome} \\
\hline
\multirow{5}{*}{\begin{tabular}[c]{@{}c@{}}Class \\ membership \\ model\end{tabular}}  & State characteristics  &   \checkmark    & \checkmark  &   \checkmark   &  &                                   \\
 & County characteristics &  \checkmark  &  \checkmark   &  \checkmark   &      &   \\
& County-level characteristics of bank distress    &                        &          \checkmark              &                        &                                &                                   \\
& State-level political economy characteristics      &                        &                        &      \checkmark                  &                                &                                   \\
& Insurer characteristic        &   &       &    &   \checkmark  &       \\
                        \hline
\multirow{2}{*}{\begin{tabular}[c]{@{}c@{}}\small Resolution type versus \\ bank level covariates\end{tabular}} & Data on resolution types  &  \checkmark   & \checkmark  & \checkmark  &   &\checkmark \\
& Bank and S\&L-level characteristics   &  \checkmark   &  \checkmark   &\checkmark &     \checkmark      &\\
\hline
\end{tabular}}
\end{table}

Table \ref{tab:data_role} summarizes the role of the aforementioned seven categories of data. For instance, the data on resolution types and bank level financial characteristics are part of the second layer of our hierarchical model (see Figure \ref{fig:flowFinal}), which specifies the relationship between the former and the latter. The remaining five categories appear in the class membership model that constitutes the first layer of the model. In the context of the three hypotheses discussed in Section \ref{sec:hypo}, state and county characteristics quantify regional economic distress for hypothesis $H_1$ and are also considered in extended specifications for testing the remaining two hypotheses. County-level characteristics of bank distress quantify regional banking industry distress and, consequently, operationalize hypothesis $H_2$. Similarly, state-level political economy characteristics operationalize political influence and are related to hypothesis $H_3$. Finally, the insurer characteristic is a covariate in this model and controls {for the extent of funds available to insurers, which may constrain the types of resolution actions that they can undertake.} In Section \ref{sec:data_app} tables \ref{tab:data_dic}, \ref{tab:summ} and \ref{tab:summt}, respectively, provide a description of the variables available under each of these seven categories and summary statistics of the data for FDIC and FSLIC resolution decisions. 
\section{Latent class ordinal probit model
}
\label{sec:model}
We propose a latent class model for ordinal outcomes to represent the decision rules of FDIC and FSLIC in resolving failed banks and to evaluate these decision rules against recommended rules from theoretical studies. Figure \ref{fig:flowFinal} depicts our latent class model in the context of hypothesis $H_1$. {In this section we discuss the main components of the model while details related to the Bayesian method for estimating this model are provided in Section \ref{sec:estimation_app} of the supplement. In the following discussion, bank $i$  refers to a representative bank or S\&L without loss of generality. Let $y_{i}$ be the resolution method applied on bank $i$ where $y_{i}$ takes values $1,2$ and $3$ to denote resolution types I, II and III respectively. By adopting this empirical framework to estimate the decision rules of the FDIC and FSLIC, we compare their observed decisions with theoretical decision rules and the two agencies with each other. }
\\[1ex]
\textbf{Class membership model - }The class membership model in Figure \ref{fig:flowFinal} uses a latent class model to represent state-dependent rules in the two regulators' decision structure. In the context of the three hypotheses discussed in Section \ref{sec:hypo}, the  classes correspond to two distinct groups of banks that failed under heterogeneous economic (hypothesis $H_1$), banking (hypothesis $H_2$) or political conditions (hypothesis $H_3$) and were thereby subject to disparate decision rules by regulators. The class indicator $s_i$ is introduced into the model to denote assignment of bank $i$ into one of the two classes.  Section \ref{app:two_latentclasses} in the supplement provides additional details on the rationale for two latent classes.

Within each latent class, the regulator applies a class-specific decision rule on the failed bank $i$ to assign it one of the three available resolution methods, denoted $y_i$.
The distinguishing feature of the latent class model is that classes are determined with probability and not deterministically. This feature is relevant to the current problem since the true assignment of banks into distinct classes by regulators is not observable as the data record the final decision made by the regulators but not the rationale that motivated each decision. Specifically, $y_i$ is observed but $s_i$ is not. As a result, the probabilistic assignment of banks into classes addresses the researcher's uncertainty on class assignments by the regulatory agencies. 

We model the regulator's problem of assigning bank $i$ to one of the two latent classes as a binary discrete choice problem with a latent outcome $s_{i}$ {and rely on the random utility representation of this model based on the framework developed by \cite{marschak1974binary}.} 
To apply the random utility representation to this discrete choice problem, we introduce a continuous latent variable $l_{i}$, which represents the difference in utilities or value to the regulator from assigning bank $i$ to latent class 2 relative to latent class 1. {See Remarks \ref{rem:rand_utility} and \ref{rem:latent_var} for more details regarding the random utility framework and this latent variable representation.} We express $l_{i}$ as,  
\begin{equation}
\label{eq:elle}
	l_{i} = \bm{w}_{i}'\bm{\alpha} + \nu_{i},
\end{equation}
where $\bm{w}_{i}$ is a $p-$dimensional vector of covariates {that includes a constant term $1$},  $\bm{\alpha}$ are parameters and $\nu_i\sim\mathcal N(0,1)$ is the error term. The covariates $\bm w_i$ in Equation \eqref{eq:elle} and Figure \ref{fig:flowFinal} are determined by the three hypotheses of interest described in Section \ref{sec:hypo} and {are derived from recommendations in theoretical papers on bank resolutions \citep{cordella2003bank, acharya2007too, deyoung2013theory}}. {In testing the hypothesis for the presence of distinct decision rules in high and low economic distress ($H_{1}$), the covariates $\bm w_i$ consist of economic indicators such as the unemployment rate, housing starts and per capita GDP growth pertaining to the state and the county in which a bank operates. To determine whether regulators applied different decision rules based on the health of the banking industry ($H_{2}$), we use measures such as the number of previous bank closures and the share of assets in distressed banks in a county where a bank  is headquartered in our covariate vector. To test whether political influence induced regulators to apply differential resolution rules ($H_{3}$), we include the percentage of Congressional representatives from the bank's headquartered state who voted for bills in ways that were favorable to the banking industry.} 
 Finally, the relationship between the discrete variable $s_{i}$ and the continuous variable $l_{i}$ is expressed via the following threshold crossing framework,
$s_{i}=1$ if $l_{i}\leq 0$ and $s_i=2$, otherwise.
\\[1ex]
\textbf{Ordinal model of resolution decisions - }
{Within each latent class, the second layer of our model in Figure \ref{fig:flowFinal} specifies the relationship
between the primary outcome of interest $y_i$, and bank-specific covariates $\bm x_i$.} We model $y_i$ 
as an ordered variable by specifying the three resolution methods of Section \ref{sec:resoultion} as ordered categories. Previous studies have shown that these resolution methods resulted in progressively more severe effects on economic outcomes \citep{ashcraft2005banks} and on the level of liquidity \citep{deyoung2013theory}. 
In particular, \cite{ashcraft2005banks} points out that each of the three resolution categories entail an increasingly severe breakdown of relationships between the bank and its customers. The provision of Type I assistance allows a bank to continue functioning in its present form. A Type II acquisition or purchase results in certain loan and deposit relationships continuing within the acquiring bank's books. A Type III liquidation and deposit payout results in the termination of all banking relationships.
The specification of resolution methods as an ordered outcome variable also allows for a decision structure in which the regulators order banks by their franchise value, {which is the present discounted value of a bank's future stream of profits and incorporates the value of its customer relationships and resulting informational advantages.} 
{Regulators} assign Type I resolutions to the most valuable and Type III resolutions to the least valuable banks {in terms of their franchise value}. Such an ordering of banks and S\&Ls by franchise value is consistent with a cost-minimization objective, which was relevant to both FDIC and FSLIC since they were required to preserve their insurance funds by controlling their costs of resolution \citep{fdicres}. 

Within latent class $s_{i}$, the regulator's utility function $z_{i,s_{i}}$ determines the final resolution method applied on bank $i$ where,
\begin{equation}
\label{eq:zeq}
z_{i,s_{i}} = \begin{cases}
	\bm x_{i}'\bm\beta_{1} + \epsilon_{i,1},&\text{if}~s_i = 1\\
	\bm x_{i}'\bm\beta_{2} + \epsilon_{i,2},&\text{if}~s_i = 2
\end{cases}.
\end{equation}
Here $z_{i,s_{i}}$ is the utility that the regulator derives from preserving bank $i$'s franchise value under the random utility interpretation of \cite{marschak1974binary}. 
The $q-$dimensional covariate vector $\bm x_i$ in Equation \eqref{eq:zeq} and Figure \ref{fig:flowFinal} consists of {a constant term $1$ and} bank $i$'s financial characteristics {listed in \cite{balla2019comparison}}, salient among which are its size, the quality of its assets and composition of risky asset classes. These covariates are constructed from banks' financial statements. {
In addition, we include an indicator to denote whether the bank was headquartered in a state that permitted banks to open branches and acquire banks across state-lines. Broadly, the components of $\bm{w}_{i}$ and $\bm{x}_{i}$ differ in one important respect--the covariates $\bm{w}_{i}$ in the class membership model contain characteristics that are outside the control of banks' decisions, whereas covariates $\bm{x}_{i}$ in the ordinal probit model for resolution are measures that are predominantly the result of banks' prior actions. See Section \ref{app:covariates} in the supplement for more details.} 

Note that in Equation \eqref{eq:zeq}, $\bm x_{i}'\bm\beta_{s_i}$ and $\epsilon_{i,s_i}$ represent, respectively, the observable and unobservable components of utility \citep{train2009discrete}, and we specify a $\mathcal{N}(0,\sigma^{2}_{s_i})$ distribution for the unobserved component. The relationship between the observed outcome $y_{i}$ and the latent utility $z_{i,s_{i}}$ is represented using the following threshold-crossing framework,
\begin{equation}
	\label{eq:yi}
y_{i}=
\begin{cases}
	3: \hspace{5pt}\text{Type III}, & \text{if}\ -\infty<z_{i,s_{i}}\leq\gamma_{1,s_{i}} \\
	2: \hspace{5pt}\text{Type II}, & \text{if}\ \gamma_{1,s_{i}}<z_{i,s_{i}}\leq\gamma_{2,s_{i}} \\
	1: \hspace{5pt}\text{Type I}, & \text{if}\ \gamma_{2,s_{i}}<z_{i,s_{i}}\leq\infty \\
\end{cases}.
\end{equation}
The regulator selects a resolution method that preserves more of the bank's franchise value as $z_{i,s_{i}}$ crosses a progressively larger threshold. When $z_{i,s_{i}}$ is below the lowest threshold, $\gamma_{1,s_{i}}$, bank $i$ loses all its franchise value as the regulator's utility level corresponds to liquidation under a Type III resolution. 

{Denote $\bm \Theta=\{\bm\beta_{1}, \bm\beta_{2},\sigma_{1}^{2},\sigma_{2}^{2}, \bm\alpha\}$. In Section \ref{sec:estimation_app} of the supplement we present a Bayesian method to estimate $\bm \Theta$ that relies on a novel collapsed Gibbs sampler algorithm which
provides a technique for efficient sampling from the posterior distribution relative to standard approaches by reducing the autocorrelations across successive draws.}
\section{Bank resolutions by the FDIC}
\label{sec:results_fdic}
This section provides an assessment of the FDIC's resolution decisions over the period 1984-1992 by evaluating the agency's decision rules against recommended rules from theoretical studies. We perform this assessment by interpreting the results from the latent class ordinal model developed in Section \ref{sec:model} to test the three hypotheses 
summarized in Section \ref{sec:hypo}. Sections \ref{res1}, \ref{ds} and \ref{res2}, respectively, discuss the results for hypotheses $H_1, H_2$ and $H_3$. 
The prior distributions that we consider in this analysis are as follows: $\bm\alpha \sim \mathcal{N}_p(\bm 0,3I_p), \bm\beta_{s} \sim \mathcal{N}_q(\bm 0,I_q)$ and $\sigma_{s}^{2} \sim \mathcal{IG}({\sf shape=}~4.3, {\sf scale=}~1.3)$ for $s\in\{1,2\}$. The hyperparameters for $\sigma_{s}^{2}$ are chosen to result in an uninformative prior with a mean of approximately 0.4 and prior standard deviation of 0.26. The collapsed Gibbs sampler algorithm of Section \ref{sec:estimation_app} is run for $11,000$ iterations and we use $G=10,000$ post-burn in samples for posterior inference.
\subsection{Regional economic distress and FDIC's decisions for bank resolutions}
\label{res1}
The period of this study, 1984-1992, presents a unique set of economic and banking conditions that facilitate the test for presence of the recommended resolution strategy from \cite{cordella2003bank} in the FDIC's decisions (see discussion in Section \ref{sec:resoultion}). 
We find that the FDIC's responses supported Hypothesis $H_{1}$ as the agency provided assistance to banks that failed amid economic distress with a higher probability than to banks that failed in low economic distress. Furthermore, our results reveal that within the class of high regional distress, the FDIC targeted assistance to banks with relatively healthy balance sheets {that were more likely to recover and operate as going concerns} and arranged for the sale or liquidation of the remaining banks.
\\[1ex]
\noindent\textbf{Class-membership model - }
The class membership model is represented in the first level of the decision structure in Figure \ref{fig:flowFinal}. We perform a Bayesian model comparison, described in Section  \ref{mcs} of the supplement, to select the specification of the class-membership model that is most decisively supported by the data. The covariates in the resolution type model, the second level of the hierarchy in Figure \ref{fig:flowFinal}, are constant across all the specifications considered. 

Table \ref{tab:cmrd} summarizes the covariate effects from the class membership model for four specifications that include indicators of state and county-level economic performance along with controls for {institutional features} underlying the resolution decision. The expression for the covariate effects are derived in Section \ref{coeffform} of the supplement. The values of log marginal likelihood reported in the last row of Table \ref{tab:cmrd} point to specification (3) as the selected model as it has the highest posterior odds among the four candidate specifications. This selected specification highlights a statistically important role for state-level unemployment in assigning banks into two different classes. The other covariates that inform the assignment of banks to latent classes are county-level indicators of economic performance along with a control variable for the amount of insurance fund available per dollar of insured deposit in the banking system. 

Among alternative specifications considered in Table \ref{tab:cmrd}, specifications (1) and (2) of the model entirely consist of state and county-level indicators of economic performance and controls for county-level shares of employment by sector. Note that specification (2) is a more parsimonious setting that is nested within specification (1). Specification (4) augments specification (2) with indicators for the charter status of failed banks since chartering agencies, namely, the OCC for federally chartered banks and state banking departments for state-chartered banks, retain the final authority to enforce closure. 
The reference group in this class membership model consists of nationally chartered banks that are supervised by the OCC.      

In the following discussion, latent class 1 is labeled as the class of failures under ``High Regional Distress (HRD)" and latent class 2, as ``Low Regional Distress (LRD)". In the model for class membership in Equation \eqref{eq:qeq}, the event of success in the binary probit model (where the latent binary indicator $s_{i}$ equals 1) is represented by a bank belonging to latent class 2. Therefore, the negative signs associated with unemployment in specification (3) and the positive signs for covariate effects of GDP growth rate and housing starts in Table \ref{tab:cmrd} show that latent class 2 contains banks that failed during periods of low unemployment or periods of relatively low regional economic stress whereas banks that failed amid high regional distress belong to latent class 1. These findings support the first element of hypothesis $H_{1}$ by confirming that the FDIC distinguished across banks based on economic distress in applying its resolution decisions.  
\begin{table}
\caption{\label{tab:cmrd} Covariate effects from class-membership models for specifications of latent classes based on regional economic distress. The reported values are posterior means of the covariate effects. Posterior standard deviations are in parentheses.}
	\centering
 	\scalebox{0.8}{
\begin{tabular}{lcccc}
	\hline
	& (1)   & (2)   & \textbf{(3)}   & (4) \\
	\hline
	\textbf{State-level characteristics} &       &       &       &  \\
	Unemployment  & -0.12 (0.06) & -0.11 (0.05) & \textbf{-0.1 (0.04)} & -0.15 (0.08) \\
	Housing starts & 0.11 (0.05) & 0.09 (0.05) & \textbf{0.05 (0.05)} & 0.1 (0.05) \\
	\textbf{County-level characteristics} &       &       &       &  \\
	Per capita GDP growth & 0.04 (0.05) & 0.05 (0.05) & \textbf{0.04 (0.05)} & 0.04 (0.05) \\
	Farm, agri, mining & 0.11 (0.09) & 0.07 (0.05) & \textbf{0.06 (0.04)} & 0.07 (0.05) \\
	Manufacturing & 0.03 (0.05) & -     & \textbf{-} & - \\
	Construction & 0.02 (0.04) & -     & \textbf{-} & - \\
	Fin Serv Transport & 0 (0.07) & -     & \textbf{-} & - \\
	Government & 0.04 (0.05) & -     & \textbf{-} & - \\
	\textbf{Insurer characteristics} &       &       &       &  \\
	Dep. Ins. Fund/ Total Deposits & -     & -     & \textbf{-0.05 (0.03)} & - \\
	\textbf{Bank-level characteristics} &       &       &       &  \\
	State charter Fed member & -     & -     & \textbf{-} & 0.03 (0.07) \\
	State charter non-Fed member & -     & -     & \textbf{-} & -0.06 (0.05) \\
	\hline
	log Marginal Likelihood & -703.35 & -701.10 & \textbf{-699.79} & -700.19 \\
	\hline
\end{tabular}}%
\end{table}%
\\
\textbf{Heterogeneity in decision rules - }
The results from the second level of the decision structure in Figure \ref{fig:flowFinal} show that the average probability of the FDIC assigning a Type I resolution was statistically higher among HRD banks than among LRD banks. These findings confirm that the FDIC's decisions fully aligned with hypothesis $H_{1}$ in that the agency was more likely to provide assistance to banks when their failure was accompanied by regional economic distress.    

\begin{figure}[!t]
	\centering
	\includegraphics[width=0.75\textwidth]{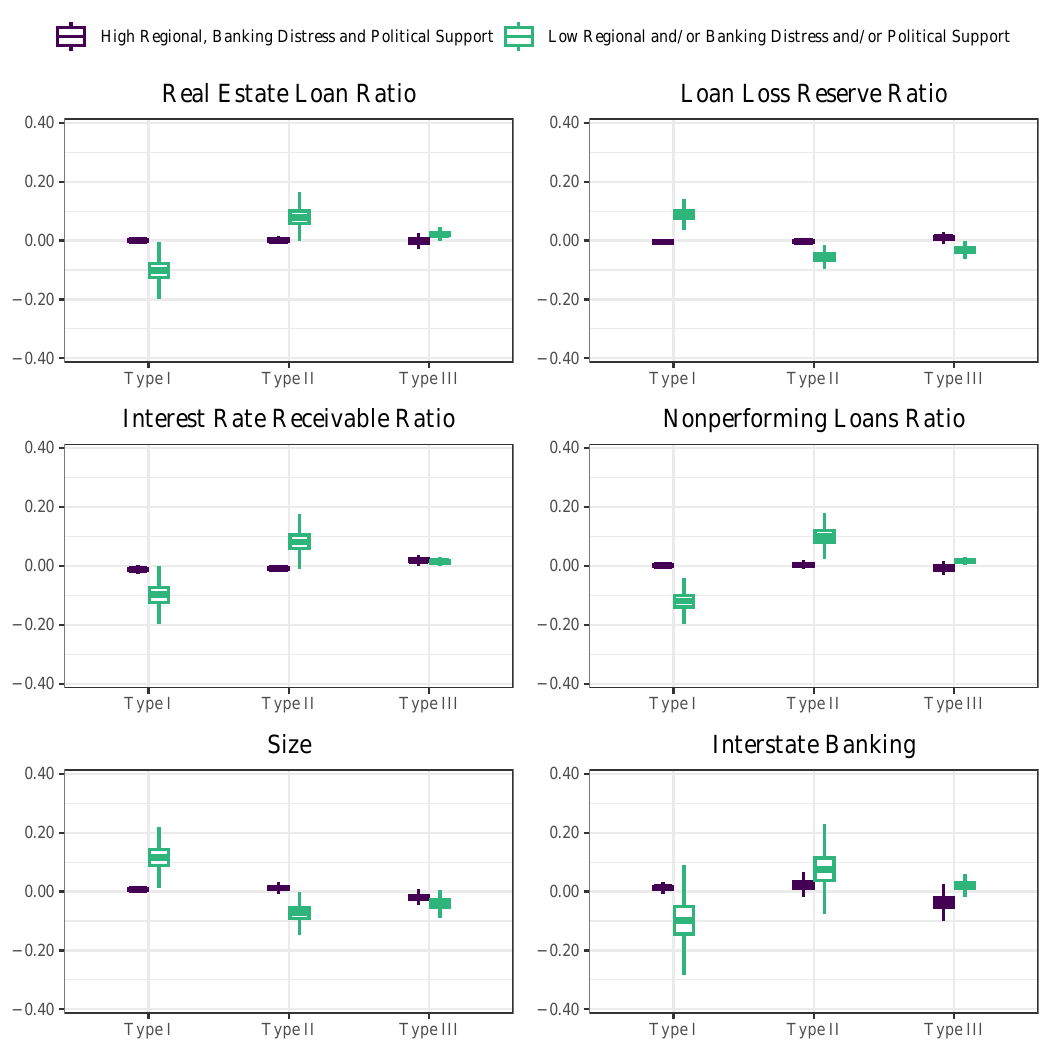}	
	\caption{Covariate effects from the models for resolution type for banks in the class of High Regional Distress (HRD) and Low Regional Distress (LRD).\label{fig:covarEff}}
\end{figure}

The average probability of each resolution type for the HRD and LRD failures is computed as follows: $\mbox{Avg. Prob}^{(g)}(Y = j \vert s) = (1/n)\sum_{i=1}^{n}P_{ij\vert s}^{(g)},~j=1,2,3$,
where $s=1$ and $s=2$ correspond to the results for the class of HRD and LRD failures respectively and $g$ is the index for the $G$ post burn-in MCMC draws. The values $P_{ij\vert s}^{(g)}$ are computed for each MCMC iteration using Equation \eqref{eq:probc} in the supplement. 
\begin{figure}[!t]
\centering
\includegraphics[width=0.85\linewidth]{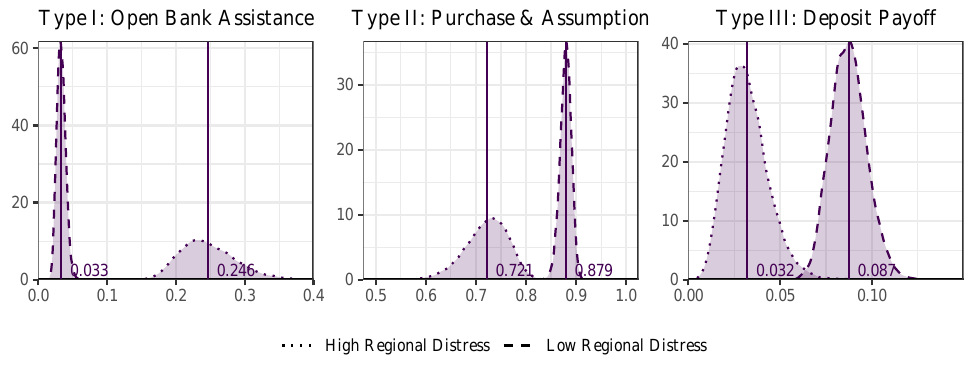}
\caption{Posterior distribution of the average probability of the FDIC assigning each resolution method within classes based on regional distress. The horizontal axis represents the probability of assigning a resolution method and the vertical axis represents the posterior density associated with that probability based on a kernel density estimate. The solid vertical lines represent the means of these posterior distributions across the $G$ MCMC draws.\label{fig:kds_rd}}
\end{figure}
Figure \ref{fig:kds_rd} provides the density of the full posterior distribution of the average probability of receiving each resolution method across the two latent classes. The average probability of receiving a Type I resolution among HRD banks was 24.6\% compared to 3.3\% for LRD banks. The fully disjoint posterior densities of the average probability of receiving a Type I resolution for the HRD and LRD classes shows that the difference between their averages is statistically important. This observation continues to hold for Type II resolutions with statistically important differences in average resolution probabilities across HRD and LRD classes at 72.1\% and 87.9\% respectively. The theoretical recommendation from \cite{cordella2003bank} does not explicitly address the decision to facilitate partial or whole acquisitions of failed banks and the findings from this estimation exercise provide new insights into the differences in the probabilities of implementing {the Type II} resolution method under varying levels of economic distress. Finally, in a further confirmation of the predictions of the theoretical model, the average probability of being liquidated under a Type III resolution was 8.7\% for LRD banks compared to 3.2\% for HRD banks. This difference is also statistically important, as evidenced by the minimal overlap in posterior densities of the two classes.  
\\[1ex]
\noindent\textbf{Resolution type - } 
The next stage of the empirical analysis centers on the results from the ordinal probit models represented in Equation \eqref{eq:probc} and depicted in the second level of the decision structure in Figure \ref{fig:flowFinal}. These models estimate separate relationships between the resolution method $y_{i}$ and bank-level financial indicators $\bm x_{i}$ in the LRD and HRD classes. If the FDIC responded differently to LRD and HRD failures for the same change in bank financial characteristics, this would manifest in different magnitudes of covariates across the two classes and provide conclusive evidence of the presence of two different decision rules implemented by the FDIC. In this section, we report the six largest covariate effects from the selected ordered response model (specification 3 in Table \ref{tab:cmrd}) in Figure \ref{fig:covarEff}. The covariate effects of the remaining variables are provided in Section \ref{sec:coveffect_banks_app} of the supplement.  

From Figure \ref{fig:covarEff}, the financial variables, Real Estate Loan Ratio and Nonperforming Loans Ratio, both exhibit qualitatively similar covariate effects. A unit standard deviation increase in Real Estate Ratio and Nonperforming Loans Ratio is associated with a reduced probability of obtaining assistance under a Type I resolution among HRD failures and an increased probability of such banks undergoing Type II and III resolutions. Since nonperforming loans provide a succinct measure of the quality of the failed bank's assets and real estate loans represent a risky asset category, these results reveal that the FDIC provided assistance under Type I to banks that had relatively healthier balance sheets even among those banks that failed amid economic distress, which is consistent with the theoretical recommendations of \cite{cordella2003bank}. 
The effects of these covariates on banks within the class of LRD failures, on the other hand, are not statistically important.

The Interest Receivable Ratio is seen to be important in the FDIC's evaluation of bank health, with elevated levels of this ratio eliciting more stringent resolution methods from the FDIC. \cite{balla2019comparison} originally identified the Interest Receivable Ratio to be highly predictive of both bank failure and loss subsequent to failure in their study. 
Accordingly, an increase in Interest Receivable Ratio among HRD failures resulted in a reduction in the probability of Type I resolution and a corresponding increase in the probability of a Type II resolution, entailing partial or whole acquisitions of the failed institutions. Banks that belonged to the LRD class of failures experienced a more severe response in the form of an increased probability of a Type III resolution and hence, complete liquidation, along with a decreased probability of the other two resolution methods. 

Larger banks were less likely to be liquidated under a Type III resolution across both latent classes. Among HRD failures, a standard deviation increase in log of assets was also associated with a decreased probability of a Type II resolution and a compensatory increase in the probability of assistance under Type I resolution. LRD failures experienced an increase in the probability of both, Type I and II resolutions concomitantly with a decrease in the probability of a Type III resolution. The increased probability of Type I resolutions associated with a larger bank reveals that the ``too-big-to-fail" doctrine was present in the decisions of FDIC during the crisis of the 1980's. 

The estimation results provide new insights into the role of Loan Loss Reserve Ratio, an accounting variable that records the amount of funds set aside to meet expected losses. An increase in Loan Loss Reserves ratio was associated with a higher probability of receiving Type I resolution in the class of HRD failures. Contrarily, among LRD failures, an equivalent increase in this measure was associated with an increase in their probability of receiving a Type III resolution and being liquidated. A possible explanation for this disparity is that in the HRD class of banks, where failures are more likely to have occurred due to systemic factors, changes in loss reserve ratios can be attributed to the deterioration in asset quality resulting from market-wide fluctuations. However, among LRD failures, the FDIC is likely to have viewed the increase in this ratio as a signal of deterioration in asset quality arising from issues idiosyncratic to the failed bank. 

Interstate is an indicator variable that identifies whether interstate banking was legal in the state in which a bank is located in the year of failure. Intuitively, interstate banking laws are likely to affect resolution outcomes as they determine the breadth of demand for assets of failed banks. 
For instance, the model in \cite{acharya2007cash} predicts that interstate banking, by expanding the set of available acquiring banks in the event of a failure, should be associated with an increase in the probability of a Type II resolution and an equivalent decline in the probability of a Type I resolution. In the bottom right panel of Figure \ref{fig:covarEff}, an increased probability of Type II resolutions is observed among both, HRD and LRD classes, even though the LRD class of failures also underwent a modest increase in the probability of Type I resolutions. 

Overall, Figure \ref{fig:covarEff} reveals that the magnitudes of the covariate effects from the selected model, specification (3), described in the preceding subsection, are larger for banks that belong to the latent class ``High Regional Distress (HRD)" relative to banks in the class labeled ``Low Regional Distress (LRD)". This pronounced difference in the effects of each covariate on the FDIC's decisions across the two classes confirms the presence of two distinct decision rules in the agency's resolution procedure. The larger covariate effects for banks in the class of HRD failures indicate that within the group of banks that failed amid regional economic distress, the FDIC ordered banks based on their financial characteristics and provided Type I resolutions to relatively healthier banks and Type II and III resolutions to relatively weaker banks. The smaller covariate effects in the LRD class suggest that the FDIC evaluated banks that failed in low economic distress on a case-by-case basis rather than appraising their relative financial strength. This approach potentially included evaluating unobservable individual circumstances, which are captured by the error term in Equation \eqref{eq:zeq}. These results reveal a smaller role for observed financial statement information in determining resolution methods within the LRD class relative to the HRD class of banks and further support the hypothesis that banks in the former group are likely to have failed due to largely idiosyncratic factors. 
\subsection{Banking industry distress and FDIC's decisions for bank resolution}
\label{ds}
We examine whether the FDIC's responses supported Hypothesis $H_{2}$, i.e.,  whether the agency provided assistance with higher probability to banks that failed amid distress within the banking industry. Our results show that the agency's decision rules qualitatively aligned with this hypothesis as it provided Type I resolutions with a marginally higher probability when the local banking industry experienced failures. However, measures of regional economic distress remained the most important determinants of membership into classes that received statistically different levels of Type I resolution. 
\begin{table}
	\caption{\label{tab:cmrbd} Covariate effects from class-membership models for specifications of latent classes based on regional and banking industry distress. The reported values are posterior means of the covariate effects. Posterior standard deviations are in parantheses.}
	\centering
	\scalebox{0.8}{
		\begin{tabular}{lrrrrrr}
			\hline
			& \multicolumn{1}{c}{(5)} & \multicolumn{1}{c}{(6)} & \multicolumn{1}{c}{(7)} & \multicolumn{1}{c}{\textbf{(8)}} & \multicolumn{1}{c}{(9)} & \multicolumn{1}{c}{(10)} \\
			\hline
			\textbf{State-level characteristics} &     &     &      &     &  & \\
			Unemployment & -     & -     & -     & \textbf{-0.07 (0.04)} & -0.08 (0.04) & -0.1 (0.05) \\
			\textbf{County-level characteristics} &     &     &      &     &  & \\
			Housing starts & -     & -     & -     & \textbf{0.03 (0.05)} & 0.04 (0.05) & 0.06 (0.06) \\
			Per capita GDP growth & -     & -     & -     & \textbf{0.04 (0.05)} & 0.04 (0.05) & 0.07 (0.06) \\
			Farm, agri, mining & -     & -     & -     & \textbf{0.03 (0.04)} & 0.04 (0.04) & 0.06 (0.05) \\
			\multicolumn{2}{l}{\textbf{Banking industry characteristics}} &     &     &      &     &   \\
			Previous closures& -0.08 (0.08) & -0.07 (0.06) & -0.08 (0.07) & \textbf{-0.02 (0.01)} & -0.02 (0.01) & -0.01 (0.02) \\
			\% Assets in distressed banks& -0.03 (0.04) & -     & -     & \textbf{-0.03 (0.02)} & -     & - \\
			\% Dep. in distressed banks  & -     & -0.03 (0.02) & -     & -     & -0.03 (0.02) & - \\
			\% distressed banks & -     & -     & -0.01 (0.01) & -     & -     & -0.02 (0.03) \\
			\textbf{Insurer characteristics} &     &     &      &    &  & \\
			Dep. Ins. Fund/ Total Dep. & -     & -     & -     & \textbf{-0.05 (0.03)} & -0.05 (0.03) & -0.05 (0.04) \\
			\hline
			log Marginal Likelihood   & -719.29 & -705.30 & -719.14 & \textbf{-697.22} & -697.78 & -701.21 \\
			\hline
	\end{tabular}}%
\end{table}%

Table \ref{tab:cmrbd} summarizes the covariate effects and log marginal likelihood from model specifications (5) through (7) that are based purely on banking industry distress and specifications (8) through (10) that incorporate a combination of banking industry and regional economic distress. The data favor the latter three specifications over the former three as evidenced by their higher marginal likelihood. Specifically, the Bayesian model selection procedure based on posterior odds selects specification (8), which defines distress in the banking industry using previous closures and the percent of assets in distressed banks within a county. Here distressed banks are those institutions whose Texas ratio, defined in Equation \eqref{eq:texasratio} of the Supplement, exceeded 100\% based on previous literature that utilize this measure \citep{cooke2015liquidity, siems2012so}. On account of the negative covariate effect of unemployment, previous closures and percent of assets in distressed banks and the positive signs for housing starts and per capita income growth, latent class 2 contains banks that failed amid relatively low regional or banking distress and banks that failed amid high regional and banking distress belong to latent class 1. As a result, in the following discussion, latent class 1 will be labeled as the class of failures under ``High Regional and Banking Distress (HRBD)" and latent class 2, as the class of failures under ``Low Regional and / or Banking Distress (LRBD)".  
\begin{figure}[!t]
	\centering
	\includegraphics[width=0.85\linewidth]{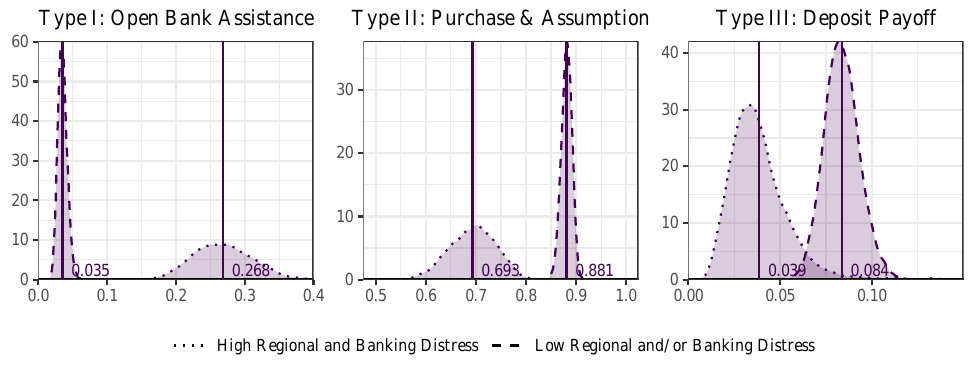}
	\caption{Posterior distribution of the average probability of the FDIC assigning each resolution method within classes based on regional and banking distress. The horizontal axis represents the probability of assigning a resolution method and the vertical axis represents the posterior density associated with that probability based on a kernel density estimate. The solid vertical lines represent the mean of these posterior distributions across the $G$ MCMC draws. \label{fig:prob4}}
\end{figure}

Figure \ref{fig:prob4} provides the density of the posterior distribution of the probability of the FDIC assigning each resolution category under the selected model, specification (8). These results qualitatively align with the recommendations from \cite{shleifer1992liquidation} and \cite{acharya2007cash} represented by Hypothesis $H_{2}$, which posited a greater reliance on public financial assistance in the form of Type I resolutions when the local banking industry experienced distress. The figure shows that the average probability of a Type I resolution was 26.8\% under banking and regional distress and 3.5\% under low regional and banking distress. These probabilities are marginally higher than the probability of Type I resolution of 24.6\% and 3.3\% under high and low regional distress respectively, depicted in Figure \ref{fig:kds_rd}. Correspondingly, the average probability of Type II resolutions under banking and regional distress was 69.3\%, which was lower than the equivalent probability under regional distress at 72.1\%. Liquidations under Type III resolutions remained largely unchanged with probabilities of 3.9\% and 3.2\% respectively under classes based on high regional and banking distress and solely regional distress. The inclusion of measures of banking distress to indicators of regional distress primarily resulted in substitutions between Type I and II resolutions. Overall, the densities based on the two models show that measures of regional distress continue to be the most important determinants of heterogeneity in the FDIC's decision rules and that measures of bank distress marginally augment the separation across the two classes. 
%
These findings show that banking industry distress in addition to regional distress likely contributed to a ``too-many-to-fail" response from the FDIC in line with predictions from the theoretical literature \citep{acharya2007too}.  
\subsection{Political economy factors and the FDIC's decisions for bank resolution}
\label{res2}
In this section, we address constraints to the FDIC's decision-making in the form of political pressures to provide Type I resolutions represented by Hypothesis $H_{3}$. We find that political factors played a limited role in the FDIC's decisions as the average probability of Type I assistance to banks increased marginally when the banking industry received political support. 
 
In Table \ref{tab:cmrp}, we present the covariate effects and log marginal likelihood based for specifications (11) through (15) based on measures of political support and specifications (16) through (20) based on a combination of regional distress and political support. To assess political support, we develop measures of congressional voting for legislation in favor of the banking industry by representatives within each state based on the definition in \cite{economides1996political}.  
In keeping with their specification, we exclude representatives who did not vote and include the percentage of Republicans in each state to control for voting along party lines. On comparing across the log marginal likelihood, the data favor specifications that include measures of regional distress over those based on only political economy factors. In particular, specification (17), which has the largest marginal likelihood, consists of statistically important negative covariate effects for both state-level unemployment and the percentage of votes for the S\&L restructuring bill (Bill 3). This bill proposed expanding the FDIC's authority to take over insuring S\&L deposits following the failure of the FSLIC. Institutions at risk of failure in the S\&L industry would have likely benefited from and lobbied for this bill as the expanded role of the FDIC would have increased their ability to obtain assistance and function as going concerns. As a result latent class 1 will be labeled as the class of failures under ``High Regional Distress and Political Support (HRDP)" and latent class 2, as the class of failures under ``Low Regional Distress and / or Political Support (LRDP)".
\begin{table}
	\caption{\label{tab:cmrp} Covariate effects from class-membership models for specifications of latent classes based on regional distress and political support. The reported values are posterior means of the covariate effects. Posterior standard deviations are in parentheses. The details underlying the bills in these model specifications are provided in Section \ref{sec:data_app}.}
	\centering
	\scalebox{0.68}{
	\begin{tabular}{lccccc}
		\hline
		& \multicolumn{1}{c}{(11)} & \multicolumn{1}{c}{(12)} & \multicolumn{1}{c}{(13)} & \multicolumn{1}{c}{(14)} & \multicolumn{1}{c}{(15)} \\
		\hline
		\textbf{State-level characteristics} &       &       &       &       &  \\
		Unemployment & \multicolumn{1}{l}{-} & \multicolumn{1}{l}{-} & \multicolumn{1}{l}{-} & \multicolumn{1}{l}{-} & \multicolumn{1}{l}{-} \\
		Housing starts & \multicolumn{1}{l}{-} & \multicolumn{1}{l}{-} & \multicolumn{1}{l}{-} & \multicolumn{1}{l}{-} & \multicolumn{1}{l}{-} \\
		\textbf{County-level characteristics} &       &       &       &       &  \\
		Per capita GDP growth & \multicolumn{1}{l}{-} & \multicolumn{1}{l}{-} & \multicolumn{1}{l}{-} & \multicolumn{1}{l}{-} & \multicolumn{1}{l}{-} \\
		Farm, agri, mining & \multicolumn{1}{l}{-} & \multicolumn{1}{l}{-} & \multicolumn{1}{l}{-} & \multicolumn{1}{l}{-} & \multicolumn{1}{l}{-} \\
		\textbf{Insurer characteristics} &       &       &       &       &  \\
		Dep. Ins. Fund/ Total Deposits & \multicolumn{1}{l}{-} & \multicolumn{1}{l}{-} & \multicolumn{1}{l}{-} & \multicolumn{1}{l}{-} & \multicolumn{1}{l}{-} \\
		\textbf{Political economy characteristics} &       &       &       &       &  \\
		\% vote for Bill 4 & \multicolumn{1}{l}{-0.01 (0.01)} & \multicolumn{1}{l}{-} & \multicolumn{1}{l}{-} & \multicolumn{1}{l}{-} & \multicolumn{1}{l}{-} \\
		\% vote for Bill 3 & \multicolumn{1}{l}{-} & \multicolumn{1}{l}{-0.02 (0.01)} & \multicolumn{1}{l}{-} & \multicolumn{1}{l}{-} & \multicolumn{1}{l}{-} \\
		\% vote for Bill 2 & \multicolumn{1}{l}{-} & \multicolumn{1}{l}{-} & \multicolumn{1}{l}{0.01 (0.01)} & \multicolumn{1}{l}{-} & \multicolumn{1}{l}{-} \\
		\% vote for Bill 1 & \multicolumn{1}{l}{-} & \multicolumn{1}{l}{-} & \multicolumn{1}{l}{-} & \multicolumn{1}{l}{0.01 (0.01)} & \multicolumn{1}{l}{-} \\
		\% vote for Bill 5 & \multicolumn{1}{l}{-} & \multicolumn{1}{l}{-} & \multicolumn{1}{l}{-} & \multicolumn{1}{l}{-} & \multicolumn{1}{l}{0.01 (0.02)} \\
		\% Republicans & \multicolumn{1}{l}{0.01 (0.01)} & \multicolumn{1}{l}{0.02 (0.01)} & \multicolumn{1}{l}{0.01 (0.01)} & \multicolumn{1}{l}{0.01 (0)} & \multicolumn{1}{l}{-0.01 (0.01)} \\
		\hline
		log Marginal Likelihood & -714.54 & -701.29 & -706.70 & -716.31 & -699.83 \\
		\hline
		&       &       &       &       &  \\
		\hline
		& \multicolumn{1}{c}{(16)} & \multicolumn{1}{c}{(17)} & \multicolumn{1}{c}{(18)} & \multicolumn{1}{c}{(19)} & \multicolumn{1}{c}{(20)} \\
		\hline
		\textbf{State-level characteristics} &       &       &       &       &  \\
		Unemployment & \multicolumn{1}{l}{-0.04 (0.02)} & \multicolumn{1}{l}{\textbf{-0.13 (0.04)}} & \multicolumn{1}{l}{-0.09 (0.05)} & \multicolumn{1}{l}{-0.1 (0.05)} & \multicolumn{1}{l}{-0.06 (0.03)} \\
		Housing starts & \multicolumn{1}{l}{-0.06 (0.03)} & \multicolumn{1}{l}{\textbf{-0.05 (0.04)}} & \multicolumn{1}{l}{0.07 (0.08)} & \multicolumn{1}{l}{0.03 (0.09)} & \multicolumn{1}{l}{-0.04 (0.08)} \\
		\textbf{County-level characteristics} &       &       &       &       &  \\
		Per capita GDP growth & \multicolumn{1}{l}{-0.01 (0.03)} & \multicolumn{1}{l}{\textbf{0 (0.04)}} & \multicolumn{1}{l}{0.04 (0.05)} & \multicolumn{1}{l}{0.02 (0.05)} & \multicolumn{1}{l}{0.03 (0.05)} \\
		Farm, agri, mining & \multicolumn{1}{l}{0.07 (0.04)} & \multicolumn{1}{l}{\textbf{0.09 (0.05)}} & \multicolumn{1}{l}{0.05 (0.05)} & \multicolumn{1}{l}{0.06 (0.04)} & \multicolumn{1}{l}{0.08 (0.06)} \\
		\textbf{Insurer characteristics} &       &       &       &       &  \\
		Dep. Ins. Fund/ Total Deposits & \multicolumn{1}{l}{-0.03 (0.03)} & \multicolumn{1}{l}{\textbf{-0.05 (0.03)}} & \multicolumn{1}{l}{-0.04 (0.04)} & \multicolumn{1}{l}{-0.05 (0.04)} & \multicolumn{1}{l}{-0.02 (0.03)} \\
		\textbf{Political economy characteristics} &       &       &       &       &  \\
		\% vote for Bill 4 & \multicolumn{1}{l}{-0.09 (0.02)} & \multicolumn{1}{l}{\textbf{-}} & \multicolumn{1}{l}{-} & \multicolumn{1}{l}{-} & \multicolumn{1}{l}{-} \\
		\% vote for Bill 3 & \multicolumn{1}{l}{-} & \multicolumn{1}{l}{\textbf{-0.14 (0.03)}} & \multicolumn{1}{l}{-} & \multicolumn{1}{l}{-} & \multicolumn{1}{l}{-} \\
		\% vote for Bill 2 & \multicolumn{1}{l}{-} & \multicolumn{1}{l}{\textbf{-}} & \multicolumn{1}{l}{-0.03 (0.04)} & \multicolumn{1}{l}{-} & \multicolumn{1}{l}{-} \\
		\% vote for Bill 1 & \multicolumn{1}{l}{-} & \multicolumn{1}{l}{\textbf{-}} & \multicolumn{1}{l}{-} & \multicolumn{1}{l}{-0.05 (0.06)} & \multicolumn{1}{l}{-} \\
		\% vote for Bill 5 & \multicolumn{1}{l}{-} & \multicolumn{1}{l}{\textbf{-}} & \multicolumn{1}{l}{-} & \multicolumn{1}{l}{-} & \multicolumn{1}{l}{0.21 (0.11)} \\
		\% Republicans & \multicolumn{1}{l}{0.05 (0.09)} & \multicolumn{1}{l}{\textbf{0.12 (0.05)}} & \multicolumn{1}{l}{0.03 (0.06)} & \multicolumn{1}{l}{0 (0.08)} & \multicolumn{1}{l}{0.2 (0.11)} \\
		\hline
		log Marginal Likelihood & -698.29 & \textbf{-693.68} & -705.50 & -700.32 & -704.52 \\
		\hline
	\end{tabular}}%
\end{table}%

\begin{figure}[!h]
	\centering
	\includegraphics[width=0.8\linewidth]{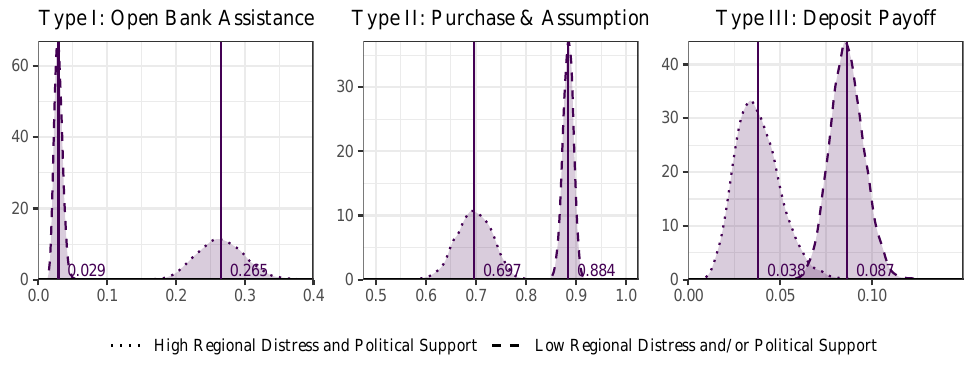}
	\caption{Posterior distribution of the average probability of the FDIC assigning each resolution method within classes based on regional distress and political economy factors. The horizontal axis represents the probability of assigning a resolution method and the vertical axis represents the posterior density associated with that probability based on a kernel density estimate. The solid vertical lines represent the mean of these posterior distributions across the $G$ MCMC draws.\label{fig:rdp}}
\end{figure}

Figure \ref{fig:rdp} plots the posterior density of the average probability of the FDIC assigning each resolution method to banks in the two classes defined by regional distress and political economy factors in specification (17). The densities reveal that in the presence of political support to the banking industry and high regional distress, the average probability of a Type I, Type II and Type III resolution is 26.5\%, 69.7\% and 3.8\% respectively. On comparing these average probabilities with the equivalent values of 24.6\%, 72.1\% and 3.2\% among bank failures that occurred amid high regional distress represented in Figure \ref{fig:kds_rd}, we find that political support for the banking industry resulted in marginally higher probability of the FDIC assigning a Type I resolution during economic distress. 
The details underlying the bills in the remaining specifications are provided in Section \ref{sec:data_app}.
\section{Savings and loans resolution by FSLIC}
\label{sec:results_fslic}
To compare the FDIC with the FSLIC, we analyze the latter's decisions through the same empirical lens and estimate the specifications described in Section \ref{sec:results_fdic} for FDIC decisions.
%
In this analysis, we continue to use the same specifications as those described in Section \ref{sec:results_fdic} for the prior distributions, their hyperparameters and the number of post-burn in MCMC samples.

\subsection{Regional economic distress and FSLIC's decisions for S\&L resolutions}
\label{sec:fslic_rd}
\begin{figure}[!t]
	\centering
	\includegraphics[width=0.85\linewidth]{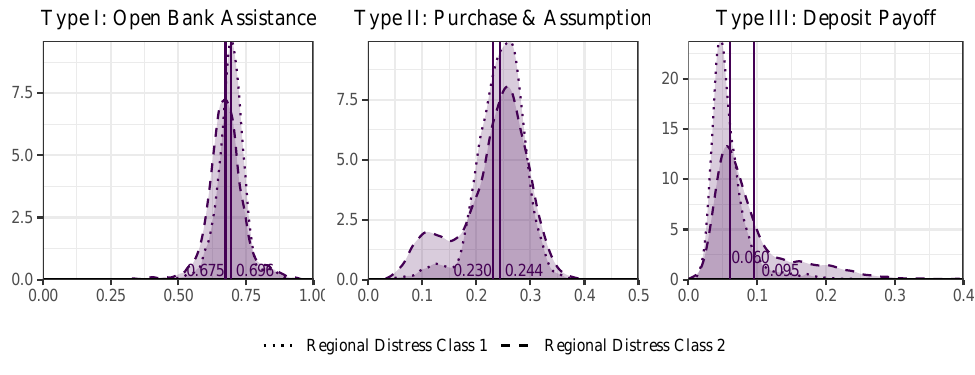}
	\caption{Posterior distribution of the average probability of the FSLIC assigning each resolution method within classes based on regional distress. The horizontal axis represents the probability of assigning a resolution method and the vertical axis represents the posterior density associated with that probability based on a kernel density estimate. The solid vertical lines represent the mean of these posterior distributions across the $G$ MCMC draws.\label{fig:rd_sl_bank}}
\end{figure}

We find that FSLIC's designation of S\&L institutions into classes based on regional economic distress is ambiguous and does not support Hypothesis $H_{1}$. 
{In particular, }the agency did not develop distinct decision rules to resolve failures that occurred amid high and low regional distress and is observed to have adopted a common decision rule for both groups of S\&L's. The implication of using a common decision rule is that institutions that failed due to weaknesses in their own risk management as well as due to economic factors are likely to have received assistance with similar probabilities. The theory underlying this hypothesis suggests that by not fully disentangling economic and idiosyncratic factors while providing assistance, the FSLIC potentially fostered moral hazard among S\&L institutions.
\\[1ex]
\noindent\textbf{Class-membership model - }
Table \ref{tab:rd_sl} summarizes the covariate effects from estimating the specifications reported in Section \ref{res1}.
\begin{table}
	\caption{\label{tab:rd_sl} Covariate effects from class-membership models of S\&L's for specifications of latent classes based on regional distress. The reported values are posterior means of the covariate effects. Posterior standard deviations are in parantheses.}
	\centering
 	\scalebox{0.8}{
	\begin{tabular}{lrrrr}
		\hline
		& \multicolumn{1}{c}{{(1$^{\dagger}$)}} & \multicolumn{1}{c}{(2$^{\dagger}$)} & \multicolumn{1}{c}{\textbf{(3$^{\dagger}$)}} & \multicolumn{1}{c}{(4$^{\dagger}$)} \\
		\hline
		\textbf{State-level characteristics} &     &     &      &      \\
		Unemployment & -0.08 (0.06) & -0.71 (0.14) & \textbf{-0.17 (0.48)} & \multicolumn{1}{r}{-0.72 (0.11)} \\
		Housing starts & -0.21 (0.1) & -0.01 (0.04) & \textbf{-0.03 (0.09)} & \multicolumn{1}{r}{-0.02 (0.04)} \\
		\textbf{County-level characteristics} &     &     &      &      \\
		Per capita GDP growth & -0.13 (0.09) & 0.02 (0.06) & \textbf{0.00 (0.09)} & \multicolumn{1}{r}{0.01 (0.05)} \\
		Farm, agri, mining & -0.09 (0.08) & 0.04 (0.04) & \textbf{-0.01 (0.09)} & \multicolumn{1}{r}{0.04 (0.03)} \\
		Manufacturing & 0 (0.07) & -     & -     & - \\
		Construction & 0.09 (0.05) & -     & -     & - \\
		Fin Serv Transport & -0.11 (0.12) & -     & -     & - \\
		Government & -0.16 (0.1) & -     & -     & - \\
		\textbf{Insurer characteristics} &     &     &      &      \\
		Dep. Ins. Fund/Total Dep. & -     & -     & \textbf{0.01 (0.08)} & - \\
		\textbf{S\&L-level characteristics} &     &     &      &      \\
		State charter & -     & -     & -     & \multicolumn{1}{l}{-0.07 (0.15)} \\
		\hline
		log Marginal Likelihood   & \multicolumn{1}{c}{-302.32} & \multicolumn{1}{c}{-305.40} & \multicolumn{1}{c}{\textbf{-302.02}} & \multicolumn{1}{c}{-305.44}\\
		\hline
	\end{tabular}}%
\end{table}%
Specification (3$^{\dagger}$) is determined to be the model selected by the data by virtue of its marginal likelihood being the largest among candidate models. The covariate effects for unemployment, per capita income and housing starts in this specification are, however, not statistically important. Accordingly, the latent classes generated by this model cannot be distinguished as representing ``high" or ``low" regional distress and are labeled as ``Regional Distress Class 1" and ``Regional Distress Class 2". 
\\[1ex]
\noindent\textbf{Heterogeneity in Decision Rules - }
Figure \ref{fig:rd_sl_bank} shows the posterior densities of the average probability of the FSLIC assigning each resolution method to S\&L's in the two latent classes. These distributions offer two main insights into the decisions of the FSLIC. First, on comparing with Figure \ref{fig:kds_rd}, it is clear that the FSLIC relied more heavily on Type I resolutions relative to the FDIC. The average probabilities of the FSLIC assigning a Type I resolution were 67.5 \% and 69.6\% in class 1 and 2 compared with probabilities of 24.6 \% and 3.3\%  of the FDIC assisting banks that failed in high and low regional distress respectively. Second, the FSLIC recognizably deviated from the recommended resolution strategy developed in \cite{cordella2003bank} since the posterior densities for the two classes overlap across all three resolution methods. Accordingly, the average probabilities of receiving a Type I resolution are not statistically different across the two classes. This finding signifies that the FSLIC did not distinguish between S\&L institutions that failed amid high and low economic distress in assigning Type I assistance.   
\\[1ex]
\noindent\textbf{Resolution Type - }
\begin{figure}[!t]
		\centering
	\includegraphics[width=0.75\linewidth]{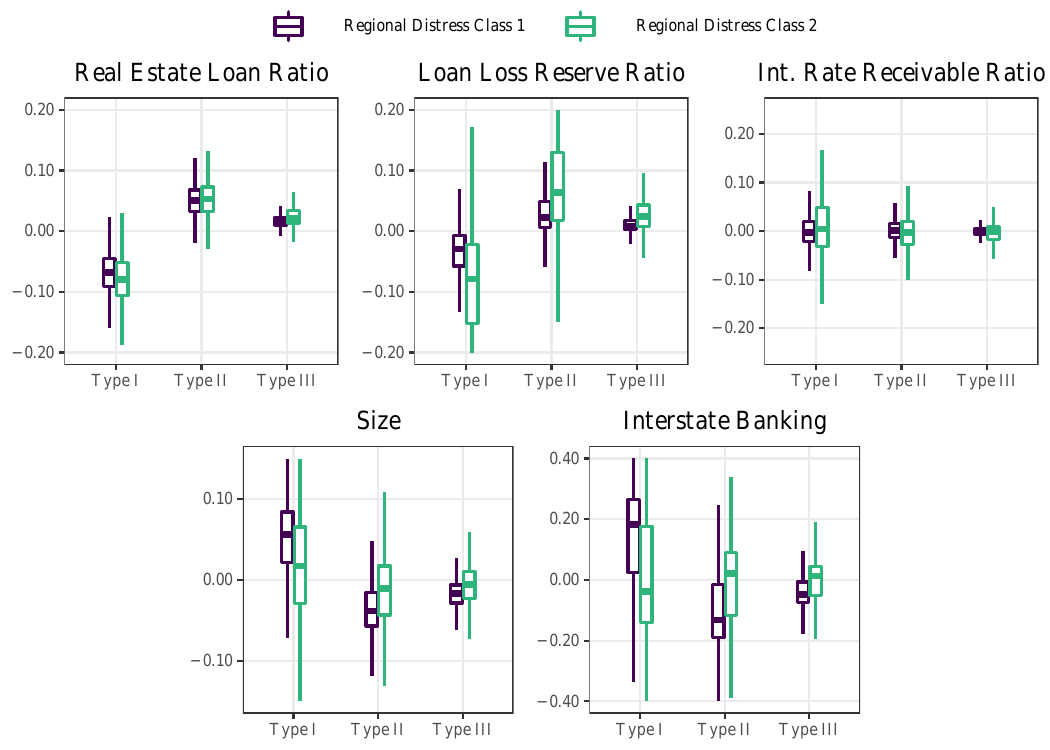}
	\caption{Covariate effects from the models for resolution type for S\&L's in the classes based on regional distress.\label{fig:combined_sl}}	
\end{figure}
The box plots for covariate effects from specification (3$^\dagger$) in Figure \ref{fig:combined_sl} show that the FSLIC adopted a common decision rule in resolving institutions in the two classes. The covariate effects of S\&L financial characteristics are homogeneous across the two latent classes. This contrasts with the covariate effects on the FDIC's responses in Figure \ref{fig:covarEff}, which are statistically different across the classes of banks that failed amid high and low regional distress. In both latent classes in the FSLIC's decision structure, a standard deviation increase in Real Estate Loan Ratio is seen to result in an average decline of around  7\% in the probability of receiving a Type I resolution and a corresponding increase in the probability of receiving the other two resolution methods. Since real estate prices collapsed across several regions during this period, the FSLIC likely viewed institutions with excess concentrations in real estate lending as being vulnerable to  defaults and losses and thereby, incapable of being revived through assistance. In both latent classes, the FSLIC responded to Larger Loan Loss Reserve Ratios by withholding assistance and instead facilitating their acquisition or liquidating them. This suggests that the FSLIC viewed larger loss reserves as a signal of greater deterioration in the asset quality of the failed institution.

The FSLIC did not respond to changes in the Interest Rate Receivable Ratio, which has been shown to be correlated with the losses resulting from the failed institution to the resolution agency \citep{balla2019comparison}. The average change in the probability of assigning each resolution method is close to zero in response to a standard deviation increase in this measure. This finding suggests that the FSLIC did not distinguish across institutions based on their shares of loans on which interest was receivable, namely, loans with payments past due. The FSLIC's decisions with respect to bank size were similar to those of the FDIC. Larger institutions were more likely to be assisted by the agency and less likely to be sold or liquidated across both latent classes, thereby reflecting the ``too-big-to-fail" doctrine in the agency's decisions. The FSLIC assigned Type I assistance to a greater extent in states where interstate banking was permitted, but this effect was solely present in class 1. Average covariate effects were close to zero for all three resolution types in class 2. The variable ``Interstate banking" pertains only to the deregulation within the banking industry as the S\&L industry was not subject to such restrictions \citep{roster1985modern}. The FSLIC likely provided assistance with greater probability in states that permitted interstate banking as heightened competition from banks across state borders likely weighed on the health of the S\&L industry in those states. The covariate effects of the remaining variables are provided in Section \ref{sec:coveffects_fslic_app} of the supplement.   

{Overall, the FSLIC's decision rules did not differ by the extent of regional distress that accompanied the failure of S\&L's. This finding suggests that the FSLIC did not adhere to the recommendation from theoretical studies of assigning S\&L's to distinct decision rules depending on whether the failure was likely related to broader economic factors or the institution's idiosyncratic weaknesses.} 
\subsection{S\&L industry distress and FSLIC's decisions for S\&L resolution}
\label{sldist}
The FSLIC's resolution decisions do not support Hypothesis $H_{2}$ 
{as the agency} did not distinguish between institutions that failed amid elevated levels of distress in the industry from those that failed in more benign industry conditions in assigning resolution decisions. As a result, the probability of the FSLIC assigning a Type I resolution was not statistically different across the two groups of S\&L's.
\begin{table}
	\caption{\label{tab:rdb_sl} Covariate effects from class-membership models for specifications of latent classes based on regional and S\&L industry distress. The reported values are posterior means of the covariate effects. Posterior standard deviations are in parantheses.}
	\centering
	\scalebox{0.8}{
	\begin{tabular}{lrrrrrr}
		\hline
		&
		\multicolumn{1}{c}{(5$^{\dagger}$)} & \multicolumn{1}{c}{(6$^{\dagger}$)} & \multicolumn{1}{c}{(7$^{\dagger}$)} & \multicolumn{1}{c}{\textbf{(8$^{\dagger}$)}} & \multicolumn{1}{c}{ {(9$^{\dagger}$)}} & \multicolumn{1}{c}{(10$^{\dagger}$)} \\
		\hline
		\textbf{State-level characteristics} &     &     &      &     &  & \\
		Unemployment & -     & -     & -     & \textbf{-0.19 (0.23)} & 0.06 (0.05) & 0.03 (0.09) \\
		\textbf{County-level characteristics} &     &     &      &     &  & \\
		Housing starts & -     & -     & -     & \textbf{-0.08 (0.08)} & 0.10 (0.07) & 0.07 (0.1) \\
		Per capita GDP growth & -     & -     & -     & \textbf{-0.10 (0.09)} & 0.13 (0.07) & 0.17 (0.1) \\
		Farm, agri, mining & -     & -     & -     & \textbf{-0.05 (0.07)} & 0.06 (0.05) & 0.04 (0.05) \\
		\multicolumn{2}{l}{\textbf{S\&L industry characteristics}} &     &     &      &     &   \\
		Previous closures & 0.12 (0.18) & 0.11 (0.18) & 0.07 (0.12) & \textbf{-0.03 (0.07)} & 0.03 (0.08) & 0.03 (0.05) \\
		\% Assets in distressed S\&L's & -0.01 (0.1) & -     & -     & \textbf{0.02 (0.07)} & -     & - \\
		\% Dep. in distressed S\&L's  & -     & -0.02 (0.09) & -     & -     & -0.05 (0.02) & - \\
		\% distressed S\&L's & -     & -     & -0.02 (0.08) & -     & -     & -0.09 (0.08) \\
		\textbf{Insurer characteristics} &     &     &      &    &  & \\
		Dep. Ins. Fund/ Total Dep. & -     & -     & -     & \textbf{0.03 (0.05)} & {-0.03 (0.02)} & -0.01 (0.06) \\
		\hline
		log Marginal Likelihood & \multicolumn{1}{c}{-301.40} & \multicolumn{1}{c}{-300.83} & \multicolumn{1}{c}{-300.95} & \multicolumn{1}{c}{\textbf{-297.15}} & \multicolumn{1}{c}{-300.69} & \multicolumn{1}{c}{-300.33} \\
		\hline
	\end{tabular}}%
\end{table}

Table \ref{tab:rdb_sl} reports the covariate effects and log marginal likelihood from specifications (5$^\dagger$) through (7$^\dagger$) that are exclusively based on measures of distress in the S\&L industry as well as from (8$^\dagger$) through (10$^\dagger$), which augment the specifications based on industry distress with measures of regional distress.  The definition of a distressed S\&L in the specifications reported in the table is an institution whose Texas ratio (Equation \eqref{eq:texasratio} in supplement \ref{sec:data_app}) exceeded 100\% and is consistent with the definition of distressed banks in Section \ref{ds}. 

Bayesian model comparison identifies specification (8$^\dagger$) as the selected model since it exhibits the highest marginal likelihood, and consequently the largest posterior odds relative to other specifications. However, neither the covariates pertaining to industry distress, namely, previous closures and the percentage of assets in distressed institutions, nor the measures of regional economic performance such as unemployment, housing starts and per capita income growth are statistically important in this specification. The two latent classes do not necessarily represent differences based on local economic or S\&L industry characteristics. Accordingly, the two classes are labeled as ``Regional and Banking Distress Class 1" and ``Regional and Banking Distress Class 2". 
\begin{figure}[!t]
	\centering
	\includegraphics[width=0.85\linewidth]{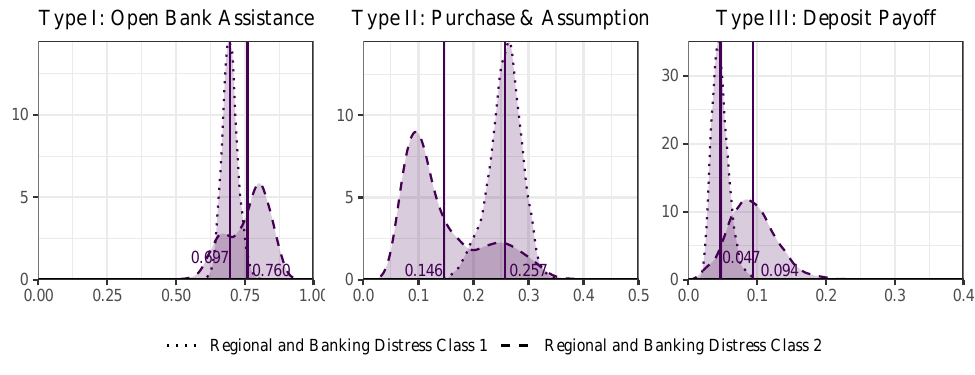}	
	\caption{Posterior distribution of the average probability of the FSLIC assigning each resolution method within classes based on regional and S\&L industry distress. The horizontal axis represents the probability of assigning a resolution method and the vertical axis represents the posterior density associated with that probability based on a kernel density estimate. The solid vertical lines represent the mean of these posterior distributions across the $G$ MCMC draws. \label{fig:rdb_sl}}
\end{figure} 

In Figure \ref{fig:rdb_sl}, we find that the average probability of the FSLIC assigning a Type I resolution are 69\% and 76\% in latent classes 1 and 2 respectively. However, the average probabilities of a Type I resolution are not statistically different across the two classes. 
The average probability of a Type II resolution is 25.7\% in class 1 and 14.6\% in class 2. Despite lesser overlap across the two densities for Type II resolutions relative to Type I resolutions, the average probabilities of the FSLIC assigning a Type II resolution are not statistically different across the two classes of S\&L's. Finally, the posterior densities of the probability of a Type III resolution entirely overlap in classes 1 and 2 with averages of 4.7\% and 9.4\% respectively, and are not statistically different from each other. 

One of the implications of the theory from \cite{acharya2007too,acharya2007cash} on which Hypothesis $H_{2}$ is based is that the probability of Type II resolutions are likely to be statistically higher among institutions that failed in high industry distress relative to failures in low industry distress. Widespread distress in the S\&L industry stymies the demand for failed S\&L's, which results in fewer Type II resolutions and thereby creates the necessity for assistance in the form of Type I resolutions. Since the two latent classes are not explicitly based on industry distress, the FSLIC's resolution decisions are not consistent with this effect of S\&L industry distress and thereby do not support any of the implications of Hypothesis $H_{2}$.
\subsection{Political economy factors and the FSLIC's decisions for S\&L resolution}
\label{sec:h3_fslic}
The FSLIC's resolution decisions support Hypothesis $H_{3}$, which states that the agency was more likely to assign Type I assistance to institutions that received political support. This finding is a departure from results in previous sub-sections, which did not support hypotheses based on assigning distinct decision rules for institutions that failed in varying levels of regional or industry distress. Among institutions that failed in the presence of political support, however, the probability of receiving a Type I resolution was statistically higher relative to institutions that failed in the absence of such support. 

The specifications relating to $H_{3}$ include measures of congressional voting on bills relating to the banking and S\&L industry in line with the specifications developed in Section \ref{res2}. The estimation of latent class models for S\&L resolutions is constrained by the presence of only 15 Type III resolutions (see Table \ref{tab:summt} in supplement Section \ref{sec:data_app}) and therefore a subset of specifications are estimable. The measures of voting evaluated in this section represent the result of lobbying by industry groups as well as elected representatives' concern for the health of financial institutions in their constituencies. A range of studies documented widespread lobbying by the S\&L industry to influence representatives on legislation pertaining to regulation in the period leading up to the S\&L crisis \citep{mason2004buildings, lowy1991high}. However, the extent to which the FSLIC's decisions were persuaded by such lobbying ventures has not been formally evaluated. We find that political economy factors played a prominent role in the decisions of the FSLIC.
\begin{table}
	\caption{\label{tab:rdp_sl} Covariate effects from class-membership models of S\&L's for specifications of latent classes based on regional distress and political support. The details underlying the bills in these model specifications are provided in Section \ref{sec:data_app}.}
		\centering
  	\scalebox{0.7}{
	\begin{tabular}{lcccc}
		\hline
		& \multicolumn{1}{c}{($12^\dagger$)} & \multicolumn{1}{c}{($13^\dagger$)} & \multicolumn{1}{c}{($14^\dagger$)} & \multicolumn{1}{c}{($16^\dagger$)} \\
	\hline
		\textbf{State-level characteristics} &       &       &       &  \\
		Unemployment  & -     & -     & -     & -0.16 (0.13) \\
		Housing starts & -     & -     & -     & -0.03 (0.09) \\
		\textbf{County-level characteristics} &       &       &       &  \\
		Per capita GDP growth & -     & -     & -     & 0.13 (0.12) \\
		Farm, agri, mining & -     & -     & -     & 0.03 (0.07) \\
		\textbf{Insurer characteristics} &       &       &       &  \\
		Dep. Ins. Fund/ Total Deposits & -     & -     & -     & 0 (0.07) \\
		\textbf{Political economy characteristics} &       &       &       &  \\
		\% vote for Bill 4 & -     & -     & -     & -0.22 (0.12) \\
		\% vote for Bill 3 & 0.36 (0.22) & -     & -     & - \\
		\% vote for Bill 2 & -     & -0.25 (0.1) & -     & - \\
		\% vote for Bill 1 & -     & -     & -0.36 (0.09) & - \\
		\% vote for Bill 5 & -     & -     & -     & - \\
		\% Republicans & 0.08 (0.15) & 0.24 (0.13) & 0.3 (0.09) & 0.18 (0.11) \\
	\hline
		log Marginal Likelihood & -299.08 & -294.01 & -300.04 & -301.15 \\
	\hline
		&       &       &       &  \\
	\hline
		& \multicolumn{1}{c}{($17^\dagger$)} & \multicolumn{1}{c}{($\mathbf{18^\dagger}$)} & \multicolumn{1}{c}{($19^\dagger$)} & \multicolumn{1}{c}{($20^\dagger$)} \\
	\hline
		\textbf{State-level characteristics} &       &       &       &  \\
		Unemployment  & -0.12 (0.07) & \textbf{-0.03 (0.06)} & -0.02 (0.06) & -0.2 (0.14) \\
		Housing starts & 0.04 (0.07) & \textbf{0.03 (0.07)} & 0.03 (0.07) & 0.01 (0.08) \\
		\textbf{County-level characteristics} &       &       &       &  \\
		Per capita GDP growth & 0.06 (0.06) & \textbf{0.21 (0.1)} & 0.08 (0.1) & 0.07 (0.09) \\
		Farm, agri, mining & -0.02 (0.04) & \textbf{-0.04 (0.05)} & -0.02 (0.06) & 0.01 (0.05) \\
		\textbf{Insurer characteristics} &       &       &       &  \\
		Dep. Ins. Fund/ Total Deposits & -0.09 (0.06) & \textbf{-0.04 (0.06)} & 0.03 (0.05) & 0.09 (0.07) \\
		\textbf{Political economy characteristics} &       &       &       &  \\
		\% vote for favor Bill 4 & -     & \textbf{-} & -     & - \\
		\% vote for Bill 3 & 0.36 (0.11) & \textbf{-} & -     & - \\
		\% vote for Bill 2 & -     & \textbf{-0.14 (0.05)} & -     & - \\
		\% vote for Bill 1 & -     & \textbf{-} & -0.25 (0.1) & - \\
		\% vote for Bill 5 & -     & \textbf{-} & -     & 0.25 (0.08) \\
		\% Republicans & 0.16 (0.04) & \textbf{0.26 (0.08)} & 0.2 (0.09) & 0.05 (0.08) \\
	\hline
		log Marginal Likelihood & -297.52 & \textbf{-288.06} & -299.10 & -300.40 \\
	\hline
	\end{tabular}}%
\end{table}%

In Table \ref{tab:rdp_sl}, the selected model is specification (18$^\dagger$) by virtue of its larger marginal likelihood relative to all other specifications.    This model determines latent classes based on the percent of votes in favor of a bill to reform the federal deposit insurance system and to restore civil penalties for criminal offenses involving financial institutions (Bill 2). Since the bill introduced punitive measures against the financial industry, political support for the industry is measured using votes against the bill. The covariate effects for per capita income growth, the percent vote in favor of the proposed bill and the control for the share of Republican representatives are statistically important in the selected model. Since latent class 2 consists of institutions that failed in counties with a high per capita income growth and a low share of votes in favor of the bill, this class is labeled as ``High Political Support and Low Regional Distress" and latent class 1 is the class of ``Low Political Support and High Regional Distress". 
\begin{figure}[!t]
	\centering
	\includegraphics[width=0.85\linewidth]{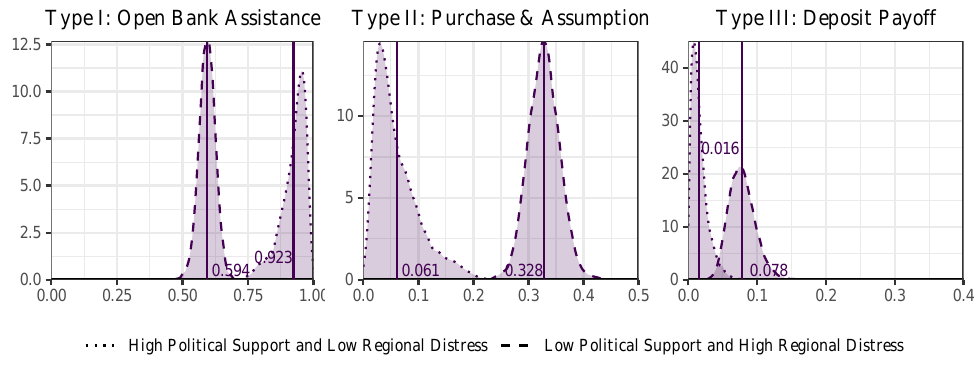}		
	\caption{Posterior distribution of the average probability of the FSLIC assigning each resolution method within classes based on regional distress and political economy factors. The horizontal axis represents the probability of assigning a resolution method and the vertical axis represents the posterior density associated with that probability based on a kernel density estimate. The solid vertical lines represent the mean of these posterior distributions across the $G$ MCMC draws. \label{fig:rdp_sl}}
\end{figure} 

Figure \ref{fig:rdp_sl} plots the posterior density of the probability of receiving each resolution method under specifications (18$^\dagger$). Where previously, the densities in figures \ref{fig:rd_sl_bank} and \ref{fig:rdb_sl} overlapped considerably, the inclusion of measures of political support notably generates distinct latent classes with minimal overlap. This shows that the difference between the classes of S\&L's that failed in the presence of high regional distress and low political support, and those that failed in a climate of low regional distress but high degree of political support is statistically important. The FSLIC assigned Type I resolutions to institutions that failed in regions with high political support but low regional distress with average probability of 92.3\% and to institutions that failed in high regional distress and low political support with a probability of 59.4\%. This finding suggests that political support outweighed the effects of economic forces in determining the eligibility of failed S\&L's for assistance. Notably, the FSLIC was also less likely to liquidate institutions in regions that received political support under Type III resolutions. The FSLIC liquidated institutions that failed in the presence of political support and low distress with a probability of only 1.6\%, but those in regions of high distress and low political support with a probability of 7.8\%. Finally, the FSLIC facilitated Type II transactions with a probability of 6.1\% and 7.8\% to classes representing higher and lower levels of political support respectively. Overall, the probabilities of assigning each resolution method was statistically different across the two latent classes. 

The FSLIC's resolution decisions were more likely to have been driven by political factors as predicted by Hypothesis $H_{3}$ rather than by the extent of economic or industry distress accompanying the failure of S\&L's. 
Furthermore, the FSLIC deviated from theoretically recommended decision rules in hypotheses $H_{1}$ and $H_{2}$ that address moral hazard and financial stability concerns. These findings suggest that the FSLIC's decisions did not best serve the public interest during the height of the S\&L crisis. 
\section{Discussion and contemporary relevance}
\label{sec:discuss}
During banking crises, financial regulators intervene to bail out certain failed institutions and liquidate others. Regulators are expected to meet the dual objectives of preserving financial stability and discouraging moral hazard in the process of reaching such decisions. However, they may deviate from socially optimal resolution decisions in the absence of regular oversight.  The risk of transgressions creates a need for the public to regularly evaluate regulators' actions. 

An important line of inquiry in evaluating regulatory agencies involves determining whether they tailored their decision rules to distinct economic and industry conditions, such as high and low levels of regional distress. Since the true assignment of banks into distinct decision rules is unobservable, we have developed a Bayesian latent class estimation framework to detect unobserved heterogeneity in the resolution decisions of banks based on underlying economic and industry conditions. This flexible estimation approach permits inferences on a broad range of measures, including the probability of receiving different resolution methods. Bayesian model comparison exercises predicated on posterior odds inform the selection of models that best explain the decision rules of regulatory agencies. These inferences ultimately enable us to assess whether regulators distinguished across banks based on economic and industry conditions, and whether they applied statistically different decision rules across the classes of banks. 

We utilize this modeling framework to assess the responses of the FDIC and the FSLIC to bank and S\&L failures respectively during the crises of the 1980's in the two industries. Our results show that the decision rules of the FSLIC, which subsequently faced insolvency at a significant cost to taxpayers, were inconsistent with the decision rules recommended by theoretical studies that address the trade-offs between financial stability and moral hazard. The FDIC, which survived the earlier crisis as well as the financial crisis of 2008, was found to have undertaken decisions that were consistent with those recommended rules. These findings consequently validate the applicability of this approach to assess regulators.

A regular assessment of resolution authorities by lawmakers and the public can uncover gaps between observed and recommended resolution rules and reveal the source of such gaps. For instance, structural differences in the banking and S\&L industries likely contributed to the divergent resolution strategies of the FDIC and FSLIC. The S\&L industry underwent notable deregulation that expanded their business activities while being subject to fewer safety and soundness requirements, thereby resulting in a deeper crisis in the S\&L industry than in the banking industry (9\% S\&L failures relative to 2\% bank failures). Assessments of the FSLIC could have revealed the issues arising from deregulation of the S\&L industry and prompted a review of the structure of the agency, preventing both, the failure of the agency and the ensuing costs to the taxpayer.  The insights from this study can potentially guide the development of resolution strategies among newer agencies such as the Single Resolution Board under the European Central Bank. 
\bibliography{scibibTYv3}
\bibliographystyle{chicago}
\newpage
\appendix
\begin{center}
\Large\textbf{Supplementary materials -- Do financial regulators act in the public's interest? A Bayesian latent class estimation framework for assessing regulatory responses to banking crises}
\end{center}
This supplement is organized as follows. Section \ref{sec:data_app} provides additional details about the data used in this paper, Section \ref{sec:estimation_app} presents the collapsed Gibbs sampler for posterior inference in the our latent class ordinal model of Section \ref{sec:model} and Section \ref{fest} presents the corresponding full Gibbs sampler.  Section \ref{coeffform} describes the calculation of covariate effects and Section \ref{mcs} provides additional details about model comparison. Section \ref{sim} reports the results of a simulation study to assess the performance of the collapsed Gibbs sampler discussed in Section \ref{sec:estimation_app}. Sections \ref{sec:coveffect_banks_app} and \ref{sec:coveffects_fslic_app} include, respectively, additional covariate effects pertaining to the FDIC's and FSLIC's resolution decisions. Finally, Section \ref{app:additional_emp_analysis} provides additional details on the empirical analyses of sections \ref{sec:results_fdic} and \ref{sec:results_fslic}.
\section{Summary statistics and description of bills}
\label{sec:data_app}
Table \ref{tab:data_dic} provides the data dictionary and tables \ref{tab:summ}, \ref{tab:summt} present descriptive statistics of the data described in Section \ref{sec:data}. In these tables, the Texas Ratio for a bank is defined as,
\begin{equation}
	\label{eq:texasratio}
 	\mbox{Texas Ratio} = \frac{\mbox{Non-performing Assets}}{\mbox{Tangible Equity + Loan Loss Reserves}}.
 \end{equation} 
The Texas-Ratio is a measure of distress as it identifies institutions whose capital would be insufficient to absorb losses that could emanate from nonperforming assets.

In this paper, we consider the following five bills that were favorable to the banking or S\&L industry.
\begin{enumerate}
	\item Bill 1 - reform, recapitalize and consolidate the federal deposit insurance system as well as to enhance the powers of federal regulatory agencies.
	\item Bill 2 - introduce additional checks on the banking industry by proposing the restoration of civil penalties for criminal offenses involving financial institutions
	\item Bill 3 - restructure the S\&L industry and recommended that the FDIC insure deposits held at S\&L institutions in addition to commercial banks following the failure of the FSLIC. This bill also authorized the establishment of the Resolution Trust Corporation (RTC) to resolve failed S\&L institutions that had been within the purview of the FSLIC and any additional failures that arose within the next three years.
	\item Bill 4 -  The Competitive Equality Bank Act  (CEBA) of 1987 provided the FDIC with the option to establish a temporary national bank or a bridge bank for a maximum period of three years. This option served as an alternative to liquidation when acquirers were not forthcoming for purchasing a bank in the period immediately following its failure \citep{huber1988competitive}
	\item Bill 5 - disclosure of ratings assigned to banks and thrifts under the Community Reinvestment Act (CRA).
\end{enumerate}
With the exception of Bill 4, all other bills described above are components of the Financial Institutions Reform, Recovery and Enforcement Act(FIRREA) of 1989. In tables \ref{tab:summ} and \ref{tab:summt}, the category `\% vote for Bill $k$' represents the percentage of Congressional representatives from the state in which the bank (S\&L) operated who voted for bill $k$ that was favorable to the banking or S\&L industry, where $k=1,\ldots,5$.
\begin{table}
	\caption{\footnotesize \label{tab:data_dic} The data dictionary. All state level and county level characteristics pertain to those U.S. states and counties where the banks (S\&L's) failed during the crises of 1980's.}
	\centering
	\fbox{%
		\scalebox{0.7}{
	\begin{tabular}{ll}
		\textbf{Covariates} & \multicolumn{1}{l}{\textbf{Description}} \\
		\hline
		\textbf{Bank and S\&L level characteristics} &  \\
		C\&I Loan Ratio &  ratio of commercial and industrial loans to total loans\\ 
		CLD Loan Ratio &  ratio of construction and land development loans to total loans\\ 
		Real Estate Loan Ratio & ratio of real estate loans to total loans \\ 
		Loan Loss Reserves Ratio  &  ratio of reserves set aside against expected losses to total loans\\ 
		Nonperforming loans Ratio (for banks) &  ratio of loans past due and loans in non-accruing status to total assets \\
		Interest Receivable Ratio &  ratio of interest income earned but not collected to total assets \\
		Securities Ratio &   ratio of securities (e.g., Treasuries, mortgage-backed securities) to total assets\\
	Core Deposits Ratio (for banks) &ratio of the sum of transaction deposits of individuals, governments,\\ 
	& corporations,	money market deposit accounts, and\\
	& time deposits with balances less than \$100,000 relative to total liabilities  \\ 
		Earnings  (for banks) &  ratio of net income to total assets \\
		Size(Assets mlns.) & total assets in millions of dollars \\ 
		State charter Fed member (for banks) &   indicator depicting state-chartered banks that are members of\\
			& the Federal Reserve system\\
		State charter non-Fed member (for banks) &  indicator depicting state-chartered banks that are not members of\\& the Federal Reserve system\\
		State charter (for S\textbackslash{}\&L's) &  indicator depicting state-chartered S\&L's\\ 
		\hline
		\textbf{Insurer characteristic} &  \\
		Dep. Ins. Fund/ Total Deposits &  ratio of insurance funds available to total deposits \\
		\hline
		\textbf{State Characteristics} &  \\
		Interstate banking & whether interstate banking is legal in the state in which the bank (S\&L) failed \\
		Unemployment &  unemployment rate in the state in which the bank (S\&L) failed \\
		Housing starts &  number of new house constructions in the state in which the bank (S\&L) failed \\
		\hline
		\textbf{County characteristics} &  \\
		Per capita GDP growth & per capita year-on-year growth of county GDP \\ 
		Farm, Agri and Mining &  quarterly share of employment in farming, agriculture and mining\\ 
		Manufacturing & quarterly share of employment in manufacturing \\
		Construction &  quarterly share of employment in construction\\
		Fin Serv and Transport &  quarterly share of employment in financial services and transportation\\
		Government & quarterly share of employment in government \\
		\hline
		\textbf{County-level characteristics of bank distress} &  \\
		\% Assets in Banks with Texas Ratio $>$ 100\% & percent of assets in banks with Texas Ratio $>$ 100\%\\
			& relative to total banking assets in the county \\
		\% Deposits in Banks with Texas Ratio $>$ 100\% & percent of deposits in banks with Texas Ratio $>$ 100\% relative to\\
			& total bank deposits in the county\\
		\% banks with Texas Ratio $>$ 100\% & percent of banks in the county with Texas Ratio $>$ 100\%  \\
		Previous Closures &  number of bank (S\&L) closures in the county in the previous year\\
		\hline
		\textbf{State-level political economy characteristics} &  \\
		\% Republicans in 1987 &  \% of Republicans in Congress in 1987\\
		\% Republicans in 1989 & \% of Republicans in Congress in 1989 \\
		\% vote for Bill 1: enhancing reg. agencies powers & \%  of Congressional votes in favor of Bill 1 \\
		 \% vote for Bill 2: restoring civil penalties for fin. Inst. & \%  of Congressional votes in favor of Bill 2 \\
		\% vote for Bill 3: recommitting S\&L restructuring bill & \%  of Congressional votes in favor of Bill 3 \\
		\% vote for Bill 4: CEBA & \%  of Congressional votes in favor of Bill 4 \\
		\% vote for Bill 5: disclosure of CRA ratings & \%  of Congressional votes in favor of Bill 5 \\
	\end{tabular}}}%
	\label{tab:addlabel}%
\end{table}%
\begin{table}
	\caption{\footnotesize \label{tab:summ} Descriptive statistics of bank, county and state-level characteristics in the data sample for FDIC resolution decisions. See Section \ref{sec:data_app} of the supplement for a description of the bills evaluated in this analysis under the state-level political economy characteristics.}
	\centering
	\fbox{%
		\scalebox{0.8}{
			\begin{tabular}{lcccccc}
				& \multicolumn{2}{c}{Type I (OBA)} & \multicolumn{2}{c}{Type II (P\&A)} & \multicolumn{2}{c}{Type III (PO)} \\
				\hline
				& Mean  & Std. dev & Mean  & Std. dev & Mean  & Std. dev \\
				\hline
				\textbf{Bank-level characteristics} &       &       &       &       &       &  \\
				C\&I Loan Ratio & 27\%  & 13\%  & 27\%  & 16\%  & 28\%  & 16\% \\
				CLD Loan Ratio & 6\%   & 8\%   & 5\%   & 7\%   & 4\%   & 8\% \\
				Real Estate Loan Ratio & 39\%  & 17\%  & 42\%  & 21\%  & 31\%  & 20\% \\
				Loan Loss Reserves Ratio  & 6\%   & 6\%   & 4\%   & 3\%   & 5\%   & 7\% \\
				Nonperforming loans Ratio & 6\%   & 6\%   & 8\%   & 5\%   & 8\%   & 6\% \\
				Interest Receivable Ratio & 1\%   & 1\%   & 1\%   & 1\%   & 2\%   & 1\% \\
				Securities Ratio & 12\%  & 12\%  & 13\%  & 10\%  & 14\%  & 11\% \\
				Core Deposits Ratio & 63\%  & 18\%  & 73\%  & 13\%  & 74\%  & 15\% \\
				Earnings  & -3\%  & 6\%   & -3\%  & 4\%   & -4\%  & 4\% \\
				Size(Assets mlns.) & 212   & 654   & 171   & 893   & 47    & 96 \\
				State charter Fed member & 4\%   & 20\%  & 7\%   & 26\%  & 14\%  & 35\% \\
				State charter non-Fed member & 50\%  & 50\%  & 53\%  & 50\%  & 48\%  & 50\% \\
					\hline
				\textbf{Insurer characteristic} &       &       &       &       &       &  \\
				Dep. Ins. Fund/ Total Deposits & 14\%  & 11\%  & 16\%  & 23\%  & 10\%  & 20\% \\
					\hline
				\textbf{State Characteristics} &       &       &       &       &       &  \\
				Interstate branching & 0.86  & 0.34  & 0.77  & 0.42  & 0.51  & 0.50 \\
				Unemployment & 8\%   & 1\%   & 7\%   & 2\%   & 6\%   & 1\% \\
				Housing starts & 13\%  & 7\%   & 11\%  & 13\%  & 16\%  & 17\% \\
					\hline
				\textbf{County characteristics} &       &       &       &       &       &  \\
				Per capita GDP growth & 3\%   & 3\%   & 5\%   & 5\%   & 6\%   & 7\% \\
				Farm, Agri and Mining & 8\%   & 8\%   & 11\%  & 11\%  & 16\%  & 14\% \\
				Manufacturing & 11\%  & 5\%   & 11\%  & 7\%   & 8\%   & 5\% \\
				Construction & 6\%   & 1\%   & 5\%   & 2\%   & 5\%   & 2\% \\
				Fin Serv and Transport & 39\%  & 7\%   & 36\%  & 9\%   & 36\%  & 11\% \\
				Government & 15\%  & 6\%   & 16\%  & 7\%   & 16\%  & 6\% \\
					\hline
				\textbf{County-level characteristics of bank distress} &       &       &       &       &       &  \\
				\% Assets in Banks with Texas Ratio $>$ 100\% & 13\%  & 16\%  & 5\%   & 11\%  & 7\%   & 15\% \\
				\% Deposits in Banks with Texas Ratio $>$ 100\% & 12\%  & 13\%  & 5\%   & 11\%  & 7\%   & 15\% \\
				\% banks with Texas Ratio $>$ 100\% & 8\%   & 8\%   & 6\%   & 9\%   & 7\%   & 11\% \\
				Previous Closures & 4.00  & 6.27  & 2.62  & 6.15  & 1.54  & 4.72 \\
				\textbf{Count} & 118   & -     & 1175  & -     & 92    & - \\
					\hline
				\textbf{State-level political economy characteristics} &       &       &       &       &       &  \\
				\% Republicans in 1987 & 41\%  & 12\%  & 43\%  & 17\%  & 52\%  & 22\% \\
				\% Republicans in 1989 & 35\%  & 14\%  & 40\%  & 18\%  & 47\%  & 19\% \\
				\% vote for Bill 1: enhancing reg. agencies powers & 83\%  & 14\%  & 80\%  & 21\%  & 83\%  & 18\% \\
				\% vote for Bill 2: restoring civil penalties for fin. Inst. & 93\%  & 20\%  & 88\%  & 23\%  & 82\%  & 28\% \\
				\% vote for Bill 3: recommitting S\&L restructuring bill & 97\%  & 2\%   & 98\%  & 3\%   & 98\%  & 3\% \\
				\% vote for Bill 4: CEBA & 99\%  & 3\%   & 98\%  & 5\%   & 98\%  & 4\% \\
				\% vote for Bill 5: disclosure of CRA ratings & 33\%  & 17\%  & 37\%  & 23\%  & 33\%  & 26\% \\
				\textbf{Count} & 118   & -     & 1170  & -     & 92    & - \\
	\end{tabular}}}%
\end{table}%
\begin{table}
	\caption{\footnotesize \label{tab:summt} Descriptive statistics of S\&L, county and state-level characteristics in the data sample for FSLIC resolution decisions. See Section \ref{sec:data_app} of the supplement for a description of the bills evaluated in this analysis under the state-level political economy characteristics.}
	\centering
	\fbox{%
		\scalebox{0.8}{
			\begin{tabular}{lcccccc}
					& \multicolumn{2}{c}{Type I (OBA)} & \multicolumn{2}{c}{Type II (P\&A)} & \multicolumn{2}{c}{Type III (PO)} \\
				\hline
				& Mean  & Std. dev & Mean  & Std. dev & Mean  & Std. dev \\
				\hline
				\textbf{S\&L-level characteristics} &       &       &       &       &       &  \\
				C\&I Loan Ratio & 3\%   & 5\%   & 4\%   & 6\%   & 5\%   & 9\% \\
				CLD Loan Ratio & 14\%  & 20\%  & 27\%  & 25\%  & 20\%  & 25\% \\
				Real Estate Loan Ratio & 93\%  & 11\%  & 96\%  & 17\%  & 103\% & 20\% \\
				Loan Loss Reserves Ratio  & 5\%   & 8\%   & 9\%   & 11\%  & 4\%   & 5\% \\
				Interest Receivable Ratio & 1\%   & 1\%   & 2\%   & 3\%   & 2\%   & 1\% \\
				Securities Ratio & 20\%  & 15\%  & 13\%  & 9\%   & 12\%  & 10\% \\
				Size(Assets mlns.) & 471   & 1961  & 316   & 484   & 201   & 243 \\
				State charter  & 11\%  & 31\%  & 8\%   & 27\%  & 7\%   & 25\% \\
				\hline
				\textbf{Insurer characteristic} &       &       &       &       &       &  \\
				Dep. Ins. Fund/ Total Deposits & -1\%  & 18\%  & 0\%   & 22\%  & 7\%   & 22\% \\
				\hline
				\textbf{State Characteristics} &       &       &       &       &       &  \\
				Interstate branching & 0.69  & 0.46  & 0.54  & 0.50  & 0.33  & 0.47 \\
				Unemployment & 8\%   & 2\%   & 8\%   & 2\%   & 8\%   & 2\% \\
				Housing starts & 16\%  & 17\%  & 19\%  & 22\%  & 23\%  & 25\% \\
				\hline
				\textbf{County characteristics} &       &       &       &       &       &  \\
				Per capita GDP growth & 4\%   & 4\%   & 4\%   & 4\%   & 6\%   & 3\% \\
				Farm, Agri and Mining & 7\%   & 8\%   & 8\%   & 8\%   & 9\%   & 7\% \\
				Manufacturing & 14\%  & 8\%   & 13\%  & 7\%   & 14\%  & 7\% \\
				Construction & 5\%   & 3\%   & 5\%   & 2\%   & 5\%   & 1\% \\
				Fin Serv and Transport & 37\%  & 8\%   & 37\%  & 9\%   & 34\%  & 10\% \\
				Government & 15\%  & 6\%   & 16\%  & 7\%   & 20\%  & 15\% \\
				\textbf{Count} & 270   & -     & 104   & -     & 15    & - \\
				\hline
				\textbf{County-level characteristics of S\&L distress} &       &       &       &       &       &  \\
				\% Assets in Banks with Texas Ratio $>$ 100\% & 3\%   & 8\%   & 2\%   & 5\%   & 3\%   & 6\% \\
				\% Deposits in Banks with Texas Ratio $>$ 100\% & 3\%   & 8\%   & 2\%   & 5\%   & 3\%   & 6\% \\
				\% banks with Texas Ratio $>$ 100\% & 4\%   & 8\%   & 3\%   & 7\%   & 3\%   & 5\% \\
				Previous Closures & 0.24  & 0.29  & 0.27  & 0.56  & 0.76  & 0.77 \\
				\textbf{Count} & 270   & -     & 102   & -     & 15    & - \\
				\hline
				\textbf{State-level political economy characteristics} &       &       &       &       &       &  \\
				\% Republicans in 1987 & 41\%  & 11\%  & 41\%  & 15\%  & 43\%  & 10\% \\
				\% Republicans in 1989 & 35\%  & 14\%  & 40\%  & 18\%  & 47\%  & 19\% \\
				\% vote for Bill 1: enhancing reg. agencies powers & 83\%  & 14\%  & 80\%  & 21\%  & 83\%  & 18\% \\
				\% vote for Bill2: restoring civil penalties for fin. Inst. & 93\%  & 20\%  & 88\%  & 23\%  & 82\%  & 28\% \\
				\% vote for Bill 3: recommitting S\&L restructuring bill & 97\%  & 2\%   & 98\%  & 3\%   & 98\%  & 3\% \\
				\% vote for Bill 4: CEBA & 97\%  & 5\%   & 97\%  & 6\%   & 94\%  & 6\% \\
				\% vote for Bill 5: disclosure of CRA ratings & 33\%  & 17\%  & 37\%  & 23\%  & 33\%  & 26\% \\
				\textbf{Count} & 267   & -     & 103   & -     & 15    & - \\
	\end{tabular}}}%
\end{table}%
\clearpage
\section{Details on the Bayesian estimation framework for the latent class model}
\label{sec:estimation_app}
In this section we continue our discussion on the latent class model from Section \ref{sec:model} and present a collapsed Gibbs sampler for posterior inference in this model. We begin with two remarks referenced in Section \ref{sec:model}.
\begin{remark}
\label{rem:rand_utility}
{In Section \ref{sec:model} the regulator's problem of assigning bank $i$ to one of the two latent classes is modeled as a binary discrete choice problem with a latent outcome $s_i$ and relies on the random utility
representation of this model based on the framework developed by \cite{marschak1974binary}. This random utility interpretation follows the convention in econometric literature around models with discrete outcomes, but our method remains valid even in settings where a direct connection with economic concepts such as utility or choice does not exist such as the incidence of weather events or responses to clinical trials \citep{jeliazkov2012binary}.}
\end{remark}
\begin{remark}
\label{rem:latent_var}
{The latent variable $l_i$ in Equation \eqref{eq:elle} corresponds to the continuous variable that is introduced under data augmentation to facilitate the use of a full Gibbs Sampler as detailed in Section \ref{fest} \citep{albert1993bayesian, tanner1987calculation}. This representation is, however, not necessary to implement standard binary probit models or our collapsed Gibbs sampler, especially as we forgo this augmentation step. Furthermore, the variable $z_{i,s_{i}}$ in Equation \eqref{eq:zeq} is introduced under data augmentation to derive analytical forms of conditional distribution for our collapsed Gibbs sampler as well as a standard Gibbs sampler.} 
\end{remark}
\subsection{Likelihood function}
\label{sec:likelihood}
The likelihood contribution $P_{ij}$ of bank $i$ receiving resolution type $j=1,2,3$ is the sum of the likelihood contribution based on each latent class weighted by the marginal probability of belonging to each of the two latent classes,
\begin{equation}\label{eq:ovrl}
P_{ij} = \sum_{s=1}^{2} P_{ij \vert s} Q_{is},
\end{equation}
where $P_{ij \vert s}$ is the probability of $y_{i}$ taking a particular value $j$ conditional on belonging to class $s_i=s\in\{1,2\}$ and $Q_{is}$ is the corresponding probability of bank $i$ belonging to class $s$. With $\nu_{i}$ in Equation \eqref{eq:elle} distributed independently as $\mathcal{N}(0,1)$, we obtain the following binary probit representation of the class membership model,
\begin{equation}\label{eq:qeq}
Q_{is} = {\Phi( \bm w_{i}'\bm \alpha)}^{s'}\Bigl\{1-\Phi( \bm w_{i}'\bm\alpha)\Bigr\}^{1-s'}, s' = s-1, \hspace{10pt} s\in\{1,2\}.
\end{equation}
On specifying a $\mathcal{N}(0,\sigma^{2}_s)$ distribution for the unobserved component $\epsilon_{i,s}$ in Equation \eqref{eq:zeq}, the probability of $y_{i}$ taking a particular value $j$ conditional on class $s_i=s\in\{1,2\}$ from Equation \eqref{eq:yi} is,
\begin{equation}
	\label{eq:probc}
	P_{ij \vert s} = \begin{cases}
		\Phi\Big(\dfrac{\gamma_{1,s} - \bm x_{i}'\bm\beta_{s}}{\sigma_{s}}\Big),~&\text{if}~j=3\\
		\Phi\Big(\dfrac{\gamma_{2,s} - \bm x_{i}'\bm\beta_{s}}{\sigma_{s}}\Big) - \Phi\Big(\dfrac{\gamma_{1,s} - \bm x_{i}'\bm\beta_{s}}{\sigma_{s}}\Big),~&\text{if}~j=2\\
		1 - \Phi\Big(\dfrac{\gamma_{2,s} - \bm x_{i}'\bm\beta_{s}}{\sigma_{s}}\Big),~&\text{if}~j=1
	\end{cases}.
\end{equation}  
In estimating the ordinal outcome model $ P_{ij \vert s}$ we use the identification scheme in which the cut-points $\gamma_{1,1}$ and $\gamma_{1,2}$ are restricted to 0 and the cut-points $\gamma_{2,1}$ and $\gamma_{2,2}$ are restricted to 1 \citep{jeliazkov2012binary}.This identification restriction eliminates the need for estimating cut-points and allows the scale parameter $\sigma_s$ to be estimated as a free parameter in each latent class. {As discussed in \cite{jeliazkov2012binary} and \cite{greene2010modeling}, identification restrictions are required in estimating ordinal models as neither the scale nor the location of the latent variable $z_{i}$ is identified in this category of models.} Additionally, we require the following identification conditions for our latent class model:
\begin{enumerate}
    \item The true model has distinct latent classes.
     \item The covariate matrices $W_{n\times p}=(\bm w_1,\ldots,\bm w_n)^T$ and $X_{n\times q}=(\bm x_1,\ldots,\bm x_n)^T$ have full column rank.
 \end{enumerate}
 {Condition (b) typically holds in practical applications while condition (a) requires that the underlying data generating mechanism admits distinct latent classes.} Denote $\bm \Theta=\{\bm\beta_{1}, \bm\beta_{2},\sigma_{1}^{2},\sigma_{2}^{2}, \bm\alpha\}$. The likelihood function is obtained as,
$
\mathcal{L}(\bm\Theta) = \prod_{i=1}^{n}\prod_{j=1}^{3}(P_{ij})^{\mathbb{I}\{y_i=j\}},
$
where $P_{ij}$ and its components are as defined in equations \eqref{eq:ovrl}, \eqref{eq:qeq}, \eqref{eq:probc}, and $\mathbb{I}\{y_i=j\}=1$ if $y_i=j$ and $0$ otherwise.
\subsection{Augmented posterior}
\label{sec:posterior}
The augmented posterior for the parameters and latent variables in this model is obtained by augmenting the likelihood with the latent variables $\bm z=(z_{1,s_1},\ldots,z_{n,s_n})$ and $\bm {u}=(s_1,\ldots,s_n)$ using the method of \cite{albert1993bayesian}. Denote
\begin{equation*}
	\mathcal{B}_i=\begin{cases}
		(-\infty,0],~&\text{if}~y_i=3\\
		(0,1],~&\text{if}~y_i=2\\
		(1,\infty),~&\text{if}~y_i=1\\
	\end{cases}.
\end{equation*}
Using equations \eqref{eq:elle}-\eqref{eq:probc}, the resulting expression for the augmented posterior is,  
\begin{equation}
	\label{eq:aug_posterior}
f(\bm{\Theta},\bm z,\bm u \vert \bm y)  \propto \Big[\prod_{i=1}^{n}\Bigl\{\mathbb{I}\{z_{i,1}\in\mathcal{B}_i\}f_{\mathcal{N}}(z_{i,1} \vert \bm x_i'\bm\beta_{1},\sigma_{1})Q_{i1}+\mathbb{I}\{z_{i,2}\in\mathcal{B}_i\}f_{\mathcal{N}}(z_{i,2} \vert \bm x_i'\bm\beta_{2},\sigma_{2})Q_{i2}\Bigr\}\Big]h(\bm\Theta),
\end{equation}
where $\bm y=(y_1,\ldots,y_n)$, $\mathbb{I}\{z_{i,s}\in\mathcal{B}_i\}$ takes the value $1$ if $z_{i,s}\in \mathcal{B}_i$ and $0$ otherwise, $f_{\mathcal{N}}(z_{i,s} \vert \bm x_i'\bm\beta_{s},\sigma_{s})$ is the density of a normal distribution with mean $\bm x_{i}'\bm\beta_{s}$ and standard deviation $\sigma_{s}$ for $s \in\{1,2\}$ and $h(\bm\Theta)$ is the joint probability density function of the prior distribution of the parameters in $\bm \Theta$. We assign a $q-$dimensional multivariate normal prior to $\bm\beta_{s}$ that has mean $\bm \beta_{0,s}$ and covariance $B_{0,s}$, and an Inverse Gamma prior to $\sigma_{s}^{2}$ with shape parameter $v/2$ and scale parameter $d/2$. Finally, we assign a $p-$dimensional multivariate normal prior to $\bm\alpha$ with mean $\bm \alpha_{0}$ and covariance $A_{0}$. Since the priors are independent, their joint density $h(\bm\Theta)$ in Equation \eqref{eq:aug_posterior} can be represented as, 
\begin{equation*}
h(\bm\Theta) = \Bigl\{\prod_{s=1}^{2} f_{\mathcal{N}}(\bm\beta_{s} \vert \bm\beta_{0,s}, B_{0,s})f_{\mathcal{IG}}\Big(\sigma_{s}^{2} \Big\vert \dfrac{v}{2},\dfrac{d}{2}\Big)\Bigr\}f_{\mathcal{N}}(\bm\alpha \vert\bm \alpha_{0},A_{0}),
\end{equation*}
where $f_{\mathcal{IG}}(\sigma_{s}^{2} \vert v/2,d/2)$ is the density of an Inverse Gamma distribution with shape parameter $v/2$ and scale parameter $d/2$.
%
\subsection{MCMC algorithm}
\label{num2}
A standard approach to developing an MCMC algorithm results in a Gibbs sampler that draws from the full conditionals of all parameters as well as the two latent variables $\bm u$ and $\bm z$ that are present in our sampler as well as an additional latent variable $\bm l$ (described in Section \ref{fest}). {However, such a standard sampler may result in posterior samples that exhibit high levels of autocorrelation. A high degree of autocorrelation in MCMC draws is undesirable for several reasons. First, it indicates that the underlying sampling algorithm is inefficient as it may take a very long time for the sampler to explore the entire posterior distribution. Second, for high dimensional parameters an inefficient sampler will require considerable compute time and memory for generating and storing the samples. Finally, inference based on highly autocorrelated posterior samples may no longer be valid. For instance, posterior credible intervals for parameters of interest may not provide valid coverage and Bayes factors for hypothesis testing may be misleading.}

{A practical strategy for tackling high levels of autocorrelation in posterior samples is to allow the MCMC sampler to run for a very long time and then employ thinning, where all but every $k^{th}$ observation from the sample is discarded. For many practical problems, such an approach may be infeasible due to the memory constraints of the compute environment. Moreover, thinning does not address the underlying issue of an inefficient sampler even though several recent works suggest that thinning can lead to consistent parameter estimation under some mild conditions on the ergodicity of the associated Markov chain (see for instance \cite{hodgkinson2020reproducing,riabiz2022optimal}). Conventional advice for handling an inefficient sampler usually involves parameter transformation, model reformulation or using sophisticated samplers such as a collapsed Gibbs sampler \citep{liu1994collapsed} or the Hamiltonian Monte Carlo (HMC) \citep{duane1987hybrid,neal1993probabilistic,neal2011hmc,betancourt2015hamiltonian}, both of which can efficiently generate approximately independent samples from the target posterior distribution for a wide variety of problems.} 

{For a parameter $\bm \theta=(\theta_1,\ldots,\theta_p)$ of interest, a Gibbs sampler and its collapsed alternative explore the target posterior distribution of $\bm \theta$ by repeatedly sampling from the univariate distribution of $\theta_j,j=1,\ldots,p$ conditional on the data and the current values of $\{\theta_k\}_{k\ne j}$. However, a standard Gibbs sampler may take a long time to fully explore the posterior distribution besides requiring a knowledge of the aforementioned conditional distributions. The HMC, in contrast, relies on the derivatives of the posterior density function to efficiently generate approximately independent samples from the posterior distribution. In particular, the HMC introduces an auxiliary momentum parameter $\rho_j$ for each $\theta_j$ and then jointly updates $\bm \rho=(\rho_1,\ldots,\rho_p)$ and $\bm \theta$ via Hamilton’s equations that give the position $\bm \theta$ and momentum $\bm \rho$ of a physical system evolving over time. This update step requires numerically solving a two-state differential equation via the St\"{o}rmer-Verlet (``leapfrog") integrator. The leapfrog integrator is a discrete approximation to the physical Hamiltonian dynamics in which $(\bm \rho,\bm\theta)$ evolve continuously in time. In the context of the HMC, there are two main hyper-parameters associated with the leapfrog integrator that must be carefully tuned; the discretization time $\epsilon$ and the number of steps taken $L$. The no-U-turn sampler (NUTS) of \cite{hoffman2014no} is a popular extension to the HMC that automatically selects $(L,\epsilon)$ and eliminates the need to hand-tune the HMC.}

In this section we present a Collapsed Gibbs (\texttt{CG}) sampler which provides an efficient technique for sampling from the posterior distribution relative to standard sampling approaches by reducing autocorrelations across successive draws. Specifically, in our algorithm, we do not have to draw from the conditionals of $\bm u$ as we sample for the parameters of the class membership model, $\bm\alpha$, independently of this latent variable. We first present the algorithm for our \texttt{CG} sampler and then discuss the details underlying each step in the sampler.  
\\~\\
\noindent\textbf{Algorithm: Collapsed Gibbs Sampler}
\begin{enumerate}
	\item  Sample $\bm\beta_{s}$ from the distribution $\bm\beta_{s}\vert \bm z,\bm u,\sigma^{2}_{s}$ for $s\in\{1,2\}$.	
	\item Sample $\sigma^{2}_{s}$ from $\sigma^{2}_{s}\vert \bm\beta_{s}, \bm z,\bm u$ for $s\in\{1,2\}$.
	\item Sample $\bm\alpha$ from $\bm\alpha\vert\bm\beta,\bm\sigma^{2},\bm y$ where $\bm\sigma^{2} = (\sigma^{2}_{1},\sigma^{2}_{2})$ and $\bm\beta = (\bm\beta_{1},\bm\beta_{2})$.
	\item Sample $s'_{i}$ from $s'_{i} \vert \bm\alpha,\bm\beta, \bm\sigma^{2},\bm y$, where $s_i'=s_i-1$ for $i = 1,\ldots,n$.
	\item Sample $z_{i,s_{i}}$ from $z_{i,s_{i}}\vert \bm\beta, \bm\sigma^{2},\bm y,\bm u$ for $i = 1,\ldots,n$.
\end{enumerate}
For the following discussion, denote $X$ to be the $n\times q$ matrix with the vector of $q$-dimensional covariates $\bm x_i'$ from Equation \eqref{eq:zeq} in its rows. 
\\~\\
\textbf{Sampling coefficients $\bm \beta_s$ of the ordinal model - } The coefficients $\bm \beta_s$ of the ordinal model are sampled for the two latent classes, i.e., for $s\in\{1,2\}$ from their respective conditional posterior distributions. We have $\bm\beta_{s}\vert \bm z,\bm u,\sigma^{2}_{s} \sim \mathcal{N}\bigl(\hat{\bm\beta}_{s},\hat{B}_{s}\bigr)$, where
$\hat{B}_{s}=\bigl(B_{0,s}^{-1}+X_{s}'X_{s}/\sigma_{s}^{2}\bigr)^{-1}$ and $\hat{\bm\beta}_{s} = \hat{B}_{s}\bigl(B_{0,s}^{-1}\bm\beta_{0,s} + X_{s}'\bm z_{s}/\sigma_{s}^{2}\bigr)$. 
Here $X_s$ is the $n_s\times q$ submatrix of $X$ that includes those rows of $X$ for which $s_i=s$ and $n_{s}$ is the number of observations in class $s$ which is updated in every MCMC iteration. Similarly, $\bm z_s$ denotes the length $n_s$ subvector of $\bm z$ that includes those elements of $\bm z$ for which $s_i=s$. In this sampling step, the computations involving $X_s$ are efficient as they only require working with matrices of reduced dimension $n_s\times q$, without having to preserve the full $n\times q$ matrix $X$.
\\~\\
\noindent\textbf{Sampling the variance $\sigma^{2}_s$ of the ordinal model - }
The variances are sampled using the conditionals $\sigma^{2}_{s} \vert \bm z,\bm u, \bm\beta_s \sim \mathcal{IG}({\sf shape=}~\hat{\nu}_{s},{\sf scale=}~\hat{d}_{s})$ for $s\in\{1,2\}$, where $\hat{\nu}_s = (\nu+n_{s})/2$ and
$\hat{d}_s = \bigl\{d+(\bm z_{s}-X_{s}\bm\beta_{s})'(\bm z_{s}-X_{s}\bm\beta_{s})\bigr\}/2$. Here $X_{s}$ and $\bm z_{s}$ are retained from the previous step. 
\\~\\
\noindent\textbf{Sampling coefficients $\bm \alpha$ of the class membership model - }
The coefficients $\bm\alpha$ of the class membership model are sampled from $\bm\alpha \vert \bm\beta, \bm\sigma^{2}, \bm y$, marginally of $\bm u$, by using a Metropolis Hastings (MH) step with a tailored proposal distribution. We use a $p-$dimensional $t$ distribution with location parameter $\hat{\bm \alpha}$, covariance matrix $\mathcal V$ and degrees of freedom $v$ as the tailored proposal distribution where $\hat{\bm \alpha}=\argmax_{\bm \alpha}f(\bm y\vert \bm \alpha, \bm \beta, \bm \sigma^{2})f_{\mathcal N}(\bm \alpha|\bm \alpha_0,A_0)$, $\mathcal V$ is the inverse of the negative Hessian of $\log\{f(\bm y\vert \bm \alpha, \bm \beta, \bm \sigma^{2})f_{\mathcal N}(\bm \alpha|\bm \alpha_0,A_0)\}$ evaluated at $\hat{\bm\alpha}$ and
\begin{equation*}
	f(\bm y\vert \bm \alpha, \bm \beta, \bm \sigma^{2}) = \prod_{i=1}^{n}\Big[ \Bigl\{1-\Phi(\bm w_{i}'\bm\alpha)\Bigr\}P_{y_{i}\vert 1} + \Phi(\bm w_{i}'\bm\alpha)P_{y_{i}\vert 2}\Big],
\end{equation*} 
with
\begin{equation}
	\label{eq:probcy}
	P_{y_i \vert s} = \begin{cases}
		\Phi\Big(\dfrac{\gamma_{1,s} - \bm x_{i}'\bm\beta_{s}}{\sigma_{s}}\Big),~&\text{if}~y_i=3\\
		\Phi\Big(\dfrac{\gamma_{2,s} - \bm x_{i}'\bm\beta_{s}}{\sigma_{s}}\Big) - \Phi\Big(\dfrac{\gamma_{1,s} - \bm x_{i}'\bm\beta_{s}}{\sigma_{s}}\Big),~&\text{if}~y_i=2\\
		1 - \Phi\Big(\dfrac{\gamma_{2,s} - \bm x_{i}'\bm\beta_{s}}{\sigma_{s}}\Big),~&\text{if}~y_i=1
	\end{cases},
\end{equation}
for $s\in\{1,2\}$. This MH step enhances the efficiency of the overall algorithm by circumventing the need for additional data augmentation through the latent variable $l_i$ from Equation \eqref{eq:elle}.
%

The proposed draw $\bm\alpha^{\dagger}$ from this proposal is accepted with probability,
\begin{equation*}
	\Upsilon_{MH}(\bm\alpha, \bm\alpha^{\dagger}) = \min\Bigg\{1,\dfrac{f(\bm\alpha^{\dagger} \vert \bm\beta, \bm\sigma^{2}, \bm y)q(\bm\alpha \vert\bm\beta,\bm\sigma^{2}, \bm y)}{f(\bm\alpha \vert \bm\beta, \bm\sigma^{2}, \bm y)q(\bm\alpha^{\dagger} \vert \bm\beta, \bm\sigma^{2}, \bm y)}\Bigg\},	
\end{equation*}
where $q(\bm\alpha \vert\bm\beta,\bm\sigma^{2}, \bm y)$ is the density of the tailored proposal distribution and the expression $f(\bm\alpha \vert \bm\beta, \bm\sigma^{2}, \bm y)$ in the display above is proportional to the product of $f(\bm y\vert \bm \alpha, \bm \beta,\bm \sigma^{2})$ and the prior probability density of $\bm \alpha$.
\\~\\
\noindent\textbf{Sampling the class membership indicator $\bm u$ - }
The vector $\bm u$ of class membership indicators $s_{i}$ identifies the latent class $s\in\{1,2\}$ to which each observation $i$ belongs. These indicators are sampled from a Bernoulli distribution by introducing the binary variable $s'_{i} = s_{i}-1$, where $s_{i}' \vert \bm\alpha, \bm\beta, \bm\sigma^{2}, \bm y \sim Bern(K_{i})$ for $i = 1,\ldots,n$ and,
\begin{equation*}
	K_{i} = \dfrac{\Phi(\bm w_{i}'\bm\alpha)P_{y_{i}\vert 2}}{\Phi({\bm w_{i}'\bm\alpha})P_{y_{i}\vert 2} + (1-\Phi({\bm w_{i}'\bm\alpha}))P_{y_{i}\vert 1}}.
\end{equation*}
The values $P_{y_{i}\vert 1}$ and $P_{y_{i}\vert 2}$ are retained from the previous step and are computed using Equation \eqref{eq:probcy}.
\\~\\
\noindent\textbf{Sampling the latent variable $\bm z$ - }
The sampling of continuous latent variables $z_{i,s_{i}}$ is based on the data augmentation step from \cite{albert1993bayesian}, resulting in $z_{i,s_{i}} \vert \bm\beta, \bm\alpha, \bm\sigma^{2}, \bm y$ having a truncated normal distribution with mean $x_{i}'\bm\beta_{s_{i}}$, variance $\sigma_{s_{i}}^{2}$ and truncated between $(\gamma_{y_{i}-1,s_i},\gamma_{y_{i},s_i})$ 
for $i = 1,\ldots,n$. {The second subscript $s_{i}$ in $(\bm\beta_{s_{i}},\sigma_{s_{i}}^{2})$ is added to establish that the sampling scheme augments just the continuous outcomes associated with the class $s_i$ to which each observation belongs and does not require the augmentation based on the counterfactual latent class}. This approach minimizes storage requirements and permits the sampling of the entire vector $\bm z$ in one step. 

In the \texttt{CG} sampler described above, the discrete latent variable $\bm u$ is marginalized out of the conditional distribution for $\bm\alpha$. This novel approach to marginalization results in a sharper decline in autocorrelations across successive lags of sample draws. Consequently, the draws from this algorithm are close to independently and identically distributed early in the chain. In Section \ref{sec:sim_comparenuts} we present a simulation study to compare \texttt{CG} and \texttt{NUTS} for inference in the latent class ordinal probit model of Section \ref{sec:model} and demonstrate that the reduction in autocorrelations gained from our \texttt{CG} sampler is substantial when compared to those from the full Gibbs sampler introduced in Section \ref{fest}. 
\subsection{Estimation of model with $J>3$ values of the ordered outcome}
\label{cpest}
The Bayesian latent class estimation framework of Sections \ref{sec:model} relies on the ordered outcome variable $y_i$, the resolution method, taking three values. However, in several practical applications $y_i$ can take $J>3$ values. For instance, consumer ratings of products and credit ratings assigned to firms are ordered outcomes that typically span over five or more categories. In this section, we provide an extension of our latent class model that allows the ordered outcome variable $y_i$ to take $J>3$ values and develop a \texttt{CG} sampler for posterior inference in that model.

The sampling algorithm is based on the identification scheme used in Section \ref{sec:likelihood} where we set $\gamma_{1,s} = 0$ and $\gamma_{J-1,s} = 1$ for $s\in\{1,2\}$. In order to ensure that the ordering of the $J-1$ cut-points, namely $\gamma_{1,s}<\cdots<\gamma_{J-1,s}$, is preserved without having to resort to the introduction of computationally intensive constraints into the estimation procedure, the following transformation proposed in \cite{chen2000bayesian} is used,
\begin{equation*}
	\delta_{j,s} = \log{\dfrac{(\gamma_{j,s}-\gamma_{j-1,s})}{(1-\gamma_{j-1,s})}},~2\leq j \leq J-2,~s\in\{1,2\}.
\end{equation*} 
We propose a \texttt{CG} sampler algorithm that uses a MH step to sample $\bm\beta_s$ and $\bm\delta_s=(\delta_{2,s},\ldots,\delta_{J-2,s})$ in one block along the lines of the examples provided in \cite{chib2001marginal}. A multivariate normal prior is assigned to $\bm\delta_{s}$ that has mean $\bm \delta_{0,s}$ and covariance matrix $D_{0,s}$. 
\\~\\
\textbf{Algorithm: Collapsed Gibbs Sampler for model with cut-points}
\begin{enumerate}
	\item  Sample $\bm\beta_{s}$ and $\bm\delta_{s}$ jointly from $(\bm\beta_{s}, \bm\delta_{s}) \vert \bm y,s, \sigma^2_s$ for $s\in\{1,2\}$.	
	\item Sample $\sigma^{2}_{s}$ from $\sigma^{2}_{s}\vert \bm\beta_{s}, \bm z,\bm u$ for $s\in\{1,2\}$.
	\item Sample $\bm\alpha$ from $\bm\alpha\vert\bm\beta,\bm\sigma^{2},\bm y$ for where $\bm\sigma^{2} = (\sigma^{2}_{1},\sigma^{2}_{2})$ and $\bm\beta = (\bm\beta_{1},\bm\beta_{2})$.
	\item Sample $s'_{i}$ from $s'_{i} \vert \bm\alpha,\bm\beta, \bm\sigma^{2},\bm y$ for $i = 1,\ldots,n$ and $s_{i}'=s_i-1$.
	\item Sample $z_{i,s_{i}}$ from $z_{i,s_{i}}\vert \bm\beta, \bm\sigma^{2},\bm y,\bm u$ for $i = 1,\ldots,n$.
\end{enumerate} 
Steps (b)--(e) are identical to the algorithm described in Section \ref{num2}. Step (a) of this algorithm is described below.
\\~\\
\noindent\textbf{Sampling coefficients $\bm \beta_s$ and cut-points $\bm\delta_s$ of the ordinal model - }
Sample $(\bm\beta_{s}, \bm\delta_{s}) \vert \bm y,s, \sigma^2_s$ by drawing $(\bm\beta_{s}^{\dagger}, \bm\delta_{s}^{\dagger})$ from a tailored  proposal distribution. We use a $q+J-3$ dimensional $t$ distribution with location $(\hat{\bm \beta}_s,\hat{\bm \delta}_s) \coloneqq \argmax_{(\bm \beta_s,\bm\delta_s)} f(\bm y \vert \bm\beta_s, \bm\delta_s, \sigma^{2}_s, s)f_{\mathcal{N}}(\bm\beta_{s} \vert \bm\beta_{0,s}, B_{0,s})f_{\mathcal{N}} (\bm\delta_{s} \vert \bm\delta_{0,s}, D_{0,s})$, covariance matrix $\mathcal V$ being the inverse of the negative hessian of the logarithm of the maximand evaluated at $(\hat{\bm \beta}_s,\hat{\bm \delta}_s)$ and degrees of freedom $v$. Here, $$f(\bm y \vert \bm\beta_s, \bm\delta_s, \sigma^{2}_s, s)=\prod_{i=1}^{n}\Big(P_{y_i|s_i}\Big)^{\mathbb I\{s_i=s\}}$$ for $s\in\{1,2\}$ and
	\begin{equation*}
		P_{y_i \vert s} = \begin{cases}
			\Phi\Big(\dfrac{- \bm x_{i}'\bm\beta_{s}}{\sigma_{s}}\Big),~&\text{if}~y_i=1\\
			1 - \Phi\Big(\dfrac{1 - \bm x_{i}'\bm\beta_{s}}{\sigma_{s}}\Big),~&\text{if}~y_i=J\\
			\Phi\Big(\dfrac{\gamma_{y_i,s} - \bm x_{i}'\bm\beta_{s}}{\sigma_{s}}\Big) - \Phi\Big(\dfrac{\gamma_{y_i-1,s} - \bm x_{i}'\bm\beta_{s}}{\sigma_{s}}\Big),~&\text{if}~y_i\in\{2,\ldots,J-1\}
		\end{cases}.
\end{equation*}
The proposed draw $(\bm\beta_{s}^{\dagger}, \bm\delta_{s}^{\dagger})$ is accepted with probability $\Upsilon_{MH}\{(\bm\beta_{s}, \bm\delta_{s}), (\bm\beta_{s}^{\dagger}, \bm\delta_{s}^{\dagger})\}$
\begin{eqnarray}
		=\min\Bigl\{1,\dfrac{f(\bm y \vert \bm\beta_s^{\dagger}, \bm\delta_s^{\dagger}, \sigma_s^{2}, s)f_{\mathcal{N}}(\bm\beta_{s}^{\dagger} \vert \bm\beta_{0,s}, B_{0,s})f_{\mathcal{N}} (\bm\delta_s^{\dagger} \vert \bm\delta_{0,s}, D_{0,s})q(\bm\beta_{s}, \bm\delta_{s} \vert \bm y,s, \sigma_s^{2})}{f(\bm y \vert \bm\beta_s, \bm\delta_s, \sigma_s^{2},s)f_{\mathcal{N}}(\bm\beta_{s} \vert \bm\beta_{0,s}, B_{0,s})f_{\mathcal{N}} (\bm\delta_s \vert \bm\delta_{0,s}, D_{0,s})q(\bm\beta_{s}^{\dagger},\bm\delta_{s}^{\dagger} \vert \bm y,s, \sigma_s^{2})}\Bigr\}\nonumber,
	\end{eqnarray}
	where $q(\bm\beta_{s}, \bm\delta_{s} \vert \bm y,s, \sigma_s^{2})$ is the density of the tailored proposal distribution.
\section{Full Gibbs sampler}
\label{fest}
Here we present the full Gibbs sampler algorithm for posterior inference in the latent class ordinal model discussed in Section \ref{sec:model}.
\\~\\
\textbf{Algorithm: Full Gibbs Sampler}
\begin{enumerate}
	\item  Sample $\bm\beta_{s}$ from the distribution $\bm\beta_{s}\vert \bm z,\bm u,\sigma^{2}_{s}$ for $s=1,2$.	
	\item Sample $\sigma^{2}_{s}$ from $\sigma^{2}_{s}\vert \bm\beta_{s}, \bm z,\bm u$ for $s=1,2$.
	\item \begin{enumerate}
		\item Sample $\bm\alpha$ from $\bm\alpha \vert \bm u, \bm\beta,\bm\sigma^{2},\bm y$, where $\bm\beta=(\bm \beta_1,\bm \beta_2)$ and $\bm \sigma^2=(\sigma_1^2,\sigma_2^2)$. 
		\item Sample $l_{i} \vert \bm\alpha, \bm u$, where for $i = 1,\ldots,n$. 
	\end{enumerate}
	\item Sample $s'_{i}$ from $s'_{i} \vert \bm\alpha,\bm\beta, \bm\sigma^{2},\bm y$ for $i = 1,\ldots,n$.
	\item Sample $z_{i,s_{i}}$ from $z_{i,s_{i}}\vert \bm\beta, \bm\sigma^{2},\bm y,\bm u$ for $i = 1,\ldots,n$.
\end{enumerate}
Steps (a), (b), (d) and (e) are identical to the algorithm described in Section \ref{num2}. Step (c) of this algorithm is described below.
\\~\\
\noindent\textbf{Sampling coefficients $\bm \alpha$ of the Class Membership Model - }
(i) The coefficients $\bm \alpha$ of the class membership model are sampled from the full conditional $\mathcal{N}(\hat{\bm\alpha}_0,\hat{A})$ where $\hat{A} = (A_{0}+W'W)^{-1}$, $\hat{\bm\alpha}_0 = \hat{A}(A_{0}^{-1}\bm\alpha_{0} + W'l)$ and $W$ is the $n\times p$ matrix that has the vector of $p$-dimensional covariates $\bm w_i'$ from Equation \eqref{eq:elle} in its rows.

(ii) The latent variable $l_i$ is sampled using the data augmentation approach in \cite{albert1993bayesian} and drawing from the full conditional distribution, which is a truncated normal distribution with mean $\bm w_i'\bm \alpha$, variance $1$ and region of truncation $\mathcal C_i$, where
\begin{equation*}
	\mathcal{C}_{i}=
	\begin{cases}
		(0,\infty), & \text{if}\ s_{i}=2 \\
		\left(-\infty,0\right] & \text{if}\ s_{i}=1\\
	\end{cases}.
\end{equation*}
\section{Covariate Effects}
\label{coeffform}
An intuitive interpretation of the relationship between covariates and outcomes is provided by covariate effects in models with discrete dependent variables. Here, we discuss the calculation of covariate effects for the resolution-type model, the second level model in Figure \ref{fig:flowFinal}. Consider any covariate, $x_{k} \in \mathcal{X}=(x_1,\ldots,x_q)$, whose effects on the outcome $y$ marginally of the other variables in $\mathcal{X}$ is of interest. Denote $\mathcal{X}_{\setminus k}$ to be the $q-1$ dimensional covariate vector obtained by excluding the $k^{th}$ covariate for $k=1,\ldots,q$ and let $\bm\theta_s=(\bm\alpha, \bm\beta_{s},\sigma_{s}^2)$. In our resolution-type model, the covariate effect measures the change in the marginal probability of observing $y=j$ for a given change in $x_{k}$ conditional on latent class $s$.

The framework for computing covariate effects while addressing both data variability and parameter uncertainty described in \cite{jeliazkov2018impact} has been adapted to the latent class model for ordinal outcomes. Each of these issues is overcome by averaging over the sample and the posterior distribution of parameters respectively. Equation \eqref{eq:coveff} details how the output from the MCMC algorithm developed in Section \ref{num2}, which consists of $G$ draws from the posterior distribution of parameters, is used in evaluating covariate effects for each of the two latent classes.  We have, for $s\in\{1,2\}$,
\begin{eqnarray}\label{eq:coveff}
	\begin{split}
		Pr(y = j \vert {x_{k}^{\dagger}}, s)
		- Pr(y = j \vert {x_{k}^{\ddagger}}, s) =& \int \Bigl\{Pr(y = j \vert {x_{k}^{\dagger}}, \mathcal{X}_{\setminus k}, \bm\theta_{s})\\ -& Pr(y = j \vert {x_{k}^{\ddagger}},\mathcal{X}_{\setminus k}, \bm\theta_{s})\Bigr\} f(\mathcal{X}_{\setminus k})f(\bm\theta_{s} \vert y) \mbox{d}\mathcal{X}_{\setminus k}\mbox{d}\bm\theta_{s}\\
		\approx& \frac{1}{nG}\sum_{i=1}^{n}\sum_{g=1}^{G}\Bigl\{Pr\left(y_{i} = j \vert {x_{ik}^{\dagger}}, \bm x_{i,\setminus k}, \bm\theta_{s}^{(g)}\right)\\ -& Pr\left(y_{i} = j \vert {x_{ik}^{\ddagger}},\bm x_{i,{\setminus k}}, \bm\theta_{s}^{(g)}\right)\Bigr\}
	\end{split}
\end{eqnarray}
where $\bm x_{i,\setminus k}$ denotes the $i^{th}$ row of $X$ excluding the $k^{th}$ element. The probabilities in the parentheses in the final step of Equation \eqref{eq:coveff} are obtained by using the expression in Equation \eqref{eq:probc}.
\section{Model Comparison}
\label{mcs}
Subsequent to estimating the model, the empirical objective is to then identify the specification of the model that is corroborated by the data most decisively. Accordingly, this section presents the procedure for the comparison of posterior probabilities of estimated models. This is a method of model comparison that conforms to the Bayesian principle of representing uncertainty in the form of probability statements. Specifically, in comparing models $\mathcal{M}_{i}$ and $\mathcal{M}_{j}$, the posterior odds ratio, $\mathcal{P}_{ij}$, is evaluated to select between the pair of models, where,
\begin{equation*}
	\mathcal{P}_{ij} = \frac{P(\mathcal{M}_{i} \vert \bm y)}{P(\mathcal{M}_{j} \vert \bm y)} = \frac{m(\bm y \vert\mathcal{M}_{i})}{m(\bm y \vert \mathcal{M}_{j})}\frac{P(\mathcal{M}_{i})}{P(\mathcal{M}_{j})}.
\end{equation*}
%
The first term on the right hand side of the second equality is the Bayes factor and the second term is the prior odds. The Bayes factor is the ratio of marginal likelihoods of models $i$ and $j$ and following standard convention in which the $a$ $priori$ probability of each model occurring is equal, this quantity singularly determines the evidence in favor of one model against the other. Therefore, Bayesian model selection among $L$ models, $\{\mathcal{M}_{1}, \mathcal{M}_{2},...,\mathcal{M}_{L}\}$, proceeds by comparing the marginal likelihood across these models. 

The \textit{basic marginal likelihood identity} recognized by \cite{chib1995marginal} allows for the exact evaluation of the marginal likelihood by MCMC methods. This identity expresses the marginal likelihood of model $l$ as
\begin{equation*}
	m(\bm y \vert\mathcal{M}_{l}) = \frac{f(\bm y\vert\mathcal{M}_{l},\bm{\theta}_{l})\pi(\bm{\theta}_{l} \vert \mathcal{M}_{l})}{\pi(\bm{\theta}_{l} \vert \bm y, \mathcal{M}_{l})},
\end{equation*}
where $\bm{\theta}_{l}$ is a parameter vector specific to model $l$. The computation of the marginal likelihood simply requires the evaluation of this ratio for a given $\bm\theta_{l}^{*}$, typically the posterior mean or mode. The likelihood $f(\bm y\vert\mathcal{M}_{l},\bm{\theta}_{l})$ and prior ordinate $\pi(\bm{\theta}_{l} \vert \mathcal{M}_{l})$ at $\bm\theta_{l}^{*}$ can be evaluated analytically for the latent class model with ordered outcomes. The posterior ordinate $\pi(\bm{\theta}_{l} \vert \bm y, \mathcal{M}_{l})$ at $\bm\theta_{l}^{*}$ is estimated to obtain  $\hat{\pi}(\bm{\theta}_{l}^{*} \vert \bm y, \mathcal{M}_{l})$ using methods outlined in \cite{chib2001marginal} and \cite{chib1995marginal}. 
\subsection{Evaluating the Marginal Likelihood}
\label{eml}
Here we present the algorithm to evaluate the estimated posterior ordinate $\hat{\pi}(\bm{\theta}^{*}_l \vert \bm y, \mathcal{M}_{l})$. Henceforth, we will drop the subscript $l$ for notational ease. 

In the latent class model $\mathcal{M}$ with ordered outcomes, the parameter vector $\bm\theta$, excluding any latent variables, consists of the coefficients $\bm\alpha, \bm\beta=(\bm \beta_1,\bm \beta_2)$ and the error variances $\bm\sigma^{2}=(\sigma_1^2,\sigma_2^2)$. The objective of this exercise is to evaluate the posterior ordinate at the posterior mean, $\bm\theta^{*} = (\bm\alpha^{*},\bm\beta^{*},\bm\sigma^{2*})$. The law of total probability is used to decompose the posterior ordinate at $\bm\theta^{*}$ as,
\begin{equation*}
	\pi(\bm \alpha^{*},\bm \beta^{*},\bm \sigma^{2*}\vert \bm y,\mathcal{M}) = \pi(\bm \alpha^{*}\vert \bm y,\mathcal{M})\pi(\bm \beta^{*}\vert \bm \alpha^{*},\bm y,\mathcal{M})\pi(\bm \sigma^{2*}\vert \bm \alpha^{*},\bm \beta^{*},\bm y,\mathcal{M})
\end{equation*}
The order of this decomposition has been chosen to minimize computational time and effort. The first component, $\pi(\bm \alpha^{*}\vert \bm y,\mathcal{M})$ is estimated using the method introduced in \cite{chib2001marginal}. By conditioning the other two densities on $\bm \alpha^{*}$, these ordinates can be estimated using reduced Gibbs samplers described in \cite{chib1995marginal}. 
The following reduced Gibbs sampler provides the estimated ordinate $\hat{\pi}(\bm \beta^{*}\vert \bm \alpha^{*},\bm y,\mathcal{M})$.%
\begin{enumerate}
	\item  Sample $\bm \beta_{s}$ from the distribution $\bm \beta_{s}\vert \bm z,\bm u,\bm \sigma^{2}_{s}$ for $s=1,2$.	
	\item Sample $\bm \sigma^{2}_{s}$ from $\bm \sigma^{2}_{s}\vert \bm \beta_{s}, \bm z,\bm u$ for $s=1,2$.
	\item Sample $s'_{i}$ from $s'_{i} \vert \bm \alpha^{*},\bm \beta, \bm \sigma^{2},\bm y$, where $s_i'=s_i-1$ for $i = 1,\ldots,n$.
	\item Sample $z_{i,s_{i}}$ from $z_{i,s_{i}}\vert \bm \beta, \bm \sigma^{2},\bm y,\bm u$ for $i = 1,\ldots,n$.
\end{enumerate}%
The ordinate  $\hat{\pi}(\bm \sigma^{2*}\vert \bm \alpha^{*},\bm \beta^{*},\bm y,\mathcal{M})$ is obtained by iterating over the following reduced Gibbs sampler.%
\begin{enumerate}
	\item Sample $\bm \sigma^{2}_{s}$ from $\bm \sigma^{2}_{s}\vert \bm \beta_{s}^{*}, \bm z,\bm u$ for $s=1,2$.
	\item Sample $s'_{i}$ from $s'_{i} \vert \bm \alpha^{*},\bm \beta^{*}, \bm \sigma^{2},\bm y$ for $i = 1,\ldots,n$.
	\item Sample $z_{i,s_{i}}$ from $z_{i,s_{i}}\vert \bm \beta^{*}, \bm \sigma^{2},\bm y,\bm u$ for $i = 1,\ldots,n$.
\end{enumerate} %
\section{Simulation studies}
\label{sim}
{In this section we assess the performance of the Collapsed Gibbs (\texttt{CG}) sampler of Section \ref{num2} on simulated data. Section \ref{sec:sim_comparenuts} compares \texttt{CG} with the no-U-turn sampler (\texttt{NUTS}) of \cite{hoffman2014no} on two simulation settings while Section \ref{sec:sim_commomcovars} evaluates \texttt{CG} when the latent class ordinal model of Section \ref{sec:model} involves common covariates across $\bm w_i$ and $\bm x_i$. In these numerical experiments, we continue to use the same prior distributions and hyperparameters as those specified in Section \ref{sec:results_fdic}.} 
The simulation results in this section have been corrected for label-switching using \cite{panagls}. This issue affects MCMC algorithms constructed for the estimation of finite mixture models and results in switching of class labels in the course of the chain. The class labels have been reassigned post-estimation using an ordering constraint on the intercept in both classes. %
\subsection{Comparison with \texttt{NUTS}}
\label{sec:sim_comparenuts}
We evaluate the performance of the \texttt{CG} sampler and \texttt{NUTS} for posterior inference in our latent class ordinal model. {To implement \texttt{NUTS} we use RStan \citep{rstan2024}, which is the R \citep{r2024} interface to Stan \citep{stan2024}, a popular platform for statistical modeling and high-performance statistical computation. The Markov chain Monte Carlo (MCMC) algorithm used in Stan relies on an adaptive Hamiltonian Monte Carlo algorithm that  dynamically tunes $L$, the number of leapfrog steps, using the \texttt{NUTS} algorithm of \cite{hoffman2014no}.}

We consider two simulation settings, the first of which contains latent classes with disparate means and the other, in which the means are relatively distinct. To evaluate the two algorithm's accuracy in recovering parameters, we generate differences in means by considering alternative values of coefficients $\bm\beta_{s}, s\in\{1,2\}$ across the two settings, and leave the covariates unchanged. The simulation exercise has been performed on a sample of size $n= 1200$ observations under both the settings with $p=2$ and $q=4$. In each setting, the first component of the covariates $\bm w_i,~i=1,\ldots,n$ is an intercept and the second component is drawn independently from a standard normal distribution. Similarly, in the ordinal model, the first component of the covariates $\bm x_i$ is an intercept, and the second, third and fourth components are drawn independently from $\mathcal{N}(0.5,1)$, $\mathcal{N}(0.5,1)$ and $\mathcal{N}(0,0.8^2)$ respectively. The true values of parameters in the two settings are provided in the second column of tables \ref{tab:set11noint} and \ref{tab:set2noint}. Under these specifications, the mean under the two latent classes for the two settings, averaged across $100$ Monte-Carlo repetitions, are obtained as follows:
\begin{itemize}
	\item Setting 1: $\sum_{r=1}^{100}[\sum_{i=1}^{n}\bm x_{ir}'\bm\beta_{1}\mathbb{I}\{s_{ir}=1\}/\sum_{i=1}^{n}\mathbb{I}\{s_{ir}=1\}] = -0.844$, $\sum_{r=1}^{100}[\sum_{i=1}^{n}\bm x_{ir}'\bm\beta_{2}\mathbb{I}\{s_{ir}=2\}/\sum_{i=1}^{n}\mathbb{I}\{s_{ir}=2\}] = 0.594$.
	\item Setting 2:$\sum_{r=1}^{100}[\sum_{i=1}^{n}\bm x_{ir}'\bm\beta_{1}\mathbb{I}\{s_{ir}=1\}/\sum_{i=1}^{n}\mathbb{I}\{s_{ir}=1\}] = -0.593$, $\sum_{r=1}^{100}[\sum_{i=1}^{n}\bm x_{ir}'\bm\beta_{2}\mathbb{I}\{s_{ir}=2\}/\sum_{i=1}^{n}\mathbb{I}\{s_{ir}=2\}] = -0.099$.
\end{itemize}
\begin{table}
\caption{\label{tab:set11noint} Setting 1: Total squared error (\texttt{TSE}), and the coverages of one-standard deviation and two-standard deviation posterior credibility intervals, denoted \texttt{1 SD Cred.Int.Cov.} and \texttt{2 SD Cred.Int.Cov.} respectively, for the two algorithms across $100$ Monte-Carlo repetitions. }
\centering
\begin{tabular}{lc|cccccc}
\hline
       & \multicolumn{1}{c}{} &\multicolumn{2}{c}{\texttt{TSE}} & \multicolumn{2}{c}{\texttt{1 SD Cred.Int.Cov.}} & \multicolumn{2}{c}{\texttt{2 SD Cred.Int.Cov.}} \\
       \hline
       & True values & \texttt{CG}     & \texttt{NUTS}           & \texttt{CG}                & \texttt{NUTS}              & \texttt{CG}                & \texttt{NUTS}              \\
       \hline
$\alpha_1$ & -0.3& 0.382  & {0.354} & 0.66              & 0.67             & 0.98              & 0.99             \\
$\alpha_2$ & 1.5&1.189  & {1.130} & 0.68              & 0.66             & 0.97              & 0.96             \\
\hline
$\beta_{11}$    & 0&0.374  & -              & 0.67              & -                & 0.96              & -                \\
$\beta_{21}$    &-0.9 &0.665  & {0.444} & 0.73              & 0.72             & 0.95              & 0.98             \\
$\beta_{31}$    &-0.8 &0.525  & {0.418} & 0.69              & 0.67             & 0.97              & 0.98             \\
$\beta_{41}$    & -0.6&0.453  & {0.365} & 0.67              & 0.72             & 0.97              & 0.99             \\
$\sigma_{1}^2$  & 0.25&0.226  & -              & 0.79              & -                & 0.99              & -                \\
\hline
$\beta_{12}$    & 0&0.394  & -              & 0.71              & -                & 0.97              & -                \\
$\beta_{22}$    & 0.6&0.265  & {0.200} & 0.71              & 0.79             & 0.98              & 0.99             \\
$\beta_{32}$    & 0.6&0.291  & {0.184} & 0.68              & 0.76             & 0.94              & 0.96             \\
$\beta_{42}$    & 1&0.570  & {0.384} & 0.73              & 0.69             & 0.98              & 0.98             \\
$\sigma_{2}^2$  &0.25 &0.168  & -              & 0.73              & -                & 1.00              &   \\
\hline
\end{tabular}
\end{table}

Under Setting 1, Table \ref{tab:set11noint} summarizes the total squared error (\texttt{TSE}) and the coverages of one-standard deviation and two-standard deviation posterior credibility intervals for the two algorithms across $100$ Monte-Carlo repetitions. Here, for any parameter $\theta$ and its estimate $\hat{\theta}_r$ in the $r^{th}$ repetition, \texttt{TSE} is calculated as $\sum_{r=1}^{100}(\theta-\hat{\theta}_r)^2$. The \texttt{TSE}s for \texttt{CG} and \texttt{NUTS} reveal that the latter does a relatively better job in estimating the various parameters. We note that the implementation of the latent class ordinal probit model in Stan only allows estimation of the cut points, $\gamma_{1,s}$ and $\gamma_{2,s}$, keeping the variances $\sigma_s^2$ fixed at $1$. Moreover, Stan does not allow intercepts in the two ordinal probit models. Therefore, in the data generating scheme for this setting the coefficients of the intercepts in the two ordinal probit models, $\beta_{11}$ and $\beta_{12}$, are set to zero. While the latent class model implementation in Stan excludes these two intercepts, the \texttt{CG} sampler does not have access to this oracle information. Nevertheless, the coverages of the two posterior credible intervals reveal that the \texttt{CG} sampler is able to successfully declare these coefficients to be statistically unimportant. 
\begin{figure}[!t]
	\centering
	\includegraphics[width=1.1\linewidth]{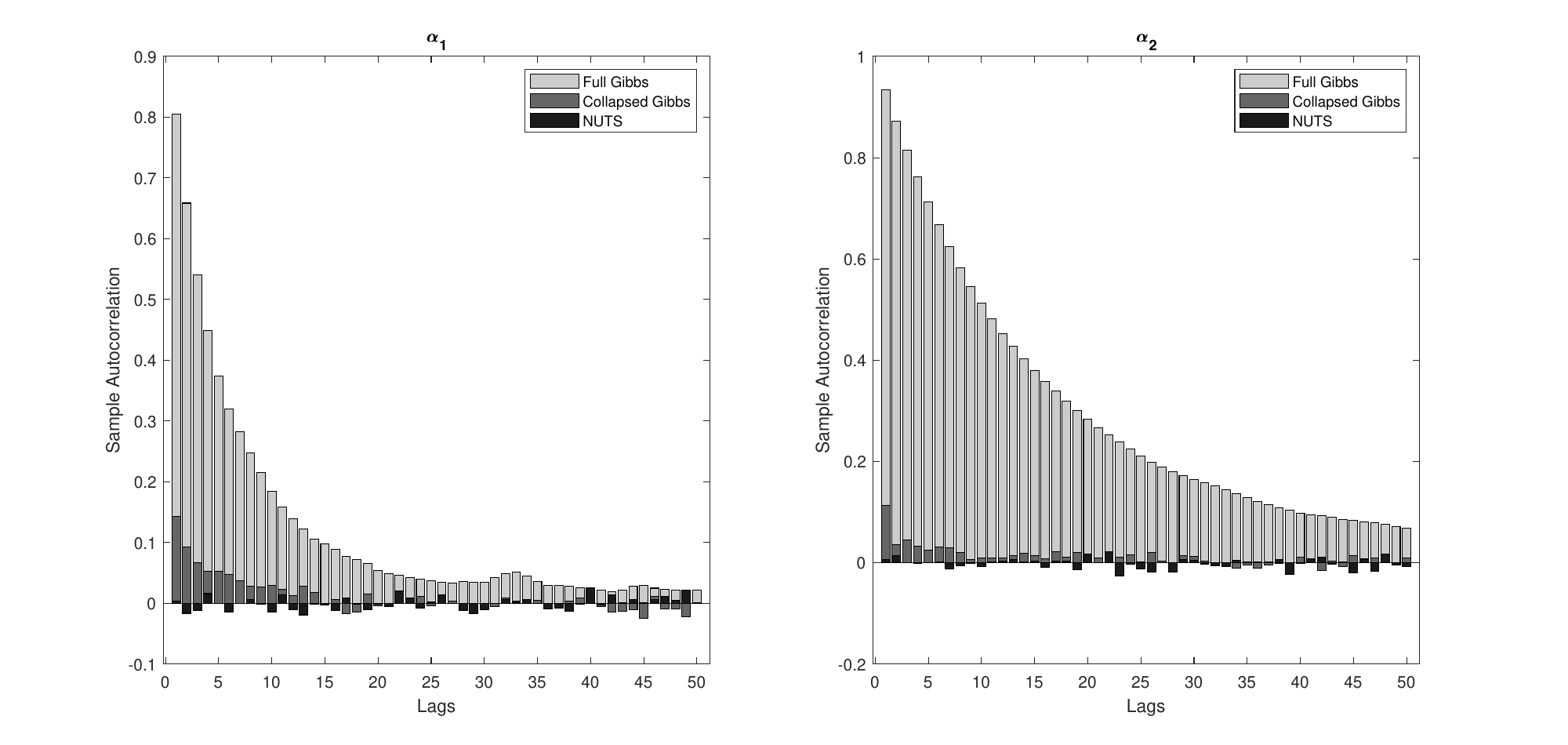}
	\caption{Autocorrelation in the posterior samples of $\alpha_1$ and $\alpha_2$ from the full Gibbs sampler (Section \ref{fest}), the \texttt{CG} sampler and \texttt{NUTS} under simulation Setting 1.}
\label{fig:acf_set11noint}
\end{figure}
The two panels in Figure \ref{fig:acf_set11noint} display the autocorrelations in the posterior samples of $\alpha_1$ and $\alpha_2$ under the full Gibbs sampler (Section \ref{fest}), the \texttt{CG} sampler and \texttt{NUTS} for Setting 1. The figure shows that the reduction in autocorrelations gained from \texttt{CG}, and particularly from \texttt{NUTS} are striking. The autocorrelations from the full Gibbs sampler are close to 1 at lower lags and decay slowly. In contrast, the autocorrelations from the \texttt{CG} sampler are negligible even at lower lags and taper to zero within 10 lags. Unsurprisingly, \texttt{NUTS} is relatively more efficient than \texttt{CG} and exhibits virtually no autocorrelation even at lower lags. This demonstrates that the proposed \texttt{CG} sampler and \texttt{NUTS} provide an efficient technique for sampling from the posterior distribution relative to standard sampling approaches by reducing autocorrelations across successive draws.
\begin{table}
\caption{\label{tab:set2noint} Setting 2: Total squared error (\texttt{TSE}), and the coverages of one-standard deviation and two-standard deviation posterior credibility intervals, denoted \texttt{1 SD Cred.Int.Cov.} and \texttt{2 SD Cred.Int.Cov.} respectively, for the two algorithms across $100$ Monte-Carlo repetitions. For some parameter estimates, the corresponding \texttt{TSE}s from \texttt{NUTS} are astronomically high and, therefore, have been reported as $``>10"$.}
\centering
\begin{tabular}{lc|cccccc}
\hline
       & \multicolumn{1}{c}{} &\multicolumn{2}{c}{\texttt{TSE}} & \multicolumn{2}{c}{\texttt{1 SD Cred.Int.Cov.}} & \multicolumn{2}{c}{\texttt{2 SD Cred.Int.Cov.}} \\
       \hline
       & True values & \texttt{CG}     & \texttt{NUTS}           & \texttt{CG}                & \texttt{NUTS}              & \texttt{CG}                & \texttt{NUTS}              \\
       \hline
$\alpha_1$ &-0.3 &5.786  & {37.077}           & 0.65              & {0.59}    & 0.98         & {0.89}         \\
$\alpha_2$ & 1.5&83.741 & {113.250}          & 0.73              & {0.63}    & 0.98         & {0.83}         \\
\hline
$\beta_{11}$    &0 &0.248  & - & {0.58}     & - & 0.97         & -                  \\
$\beta_{21}$    &-0.6 & 2.679  & {\textgreater{}10} & 0.68              & {0.57}    & 0.97         & {0.88}         \\
$\beta_{31}$   & -0.6& 2.272  & {\textgreater{}10} & {0.64}     & 0.66             & 0.94         & {0.88}         \\
$\beta_{41}$  & 0.5 & 0.945  & {\textgreater{}10} & 0.70              & 0.66             & 0.95         & {0.89}         \\
$\sigma_{1}^2$ & 0.25& 0.280  & -                         & 0.68              & -                & 0.97         & -                     \\
\hline
$\beta_{12}$  &0  & 0.331  & -                         & 0.62              & -                & 0.97         & -                     \\
$\beta_{22}$  & -0.1 & 2.025  & {\textgreater{}10} & 0.71              & 0.68             & 0.95         & {0.90}         \\
$\beta_{32}$  & -0.1 & 1.811  & {\textgreater{}10} & 0.74              & 0.67             & 1.00         & {0.93}         \\
$\beta_{42}$   &0.8 & 1.101  & {\textgreater{}10} & 0.59              & {0.58}    & 0.94         & 0.94                  \\
$\sigma_{2}^2$ &0.25 & 0.198  & -                         & 0.68              & -                & 0.97         & -  \\
\hline
\end{tabular}
\end{table}

Table \ref{tab:set2noint} summarizes the total squared error (\texttt{TSE}) and the coverages of one-standard deviation and two-standard deviation posterior credibility intervals for the two algorithms under Setting 2. In contrast to Setting 1, this setting represents a relatively difficult scenario for parameter estimation and inference. \texttt{NUTS}, in particular, returns poorer estimates of the various parameters as exhibited in its elevated \texttt{TSE}s. For some parameter estimates, the corresponding \texttt{TSE}s from \texttt{NUTS} are astronomically high and, therefore, have been reported as $``>10"$ in Table \ref{tab:set2noint}. In terms of the one standard deviation credible interval coverage, both \texttt{CG} and \texttt{NUTS} undercover for several parameters. However, \texttt{CG} is far more superior to \texttt{NUTS} when we consider the two standard deviation credible interval coverages. As discussed in Setting 1, in this setting too the coefficients of the intercepts in the two ordinal probit models, $\beta_{11}$ and $\beta_{12}$, are zero. The latent class model implementation in Stan excludes these two intercepts but the \texttt{CG} sampler does not have access to this oracle information. While the one standard deviation credible intervals from the \texttt{CG} sampler undercover $\beta_{11}$ and $\beta_{12}$, the two standard deviation posterior credible intervals exhibit the right coverage.  

{Overall, the results of this simulation exercise reveal that for the latent class ordinal probit model of Section \ref{sec:model} both the \texttt{CG} sampler and \texttt{NUTS} provide efficient alternatives for sampling from the posterior distribution relative to standard sampling approaches by reducing autocorrelations across successive draws. \texttt{NUTS}, in particular, is more efficient than \texttt{CG} and returns better parameter estimates when the two latent classes are distinct. In contrast, \texttt{CG} dominates \texttt{NUTS} in estimation accuracy when the classes are relatively less distinct.} Our  recommendation for when to use \texttt{CG} or \texttt{NUTS} mainly depends on two factors. First, we note that the implementation of the latent class ordinal probit model in Stan only allows estimation of the cut points, $\gamma_{1,s}$ and $\gamma_{2,s}$, keeping the scale parameter $\sigma_s$ fixed at $1$. Moreover, Stan does not allow intercepts in the two ordinal probit models. In contrast, the identification scheme in the \texttt{CG} sampler allows intercepts and sets $\gamma_{1,1}=\gamma_{1,2}=0$, and $\gamma_{2,1}=\gamma_{2,2}=1$. This restriction eliminates the need for estimating cut-points and allows $\sigma_s$ to be estimated as a free parameter in each latent class. In this context, we recommend using either sampler depending on the empirical application at hand and the associated parameters of interest. Second, the results of our numerical experiment suggest that the \texttt{CG} sampler has a relatively better estimation accuracy than \texttt{NUTS} when the classes are less distinct and is marginally poorer than \texttt{NUTS} for distinct classes. This is not surprising since less distinct classes may result in a highly multimodal posterior distribution and the HMC sampler underlying \texttt{NUTS} is known to mix poorly for such multimodal target distributions \citep{mangoubi2018does}. Prior knowledge on how distinct the classes are is usually not available in practice and for those applied settings, we recommend using the \texttt{CG} sampler even though \texttt{NUTS} is relatively more efficient overall.

\subsection{Common covariates}
Here we consider settings where $\bm w_i$ and $\bm x_i$ may share some covariates. In particular, we fix $n=1200, p = q = 4$ and evaluate the following scenarios:
\label{sec:sim_commomcovars}
\begin{itemize}
    \item Setting 1 Case 1: we set $\bm \alpha =(-0.3,0.7,0.1,0.6)^T,\bm\beta_1=(0.6,-0.7,-0.6,0.5)^T, \bm\beta_2=(0.1,0.6,0.2,0.8)^T$ and, $\sigma_1^2=\sigma_2^2=0.25$. The first component of $\bm w_i$ is an intercept and the second, third and fourth components are drawn independently from $\mathcal{N}(0.5,1)$, $\mathcal{N}(0.5,1)$ and $\mathcal{N}(0,1)$ respectively. Similarly, in the ordinal model, the first component of $\bm x_i$ is an intercept, and the second, third and fourth components are drawn independently from $\mathcal{N}(0.5,1)$, $\mathcal{N}(0.5,0.8^2)$ and $\mathcal{N}(0,0.8^2)$ respectively.
    \item Setting 1 Case 2: same as Setting 1 Case 1 except that in the ordinal model, the second, third and fourth components of $\bm x_i$ are equal to the corresponding components of $\bm w_i$ for all $i$.
    \item Setting 2 Case 1: we set $\bm\beta_1=(0.6,-0.6,-0.6,0.5)^T, \bm\beta_2=(0.1,-0.1,-0.1,0.8)^T$ and leave $\bm \alpha,\sigma_1^2, \sigma_2^2, \bm w_i,\bm x_i$ unchanged from Setting 1 Case 1.
    \item Setting 2 Case 2: the parameters remain unchanged from Setting 2 Case 1 but $\bm x_i$ follows the scheme described in Setting 1 Case 2.
\end{itemize}
\begin{figure}[!h]
	\centering
	\includegraphics[width=1\linewidth]{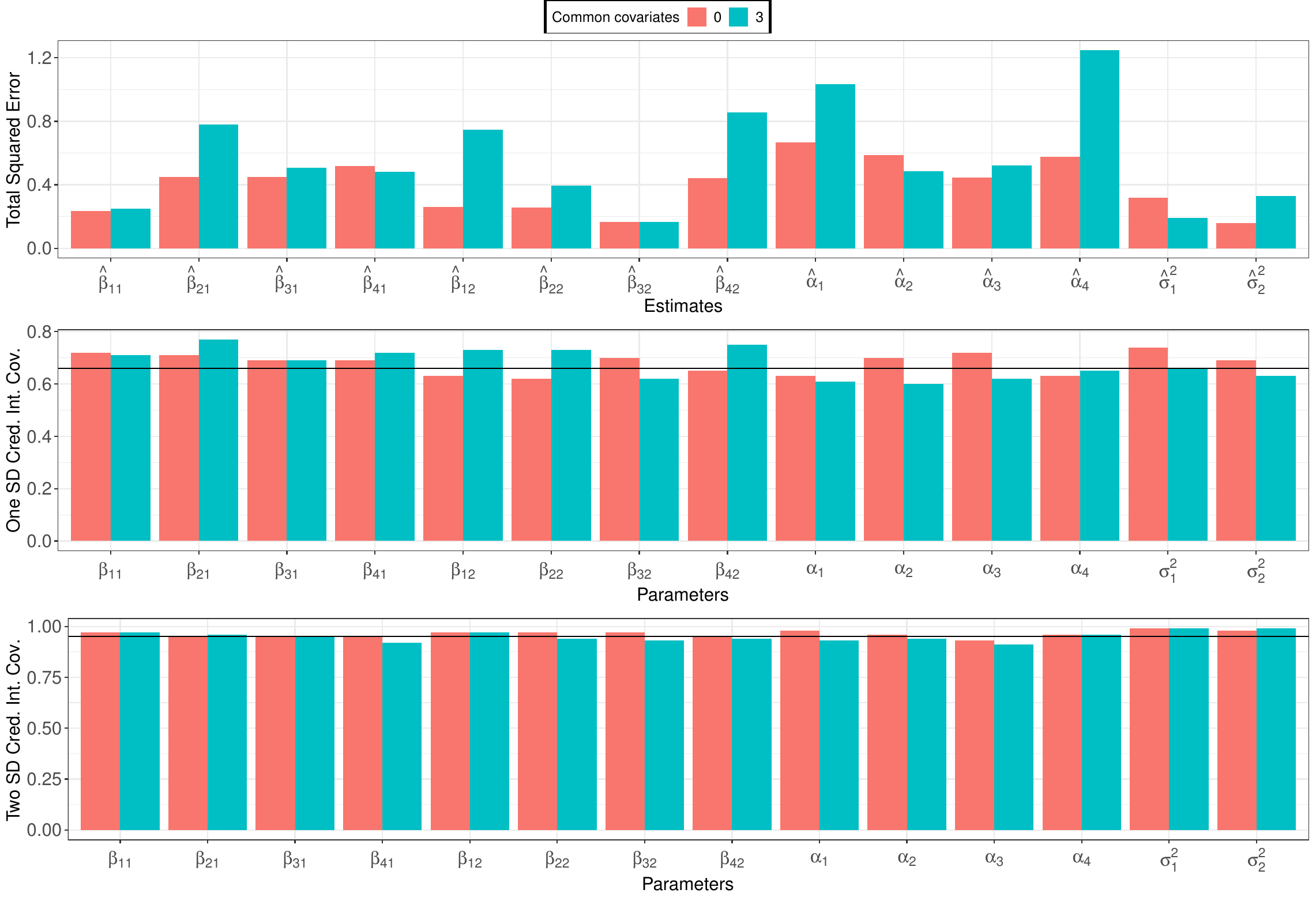}
	\caption{\label{fig:sim_sepc1} \texttt{TSE} and the coverages of one standard deviation and two standard deviation posterior credible intervals under Setting 1 across $100$ Monte-Carlo repetitions.}
\end{figure}
\begin{figure}[!h]
	\centering
	\includegraphics[width=1\linewidth]{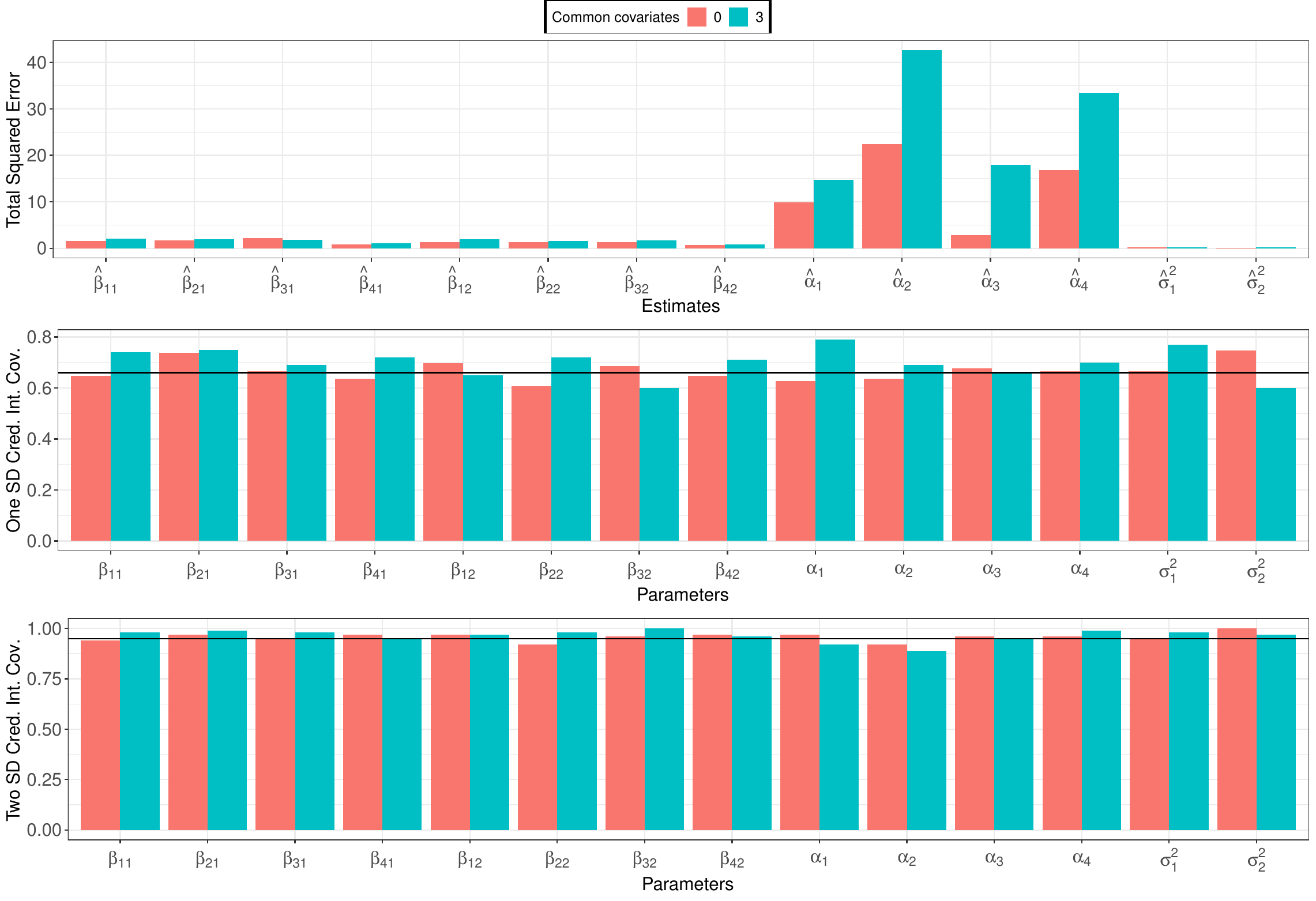}
	\caption{\label{fig:sim_sepc2}  \texttt{TSE} and the coverages of one standard deviation and two standard deviation posterior credible intervals under Setting 2 across $100$ Monte-Carlo repetitions.}
\end{figure}
Under these
specifications, Setting 1 represents a scenario where the two latent classes are relatively more distinct than Setting 2. Furthermore, the empirical applications discussed in sections \ref{sec:results_fdic} and \ref{sec:results_fslic} are an example of Case 1 where $\bm w_i$ and $\bm x_i$ do not share any covariates, while Case 2 represents the other extreme where $\bm w_i$ and $\bm x_i$ share all covariates. 

Figure \ref{fig:sim_sepc1} presents the \texttt{TSE} and the coverages of the one standard deviation and two standard deviation posterior credible intervals under Setting 1 across $100$ Monte-Carlo repetitions. Overall, the top panel reveals that estimation accuracy is relatively poorer when $\bm x_i$ and $\bm w_i$ share all covariates. This is not entirely surprising {because conditional on the class membership, the common covariates in $\bm x_i$ are now relatively less informative in predicting the ordinal outcome in this setting}. In terms of the coverages, the middle and bottom panels of Figure \ref{fig:sim_sepc1} reveal that, in general, the posterior credible intervals provide the correct coverage across the two cases, especially for the 2-standard deviation credible intervals. 

As observed under Setting 2 of Section \ref{sec:sim_comparenuts}, Setting 2 here represents a relatively difficult scenario and the estimation accuracy of $\bm \alpha$ is particularly poor under Case 2 of Setting 2. This is depicted in the top panel of Figure \ref{fig:sim_sepc2} where the inflated \texttt{TSE}'s under Case 2 reveals that the point estimate of $\bm \alpha$ returned by the \texttt{CG} sampler is relatively inferior in this setting. While the posterior credible intervals under Setting 2, in general, continue to provide close to correct coverage across the two cases, the results for Case 2 across both the settings reveal that the estimation accuracy of the proposed \texttt{CG} sampler may be impaired if $\bm x_i$ and $\bm w_i$ share a majority of their covariates in the latent class ordinal probit model of Section \ref{sec:model}. Conversely, the latent class model yields accurate results and inferences in settings such as our empirical study where the two sets of covariates are entirely disjoint.

 Latent class models, by construction, entail likelihoods that are multimodal and have limited curvature, especially in settings where the true latent classes are not substantially distinct. Simulation settings reveal that the Collapsed Gibbs sampler developed in this paper recovers accurate estimates and provides valid inferences even in the presence of such estimation challenges. More broadly, these findings highlight the particular suitability of Bayesian computational methods in estimating latent class models. Frequentist techniques that rely on hill-climbing optimization techniques would be especially vulnerable to problems such as running into local maxima and convergence in flat regions of the likelihood. Bayesian sampling methods such as the \texttt{CG} sampler introduced in this paper and \texttt{NUTS} effectively explore the parameter space using proposals that permit them to exit flat regions. Accordingly, Bayesian techniques recover underlying latent classes and support clearer substantive conclusions on unobserved heterogeneity in such models.  
\section{Regional economic distress and FDIC's decisions for bank resolutions}
\label{sec:coveffect_banks_app}
In this section, we report the covariate effects of the remaining variables from the selected ordered response model in specification 3 of Table \ref{tab:cmrd}. 
\begin{figure}[!t]
	\centering
	\includegraphics[width=0.8\linewidth]{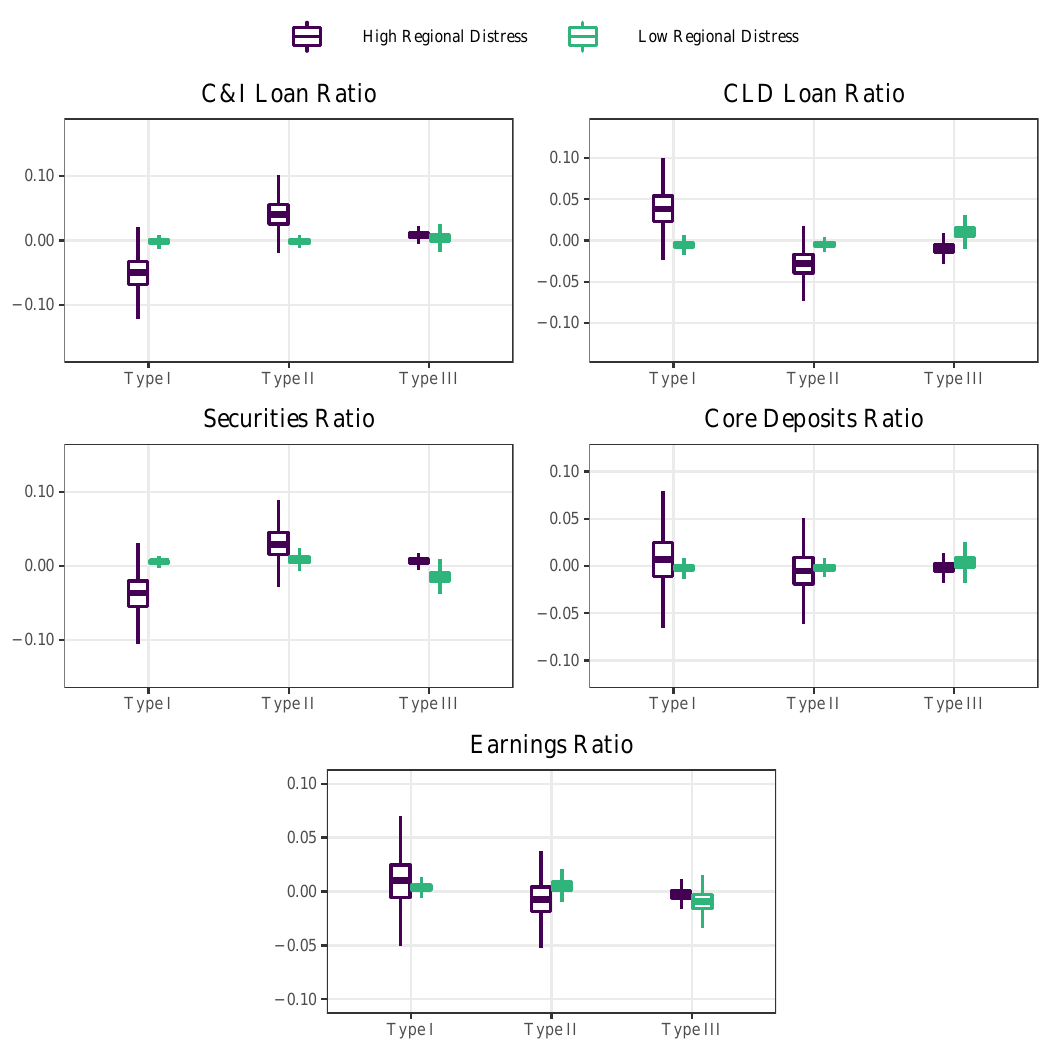}
	\caption{Additional covariate effects from the models for resolution type for banks in the class of High Regional Distress (HRD) and Low Regional Distress (LRD).\label{fig:covarEff1}}
\end{figure}

In Figure \ref{fig:covarEff1}, a standard deviation increase in Commercial and Industrial (C\&I) loan ratio is associated with an increased probability of Type II and Type III resolutions and a corresponding decline in the probability of a Type I resolution among HRD failures. These findings are consistent with the higher levels of risk attributed to increases in this ratio by the FDIC \citep{fdiccrisis}. A standard deviation increase in Construction and Land Development Loans (CLD) is associated with an increased probability of Type I resolutions among HRD failures and of Type III resolutions among LRD failures. In the period under study, C\&I loans constitute a higher concentration of bank balance sheets (27.5\% of assets) than CLD loans (4.6\% of assets). Accordingly, an increase in C\&I loans represents a more acute concentration in that loan category and elicits a more stringent response than an equivalent increase in CLD loans. 

The remaining covariates, securities ratio, core deposits ratio and earnings ratio resulted in covariate effects that were close to zero upon controlling for other balance sheet items pertaining to size, asset quality and interstate branching restrictions. An increase in each of these characteristics is associated with an increase in the bank's franchise value and is accordingly expected to result in a lower probability of liquidation under Type III resolutions.
\section{Regional Distress and S\&L Resolutions}
\label{sec:coveffects_fslic_app}
\begin{figure}[!t]
	\centering
	\includegraphics[width=0.8\linewidth]{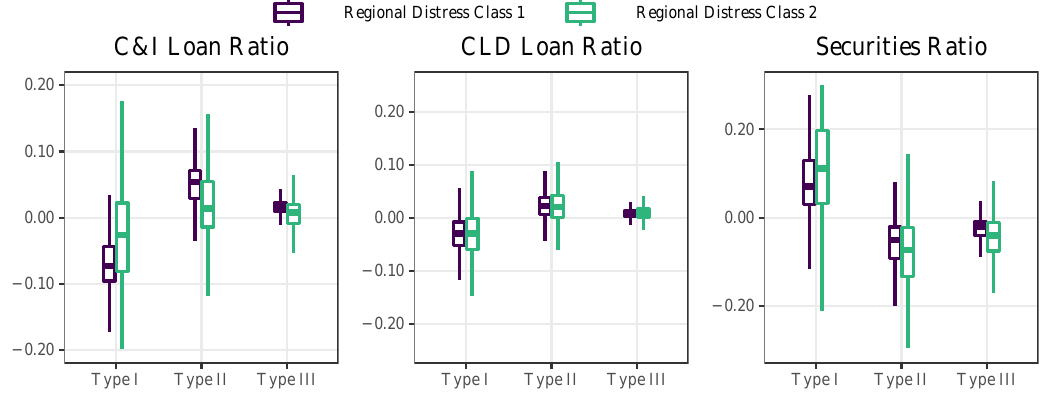}
	\caption{Additional covariate effects from the models for resolution type for S\&L's in Regional Distress Class 1 and Regional Distress Class 2.\label{fig:covarEff1sl}}
\end{figure}
In this section, we report the covariate effects of the remaining variables from the selected ordered response model in specification $3\dagger$ of Table \ref{tab:rd_sl}. In Figure \ref{fig:covarEff1sl}, across both latent classes, the probability of receiving assistance declined with a standard deviation increase in commercial and industrial (C\&I) loan ratio. S\&L's, which historically specialized in retail mortgages, were permitted to lend C\&I loans only in the mid-1980's, and thereby this category represents riskier loans as S\&L's had lesser expertise and experience in evaluating these loans compared to commercial banks. Therefore, institutions with higher shares of C\&I loans were less likely to be assisted and more likely to be sold to other S\&L's or liquidated. Similarly, the FSLIC was less likely to assist S\&L institutions with higher shares of construction and land development (CLD) loans on their balance sheets and more likely to facilitate the acquisition of such institutions or to liquidate them. These decisions reflect the high-risk nature of CLD loans within the broader category of commercial real estate loans because projects backed by such loans may become subject to construction delays, which may result in missed loan payments. In contrast to C\&I and CLD loans, S\&L's with larger shares of securities relative to total assets were more likely to be assisted and less likely to be sold or liquidated. Securities are considered to be safer and more liquid assets compared to loans, and higher shares of these assets are likely to have enabled S\&L's to retain their value through the crisis. Overall, in line with results in Figure \ref{fig:combined_sl}, we continue to find that the FSLIC's decision rules were homogeneous across the two latent classes.
\section{Additional details on the empirical analyses}
\label{app:additional_emp_analysis}
Here we discuss several additional details on the empirical analyses of sections \ref{sec:results_fdic} and \ref{sec:results_fslic}.
\subsection{Why are there two latent classes?}
\label{app:two_latentclasses}
There are two reasons for considering two latent classes in our analyses. We discuss them below.
 \begin{enumerate}
 \item \textbf{Consistency with the theoretical studies - }we have considered two latent classes to be consistent with the theoretical studies that designed decision rules for regulators to preserve the maximum economic value of the banking system during crises. Broadly, preserving the economic value of the banking system entails actions that maintain financial stability while curtailing moral hazard. In these studies, researchers found that regulators can reconcile these two competing objectives and maximize the value of the banking system when they apply distinct decision rules that vary with economic and banking conditions, and typically considered two disparate states of the world in their models to encapsulate variations in underlying conditions. For instance, in one such foundational study by \cite{cordella2003bank}, the authors derived a threshold level of economic conditions accompanying bank failures and a decision rule around the threshold. According to this rule, when the economy is worse off relative to the threshold, regulators bail out failed banks, and when conditions are more benign, regulators liquidate them. Similarly, \cite{acharya2007too} considered a state in which multiple bank failures occur simultaneously and another in which only few banks fail. They found that regulators would preserve most value by assisting banks in the former state and facilitating acquisitions of failed banks by healthy banks in the latter. \cite{deyoung2013theory} considered the effects of political pressure on regulatory decisions and derived a threshold level  below which regulators would apply disciplinary actions and otherwise, bail them out. Since we use the decision rules and equilibrium conditions derived in theoretical studies as a benchmark to assess the extent to which banking regulators served the public interest, we consider two latent classes that correspond to the dichotomous states described in these studies.
\item \textbf{Sample size constraints and parsimonious test of theoretical predictions - }in addition to adhering to the structure presented in theoretical studies, our two-class model supports a parsimonious test of theoretical predictions. Although our data sample is based on crises that precipitated the most number of depository institution failures after the Great Depression, our sample consists of 1,385 bank failures and only 389 S\&L failures. Because we consider an empirical framework where we not only compare regulators' actions with theoretical predictions, but also compare the two regulators with each other, we adopt a method that accommodates the setting with the lower of the two sample sizes. Accordingly, in the light of sample size constraints, the two-class structure permits us to test theoretical predictions while limiting the risk of over-fitting that may arise in a multi-class model.  
\end{enumerate} 
\subsection{Assignment of covariates in the class membership and ordinal probit models}
\label{app:covariates}
Broadly, we assign covariates $\bm w_i$ into the class membership model when they are associated with the \textit{decision rule} a regulator will choose, namely, whether the regulator will adopt a rule that is more forbearing and entails a higher probability of assigning assistance, or one that is more stringent, and associated with a higher probability of liquidations. Correspondingly, we assign covariates $\bm x_i$ into the ordinal outcome model when they influence the \textit{action} the regulator will undertake, namely, whether the regulator will provide assistance to the failed bank (Type I), facilitate its acquisition by another bank (Type II) or liquidate the bank and pay out its depositors (Type III). As a result of this assignment procedure, the covariates in the class membership model contain characteristics that are outside the control of banks' decisions, whereas covariates in the ordinal probit model for resolution are measures that are predominantly the result  of banks' prior actions. Accordingly, the vector $\bm w_i$ contains covariates that represent the underlying economic, banking or political conditions in which bank failures took place. The vector $\bm x_i$ consists of variables derived from banks' income statements and balance sheets representing their financial health.   
 
 The specific components of the covariates in the class membership model, the vector $\bm w_i$, depend on the hypothesis being tested and are derived from recommendations in theoretical papers on bank resolutions \citep{cordella2003bank, acharya2007too, deyoung2013theory}. Theoretical papers derive state-dependent decision rules, and provide conditions under which a particular decision rule is appropriate. As detailed in Section \ref{sec:model}, in testing the hypothesis for the presence of distinct decision rules in high and low economic distress ($H_{1}$), the covariates consist of economic indicators such as the unemployment rate, housing starts and per capita GDP growth pertaining to the state and the county in which a bank operates. To determine whether regulators applied different decision rules based on the health of the banking industry ($H_{2}$), we use measures such as the number of previous bank closures and the share of assets in distressed banks in a county where a bank  is headquartered in our covariate vector. To test whether political influence induced regulators to apply differential resolution rules ($H_{3}$), we include the percentage of Congressional representatives from the bank's headquartered state who voted for various bills that were favorable to the banking industry. 
 
 In operationalizing these hypotheses, we also considered settings where we included covariates associated with local economic conditions (pertaining to $H_{1}$) into models that tested for hypotheses that consider banking industry distress and political influence ($H_{2}$ and $H_{3}$ respectively). These specifications address guidance from regulatory agencies, which advise bank examiners to incorporate local economic conditions into their assessments while closing and resolving failed banks \citep{occ2011, fdiccrisis}. Because these guidelines do not specifically address how regulators must incorporate economic conditions into their decisions, we reconcile these guidelines with recommendations from theoretical studies and include them in the class membership model. 
 
To determine the type of resolution that regulators assigned to failed banks, we used the set of covariates in \cite{balla2019comparison}, which encapsulate banks' financial position and include variables that represent bank size, the quality of bank assets and composition of risky asset classes. The authors selected bank characteristics that are associated with failure and a higher loss to the FDIC in their covariate vector. Since banks that are in a more adverse financial position are more likely to fail and to leave a larger dent on the insurer's funds, the covariates that predict failure and losses to the FDIC also provide information about the value regulators assign to these banks and the action they are likely to take on them. Besides the covariates based on bank financial statements, we also included an indicator to denote whether the bank was headquartered in a state that permitted banks to open branches and acquire banks across state-lines. Although this variable is not directly the result of banks' actions, \cite{strahan2003real}
suggest policymakers respond to banks' lobbying efforts in enacting legislation to permit interstate banking. Moreover, \cite{acharya2007cash} find that interstate banking expands the probability of facilitating mergers (Type II) and thereby affects regulators' resolution action rather than the decision rule, which indicates that this measure befits as a covariate in the ordinal probit model.
\subsection{Model fit}
\begin{table}
  \caption{Fit statistics for models based on the FDIC's decisions on failed banks\label{tab:fit_banks}}
  \centering
  \scalebox{0.8}{
    \begin{tabular}{cccccc}
    \hline
    \multicolumn{1}{c}{} &\multicolumn{1}{c}{Model} & \multicolumn{1}{l}{log Marginal Likelihood} & \multicolumn{1}{l}{Accuracy Rate} & \multicolumn{1}{l}{RMSE} & \multicolumn{1}{l}{DIC} \\
    \hline
    &1     & -703.4 & 0.86  & 0.38  & 3874 \\
    Regional&2     & -701.1 & 0.85  & 0.38  & 3828 \\
   distress&\textbf{3}     & \textbf{-699.8} & \textbf{0.86}  & \textbf{0.38}  & \textbf{3810} \\
    &4     & -700.2 & 0.85  & 0.38  & 3822 \\
    \hline
    &5     & -719.3 & 0.85  & 0.39  & 3053 \\
    &6     & -705.3 & 0.85  & 0.39  & 3050 \\
    Banking &7     & -719.1 & 0.85  & 0.39  & 3054 \\
    industry&\textbf{8}     & \textbf{-697.2} & \textbf{0.86} &\textbf{0.37}  & \textbf{3831} \\
    distress&9     & -697.8 & 0.86  & 0.37  & 3828 \\
    &10    & -701.2 & 0.86  & 0.37  & 3898 \\
    \hline
    &11    & -714.5 & 0.85  & 0.39  & 2944 \\
    &12    & -701.3 & 0.85  & 0.39  & 2967 \\
    &13    & -706.7 & 0.85  & 0.39  & 2994 \\
    &14    & -716.3 & 0.85  & 0.39  & 3026 \\
    &15    & -699.8 & 0.85  & 0.39  & 2894 \\
    Political&16    & -698.3 & 0.87  & 0.37  & 3833 \\
    factors&\textbf{17}    & \textbf{-693.7} & \textbf{0.86}  & \textbf{0.38}  & \textbf{3823} \\
    &18    & -705.5 & 0.86  & 0.38  & 3910 \\
    &19    & -700.3 & 0.86  & 0.38  & 3801 \\
    &20    & -704.5 & 0.86  & 0.37  & 3912 \\
    \hline
    \end{tabular}}%
 \end{table}%
 Table \ref{tab:fit_banks} summarizes the measures of fit for models based on the FDIC's decisions on failed banks. The three sections of the table represent models pertaining to classes based on regional distress (models 1-4), banking industry (models 5-10) distress and political factors(models 11-20). Here, the accuracy rate is the ratio of correct predicted outcomes relative to total sample size, and the root mean square error is based on the difference between the observed and predicted outcomes. 
 
 We find that for models based on regional distress, our measures of fit are mostly consistent in selecting model 3 across all the candidates within the group--the marginal likelihood is highest, and 
 DIC is lowest for this model. We find limited variability in the accuracy rate and RMSE across specifications. In the specifications based on distress in the banking industry, we find that the marginal likelihood and RMSE select model 8, but the DIC favors specifications 5-7, which have fewer covariates. Similarly, in the specifications based on political influence on regulatory decisions, we find that the marginal likelihood selects model 17, but the DIC is lower for specifications 11-15. The accuracy ratio and RMSE remain broadly consistent across specifications and only marginally favor other models. In the event of disagreement between the marginal likelihood and the DIC, we assign a higher weight to the former because the DIC has been shown to have a higher mis-selection probability that remains bounded below by a constant while this probability converges to zero for the marginal likelihood \citep{maity2021bayesian}. In selecting model 17, we have incorporated guidance from official agencies that require examiners to include measures of regional economic conditions in their assessments of banks and in their resolution \citep{occ2011, fdicres} as this model includes state and county-level economic indicators in addition to measures of political influence.

 Table \ref{tab:fit_sl} summarizes the measures of fit for the FSLIC's decisions on failed S\&L's. The four measures of fit broadly are in alignment for models based on regional distress, as well as those based on distress in the S\&L industry and select models 3 and 8 respectively across candidate models. In the group of models based on political influence, the marginal likelihood, the accuracy ratio and RMSE all favor model 18, whereas the DIC is lowest for model 12, for which the accuracy ratio is lower and RMSE is higher than our selected model.  
 \begin{table}
   \caption{Fit statistics for models based on the FSLIC's decisions on failed S\&L's\label{tab:fit_sl}}
  \centering
  \scalebox{0.8}{
    \begin{tabular}{cccccc}
    \hline
    \multicolumn{1}{c}{} &\multicolumn{1}{c}{Model} & \multicolumn{1}{c}{log Marginal Likelihood} & \multicolumn{1}{c}{Accuracy Rate} & \multicolumn{1}{c}{RMSE} & \multicolumn{1}{c}{DIC} \\
    \hline
    &1     & -302.3 & 0.73  & 0.60  & 571.7 \\
    Regional&2     & -305.4 & 0.72  & 0.60  & 589.0 \\
    distress&\textbf{3}     & \textbf{-302.0}  & \textbf{0.73}  & \textbf{0.59}  & \textbf{511.8} \\
    &4     & -305.4 & 0.73  & 0.59  & 585.4 \\
    \hline
    &5     & -301.4 & 0.70  & 0.65  & 570.6 \\
    &6     & -300.8 & 0.72  & 0.61  & 563.6 \\
   S\&L &7     & -301.0  & 0.73  & 0.60  & 579.8 \\
    distress&\textbf{8}     & \textbf{-297.2} & \textbf{0.74}  & \textbf{0.59}  & \textbf{554.0} \\
    &9     & -300.7 & 0.73  & 0.60  & 573.5 \\
    &10    & -300.3 & 0.74  & 0.59  & 574.2 \\
    \hline
    &12    & -299.1 & 0.74  & 0.58  & 546.2 \\
    &13    & -294.0  & 0.76  & 0.58  & 548.4 \\
    &14    & -300.0  & 0.76  & 0.56  & 587.7 \\
     Political&16    & -301.2 & 0.74  & 0.59  & 571.2 \\
   factors&17    & -297.5 & 0.76  & 0.57  & 581.3 \\
    &\textbf{18}    & \textbf{-288.1} & \textbf{0.76}  & \textbf{0.56}  & \textbf{562.1} \\
    &19    & -299.1 & 0.75  & 0.57  & 595.7 \\
    &20    & -300.4 & 0.76  & 0.57  & 578.8 \\
    \hline
    \end{tabular}}%
 \end{table}%
\subsection{Interactions Across Categories of Latent Classes}
\label{sec:interact_lc}
  To understand interactions across the different classifications we used in the paper, we estimated a model that incorporates covariates pertaining to all three hypotheses and arrive at latent classes that are consistent with those that we reported in the paper. Notably, we do find evidence of predominantly stable latent classes for the FDIC as the estimates of parameters in the class membership model and probabilities of each resolution method within latent classes have remained stable across the previous specifications and the combined model. Overall, the FDIC's decisions applied distinct decision rules on underlying economic and banking conditions, as recommended by theoretical studies. In addition, political support for the banking industry also explained variations in the FDIC's decision rules. The FSLIC's decision rules, however, did not vary with economic or banking distress nor with political support for the industry. Our qualitative finding remains unchanged in that the FDIC's actions were broadly consistent with serving the public interest whereas the FSLIC's deviated from this objective. 
  
  Tables \ref{tab:cmoverall} and \ref{tab:cmoverall_t} summarize the results from including covariates pertaining to regional distress, banking or S\&L distress and political influence in the class membership model for banks and S\&L's respectively. The column labelled ``Overall" reports the results from combining covariates across the three hypotheses and the preceding columns report the results for models based on each individual hypothesis from in the paper. In Table \ref{tab:cmoverall}, unemployment rate and housing starts continue to be statistically important in determining latent classes in this overall model along with previous bank closures and the percent of Congressional votes in favor of a bill supporting the industry. Thereby, we label latent class 1 as the class of failures under ``High Regional, Banking distress and Political Support", and latent class 2 as failures under ``Low Regional and/or Banking Distress and/or Political Support". Finally, we note that the estimates of coefficients for each covariate are similar in magnitude to those from the settings pertaining to each individual hypothesis under the columns ``Regional distress", ``Banking distress" and ``Political economy".
  
  In Table \ref{tab:cmoverall_t}, we find that the covariates predicting regional and banking distress were not statistically important in explaining differences in the FSLIC's decisions. Similarly, political influence as measured by support for a bill to intensify penalties for financial institutions did not uncover differences in the FSLIC's decisions. Instead, the share of Republican representatives that is included to control for the role of political affiliation in Congressional members' votes, and is not informative in and of itself, was statistically important. Because the two latent classes are not statistically determined by covariates corresponding to any of the three hypotheses in our analysis, we label the two latent classes as ``Combined Model Class 1" and ``Combined Model Class 2". 
  
  We report the marginal likelihood as well as the number of observations for each model from the paper and the combined model in the final two rows of Tables \ref{tab:cmoverall} and \ref{tab:cmoverall_t}. We find that the overall model has a lower marginal likelihood than the one based on political factors in explaining the FDIC's decisions in Table \ref{tab:cmoverall} (the only comparable marginal likelihood as the two models have an equal number of observations). The overall model has a higher marginal likelihood than other specifications in explaining the FSLIC's decisions in Table \ref{tab:cmoverall_t} but this model is also based on fewer observations owing to limited matches across datasets. 

\begin{figure}[!t]
\centering
\includegraphics[width=0.85\linewidth]{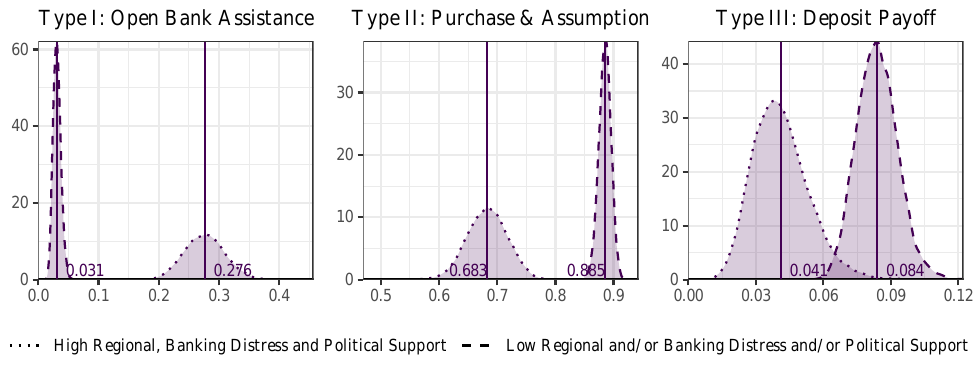}
\caption{Posterior distribution of the average probability of the FDIC assigning each resolution method within classes based on regional and banking distress, as well as political influence. The horizontal axis represents the probability of assigning a resolution method and the vertical axis represents the posterior density associated with that probability based on a kernel density estimate. The solid vertical lines represent the means of these posterior distributions across the $G$ MCMC draws.\label{fig:kds_h123}}
\end{figure}

\begin{figure}[!t]
\centering
\includegraphics[width=0.85\linewidth]{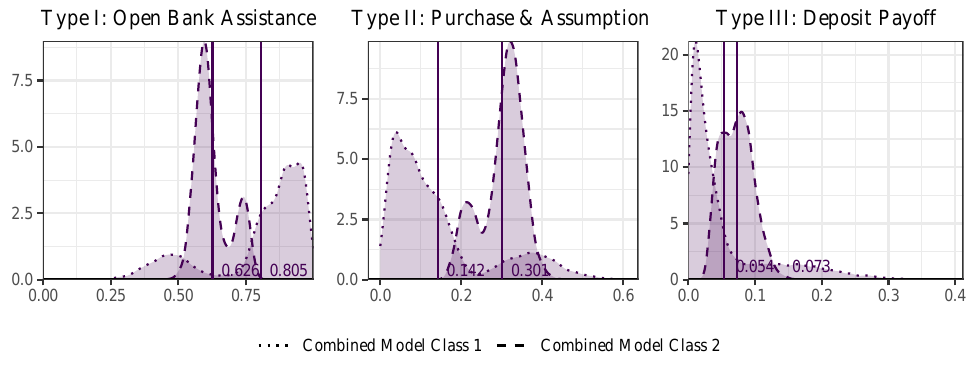}
\caption{Posterior distribution of the average probability of the FSLIC assigning each resolution method within classes based on regional and banking distress, as well as political influence. The horizontal axis represents the probability of assigning a resolution method and the vertical axis represents the posterior density associated with that probability based on a kernel density estimate. The solid vertical lines represent the means of these posterior distributions across the $G$ MCMC draws.\label{fig:kds_sl_h123}}
\end{figure}

  Figures \ref{fig:kds_h123} and \ref{fig:kds_sl_h123} respectively plot the density of the posterior distribution of the average probability of the FDIC and FSLIC assigning each resolution method to banks in the two latent classes in the model that combines regional, banking and political factors. Figure \ref{fig:kds_h123} shows that the classes that we uncover for banks continue to remain distinct based on all three factors. Moreover, the probability of Type I, II and III resolution in the latent class ``High Regional, Banking distress and Political Support" is 27.6\%, 68.3\% and 4.1\% respectively. These probabilities are similar to those reported for classes based on high regional distress in the paper, which were 24.6\%, 72.1\% and 3.2\%. For S\&L's, we continue to find limited separability across classes on incorporating covariates from all three settings in Figure \ref{fig:kds_sl_h123}. This finding is the only departure from our results in the paper as political factors were found to be statistically important in explaining differences in the FSLIC's decision rules in the paper. 

\begin{table}
	\caption{\label{tab:cmoverall} Covariate effects from class-membership models for specifications of latent classes based on regional economic distress, banking industry distress and political economy characteristics. The reported values are posterior means of the covariate effects. Posterior standard deviations are in parentheses.}
\centering
 \scalebox{0.85}{
	\begin{tabular}{lc|c|c|c}
		\hline
		& Regional distress & Banking distress & Political economy & Overall \\
		\hline
		\textbf{State-level characteristics} &       &       &       &  \\
		Unemp. & -0.1 (0.04) & -0.07 (0.04) & -0.13 (0.04) & -0.11 (0.04) \\
		Housing starts & 0.05 (0.05) & 0.03 (0.05) & -0.05 (0.04) & -0.07 (0.04) \\
		\textbf{County-level characteristics} &       &       &       &  \\
		Per capita GDP growth & 0.04 (0.05) & 0.04 (0.05) & 0 (0.04) & 0.04 (0.05) \\
		Farm, agri, mining & 0.06 (0.04) & 0.03 (0.04) & 0.09 (0.05) & 0.06 (0.05) \\
		\textbf{Banking industry characteristics} &       &       &       &  \\
		\multicolumn{1}{p{12.265em}}{Previous closures} & -0.05 (0.03) & -0.02 (0.01) & -     & -0.02 (0.01) \\
		\multicolumn{1}{p{12.265em}}{\% Assets in distressed banks} & -     & -0.03 (0.02) & -     & -0.02 (0.02) \\
		\textbf{Insurer characteristics} &       &       &       &  \\
		Dep. Ins. Fund/Total Deposits & -     & -0.05 (0.03) & -0.05 (0.03) & -0.04 (0.03) \\
		\textbf{Political economy characteristics} &       &       &       &  \\
		\% vote for Bill 3 & -     & -     & -0.14 (0.03) & -0.13 (0.04) \\
		\% vote for Republicans & -     & -     & 0.12 (0.05) & 0.10 (0.04) \\
        \hline
		log Marginal Likelihood   & -699.79 & -697.22 & -693.68 & -694.34 \\
  \hline
  Count & 1385 & 1385 & 1380 & 1380\\
		\hline
	\end{tabular}}%
\end{table}%

\begin{table}
\caption{\label{tab:cmoverall_t} Covariate effects from class-membership models for specifications of latent classes based on regional economic distress, S\&L industry distress and political economy characteristics. The reported values are posterior means of the covariate effects. Posterior standard deviations are in parentheses.}
 \centering
 \scalebox{0.85}{
	\begin{tabular}{lc|c|c|c}
		\hline
		& Regional distress & Banking distress & Political economy & Overall \\
		\hline
		\textbf{State-level characteristics} &     &     &      &      \\
		Unemp. & -0.17 (0.48) & -0.19 (0.23) & -0.03 (0.06) & -0.01 (0.07) \\
		Housing starts & -0.03 (0.09) & -0.08 (0.08) & 0.03 (0.07) & 0.04 (0.08) \\
		\textbf{County-level characteristics} &     &     &      &      \\
		Per capita GDP growth & 0 (0.09) & -0.1 (0.09) & 0.21 (0.1) & 0.12 (0.11) \\
		Farm, agri, mining & -0.01 (0.09) & -0.05 (0.07) & -0.04 (0.05) & -0.04 (0.06) \\
		\textbf{S\&L industry characteristics} &     &     &      &      \\
		Previous closures &       & -0.03 (0.07) & -     & -0.04 (0.06) \\
		\% Assets in distressed S\&L's & -     & 0.02 (0.07) & -     & -0.03 (0.07) \\
		\textbf{Insurer characteristics} &     &     &      &      \\
		Dep. Ins. Fund/Total Deposits & 0.01 (0.08) & 0.03 (0.05) & -0.04 (0.06) & -0.02 (0.06) \\
		\textbf{Political economy characteristics} &     &     &      &      \\
		\% vote for Bill 2 & -     & -     & -0.14 (0.05) & -0.07 (0.2) \\
		\% Republicans & -     & -     & 0.26 (0.08) & 0.22 (0.11) \\
		\hline
		log Marginal Likelihood   & -302.02 & -297.15 & -288.06 & -280.96 \\
  \hline
  Count & 389 & 387 & 385 & 383\\
		\hline
	\end{tabular}}%
\end{table}%

\subsection{Comparison with alternative specifications}
In this section, we consider two alternative specifications to model regulators' decision rules--the first involves interacting covariates $\bm w_i$ and $\bm x_i$ to arrive at distinct decision rules, the second considers $y_i$ to be multinomial instead of an ordered outcome. On estimating these alternative specifications, we find that the data continue to decisively select the latent class ordinal probit model based on marginal likelihood. 

Specifically, we estimate a simple ordinal probit model and a multinomial probit model--both with covariates $\bm x_i$, the broader set of covariates $(\bm x_i,\bm w_i)$ and interactions between $\bm x_i$ and an indicator for state-level recessions obtained from \cite{fdiccrisis}. We consider this indicator from a retrospective study of the crisis to avoid a proliferation of parameters from interacting the full covariate matrices $\bm x_i$ and $\bm w_i$ with each other. This specification with interaction terms implicitly assumes that researchers have knowledge of the FDIC's criterion for parsing banks through distinct decision rules. However, the indicator represents only an approximate reconstruction of the economic data the examiners may have considered in their decisions, and may not necessarily align with the thresholds they used to classify regions into groups of high and low distress. Table \ref{tab:mlval} provides the log marginal likelihood for each of these models along with our main benchmark model, specification 3 for the FDIC's resolution of failed banks.
Overall, we find that the values of log marginal likelihood in Table \ref{tab:mlval} decisively support the latent class model with a Bayes Factor that exceeds 1000 \citep{kass1995bayes} relative to the alternative specifications. 
\begin{table}
\caption{\label{tab:mlval}Log marginal likelihood from the latent class ordinal model and alternative specifications of the FDIC's resolution decisions.}
	\begin{tabular}{lr}
		\hline
		Model & \multicolumn{1}{l}{log Marginal Likelihood} \\
		\hline
        Selected latent class model (spec. 3) & -699.79\\
        Multinomial probit model & -764.13 \\
		Multinomial probit model + class membership covariates from (3) & -755.58\\ 
  	Multinomial probit model with interactions & -754.51\\ 
		Ordinal probit model  & -729.66 \\
		Ordinal probit model + class membership covariates from (3)  & -722.55 \\
		Ordinal probit model with interactions  & -729.52 \\		
		\hline
	\end{tabular}%
\end{table}%
\subsection{Model diagnostics related to the empirical applications}
 \begin{figure}[!t]
	\centering
\includegraphics[width=1\textwidth]{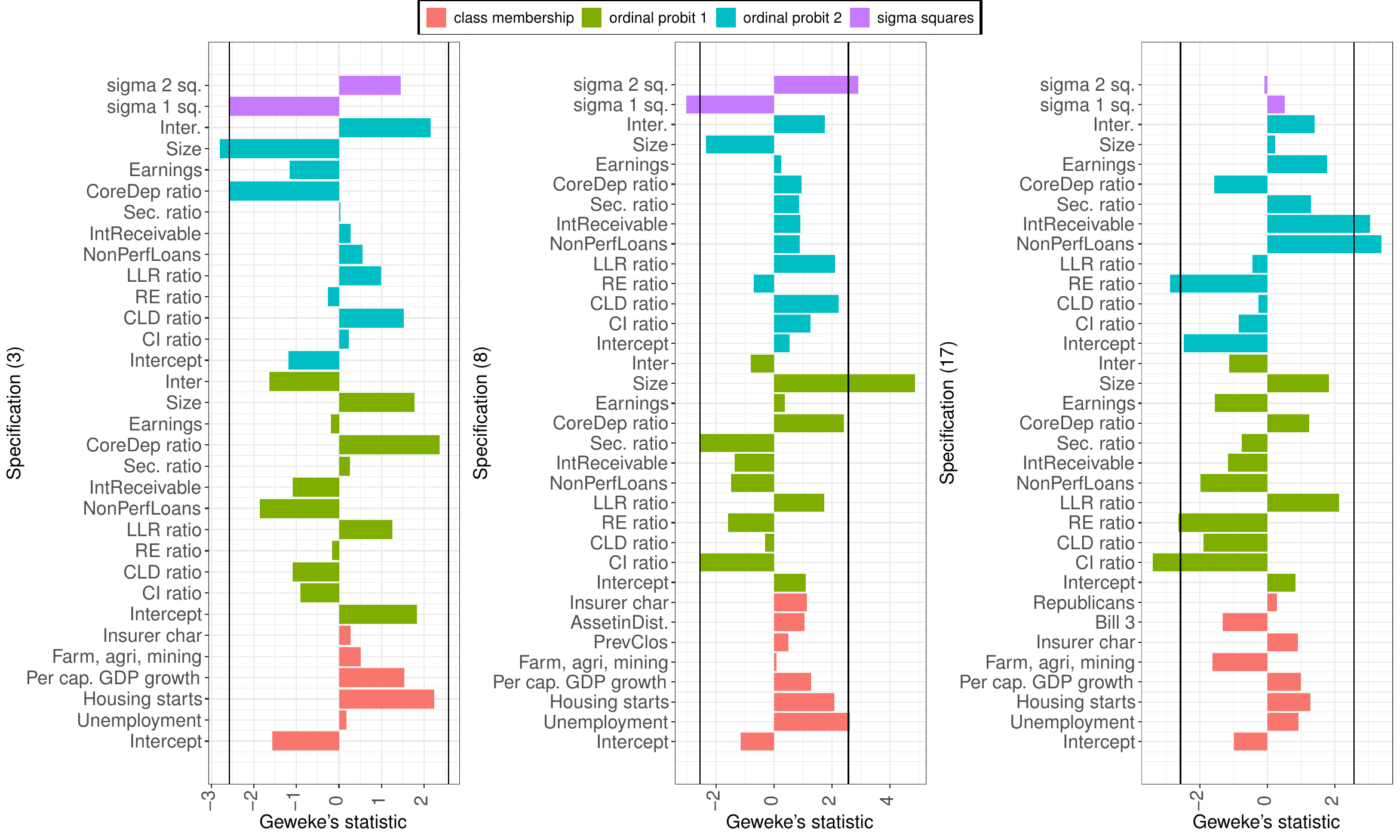}	
\caption{Banks}
 \label{fig:gweke_banks}
\end{figure} 
We rely on the univariate Geweke's convergence diagnostic \citep{geweke1992evaluating} to determine a burn-in of $1,000$ iterations for all our specifications. For each parameter, this diagnostic calculates the $Z-$score for a test of equality of means between the first $10\%$ and the last $50\%$ of the corresponding MCMC chain. Figure \ref{fig:gweke_banks} plots the $Z-$scores of the MCMC draws pertaining to each parameter of interest under the three selected specifications for FDIC. These parameters correspond to the coefficients of the different covariates in (1) the class membership model, (2) the two ordinal probit models, and (3) $\sigma_s,s\in\{1,2\}$. The two vertical bars in each of these panels indicate the $2.5^{th}$ and $97.5^{th}$ percentiles of a standard Normal distribution. With the exception of the coefficient for Size in Specification 8, and the coefficients for C\&I Loan Ratio (CI ratio), Interest Receivables (IntReceievable) and Nonperforming Loans Ratio (NonPerfLoans) for Specification 17, a burn-in of $1,000$ is overall consistent with the $Z-$scores of all other coefficients.
 \begin{figure}[!t]
	\centering
\includegraphics[width=1\textwidth]{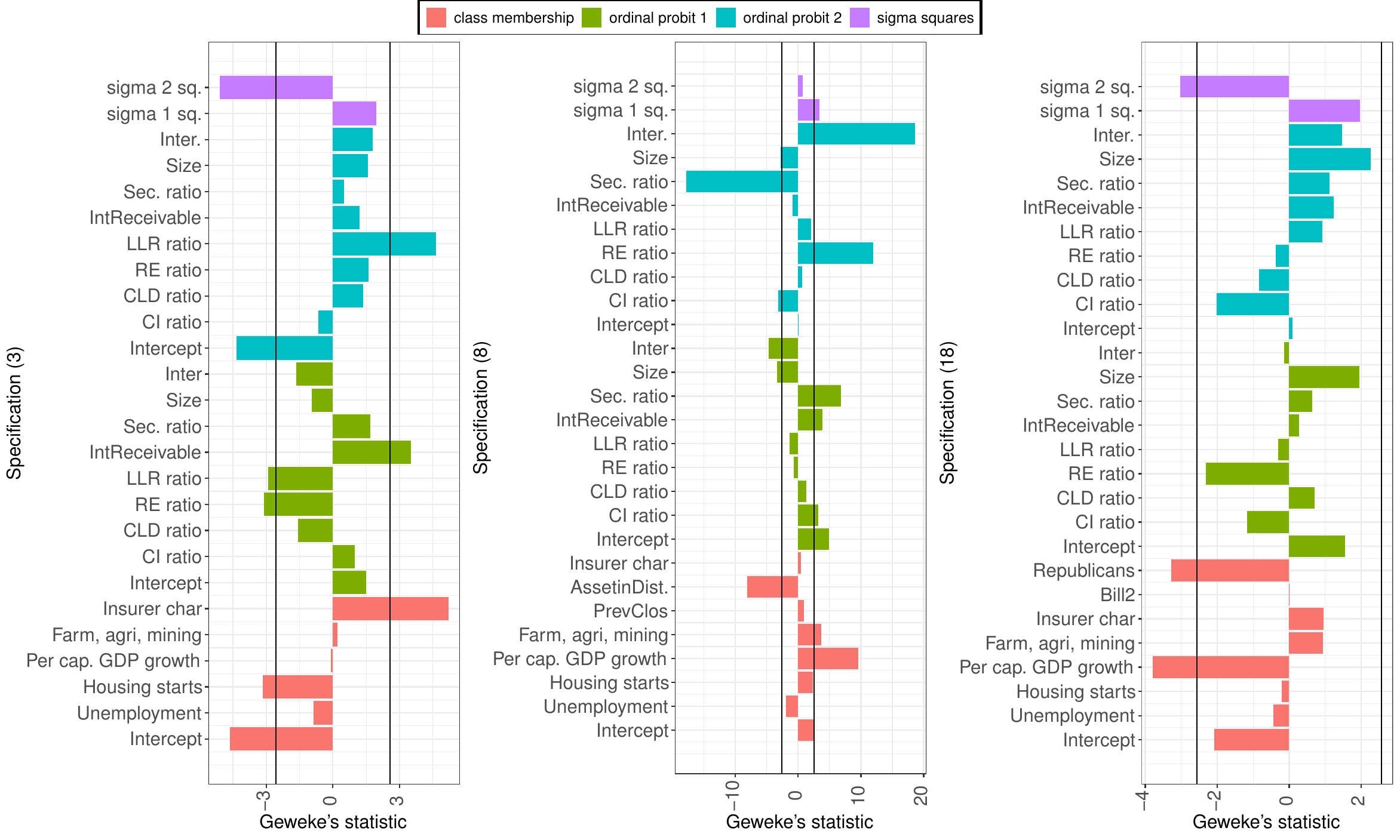}	
\caption{Thrifts}
 \label{fig:gweke_thrifts}
\end{figure} 

Figure \ref{fig:gweke_thrifts} plots the $Z-$scores of the MCMC draws pertaining to each parameter of interest under the three selected specifications for FSLIC. Here the middle panel suggests that a burn-in of $1,000$ may not be appropriate for Specification 8. The results from the left panel suggest a similar conclusion for Specification 3. These results are not surprising since in our data there are only 389 S\&L institutions, as opposed to 1,385 banks, of which just 15 institutions underwent resolution method III. With such limited data to represent one of the outcomes of the ordinal probit model, the MCMC chain may require a larger burn-in.
\begin{figure}[!h]
	\centering
\includegraphics[width=0.6\textwidth]{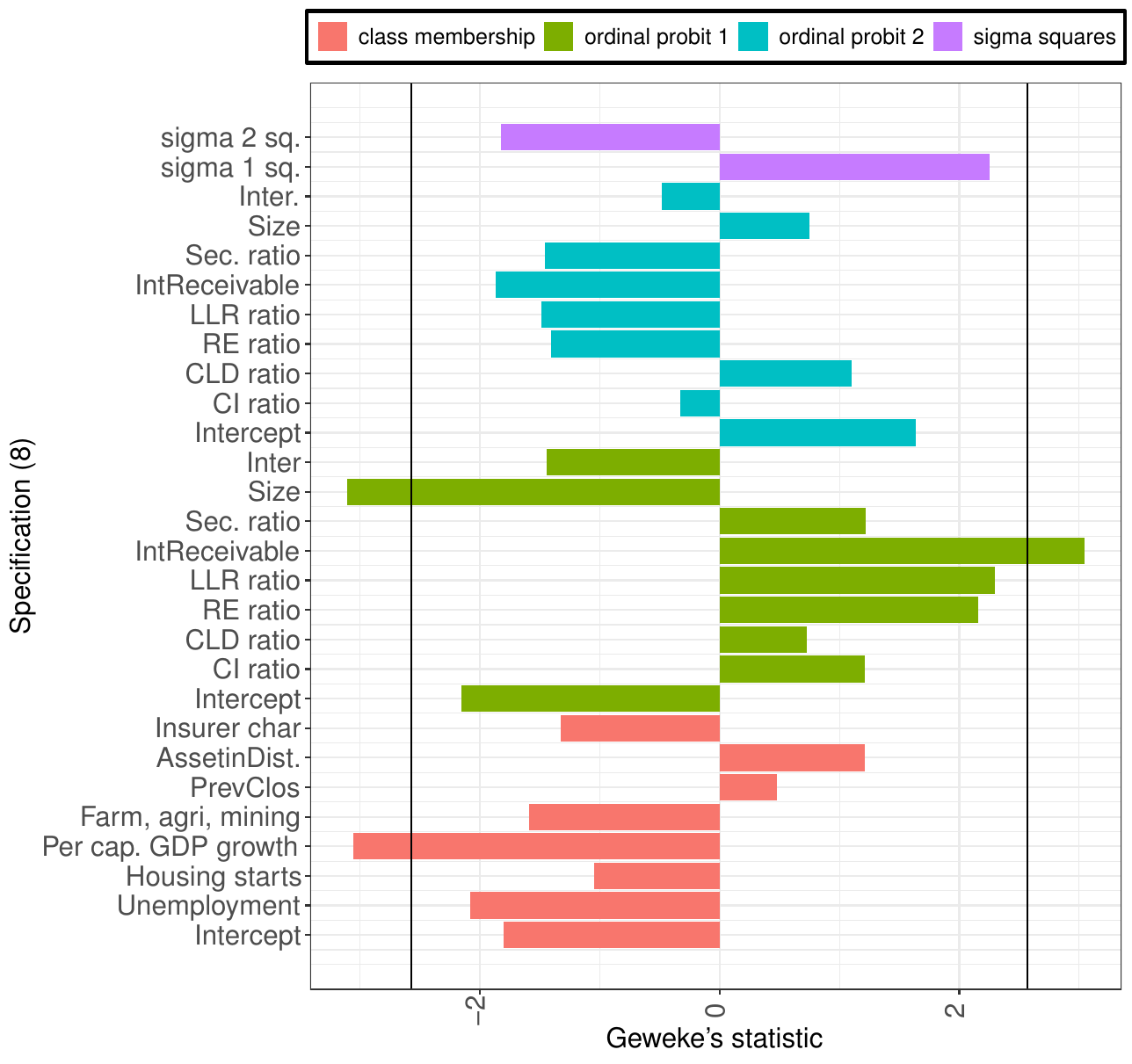}	
\caption{The horizontal bars represent univariate Geweke's convergence diagnostic for the MCMC draws pertaining to each parameter of interest under Specification 8 for FSLIC. The burn-in for these MCMC draws is $5,000$.}
 \label{fig:gweke_thriftsm8_burnin5k}
\end{figure} 
\begin{figure}[!h]
	\centering
\includegraphics[width=0.9\textwidth]{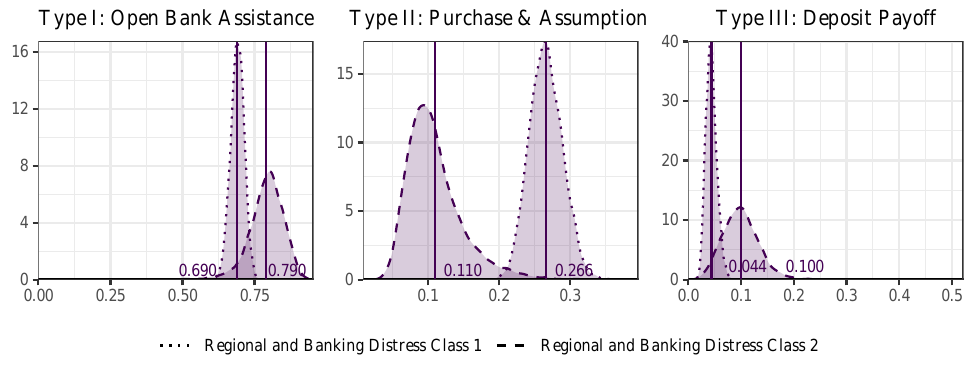}	
\caption{Posterior distribution of the average probability of the FSLIC assigning each resolution method
within classes based on regional and S\&L industry distress. The horizontal axis represents the probability of assigning a resolution method and the vertical axis represents the posterior density associated with that probability based on a kernel density estimate. The solid vertical lines represent the mean of these posterior distributions across the $G=6,000$ post burn-in MCMC draws.}
 \label{fig:posterior_thriftsm8_burnin5k}
\end{figure}
For instance, with a burn-in of $5,000$, Geweke's convergence diagnostic for Specification 8 in Figure \ref{fig:gweke_thriftsm8_burnin5k} suggests substantially improved $Z-$score magnitudes as opposed to the middle panel of Figure \ref{fig:gweke_thrifts}. However, even under this setting our overall conclusion that FSLIC’s resolution decisions do not support Hypothesis $H_2$, does not change 
 since, as observed in Figure \ref{fig:rdb_sl}, Figure \ref{fig:posterior_thriftsm8_burnin5k} reveals that the posterior densities of the average probability of the FSLIC assigning each resolution method to S\&L’s in the two latent classes continue to exhibit substantial overlap, especially in the left and right panels.

\end{document}